\documentclass{article}
\usepackage{fullpage}
\usepackage{graphicx}
\usepackage{amsmath}
\usepackage{amssymb}
\title{A Variable Flavour Number Scheme for Heavy Quark Production at small $x$}
\date{}

\begin{document}
\bibliographystyle{utphys}
\newcommand{\msbar}{\ensuremath{\overline{\text{MS}}}}
\newcommand{\DIS}{\ensuremath{\text{DIS}}}
\newcommand{\abar}{\ensuremath{\bar{\alpha}_S}}
\newcommand{\bb}{\ensuremath{\bar{\beta}_0}}
\setlength{\parindent}{0pt}

\titlepage
\begin{flushright}
NIKHEF/2006-011\\
\end{flushright}

\vspace*{0.5cm}

\begin{center}
{\Large \bf A Global Fit to Scattering Data with NLL BFKL Resummations}

\vspace*{1cm}
\textsc{C.D. White$^{a,}$\footnote{cwhite@nikhef.nl} and R.S. Thorne$^{b,}$\footnote{thorne@hep.ucl.ac.uk}} \\

\vspace*{0.5cm} $^a$ NIKHEF, Kruislaan 409, 1098 SJ Amsterdam, The Netherlands\\

\vspace*{0.5cm} $^b$  Department of Physics and Astronomy, University College London, \\
Gower Street, London, WC1E 6BT, UK
\end{center}

\vspace*{0.5cm}

\begin{abstract}
We calculate DIS-scheme splitting and coefficient functions for electromagnetic deep inelastic scattering with small $x$ resummations, as given by the NLL BFKL equation with running coupling and approximate NLL resummed impact factors. The resummations are combined with a NLO fixed order expansion, and the improved quantities thus obtained are stable at small $x$ and significantly suppressed with respect to LL results. These results are implemented in a global fit to DIS and related data, and the results compared to a NLO fixed order DIS-scheme fit and with a previous LL resummed fit. The NLL resummed fit quality is excellent, and constitutes a marked improvement over the purely fixed order approach. The input gluon, at $Q_0^2=1\,$GeV$^2$, obtained from the resummed fit is positive and slightly increasing as $x\rightarrow0$, in contrast to the result obtained at fixed order. A resummed prediction for the longitudinal structure function $F_L$ is presented. It is positive definite and growing with energy at low $x$ and $Q^2$, where the fixed order results show a significant perturbative instability. 
\end{abstract}

\vspace*{0.5cm}

\section{Introduction}
Current and forthcoming particle collider experiments involve very high energies, such that the momentum fractions $x$ of initial state partons are extremely small. For example, the HERA data on the proton structure function $F_2$ extends to $x\gtrsim 2\times10^{-5}$ \cite{H1a,ZEUSa}. At these values of the Bjorken variable, the DGLAP splitting functions used to evolve the parton distributions, and the coefficient functions used to relate the partons to measurable structure functions, are potentially unstable due to large logarithms of the form $x^{-1}\alpha_S^n\log^m(1/x)$ with $n\geq m+1$ which threaten to undermine the perturbation expansion ordered in fixed powers of $\alpha_S$. In principle one is able to resum these terms by using the BFKL equation \cite{BFKL}, an integral equation for the unintegrated gluon 4-point function whose kernel is currently known to next-to-leading order \cite{Fadin,Camici}. \\

There are several sources of evidence that small $x$ resummations may be necessary when comparing QCD with current scattering data, aside from instability in the coefficient and splitting functions for $F_2$ and $F_L$ (the fixed order results up to NNLO, and to NNNLO for the $F_2$ coefficient functions, 
can be found in \cite{Vogt_c}). Firstly, NNLO global fits seem to benefit from the addition of phenomenological higher order terms involving powers of $\log(1/x)$, whose coefficients are determined by the data \cite{MRSTerrors}. Secondly, the longitudinal structure function obtained from the reduced cross-section measured at HERA appears to be inconsistent with the theoretical prediction using NLO QCD at small $x$ \cite{ThorneFL}, indicating the importance of higher order contributions. We recently carried out a global fit to scattering data at LO in the QCD expansion, supplemented by LL resummations inclusive of running coupling corrections \cite{WT2} following the approach of \cite{Thorne}. The BFKL effects were seen to significantly improve the description of the low $x$ data when compared to a standard NLO $\msbar$-scheme global fit. However, the description of the data at moderately high $x$ was not good due to a strong prevalence of the resummation effects in this region where the DGLAP theory with no modification ought to be reliable.\\

The purpose of this paper is to extend the approach of \cite{Thorne,WT2} to NLL order in the resummation, and to implement the resummed splitting and coefficient functions alongside a NLO QCD expansion. We will see that NLL effects from the BFKL kernel and impact factors suppress the small $x$ divergence in the resummed results. This, together with the correct high $x$ behaviour from the NLO expansion, leads to an excellent description of the data. Whilst the NLL BFKL kernel is known exactly\footnote{Contributions to the kernel arising from heavy quarks are yet to be calculated.}, the NLL resummed impact factors coupling the BFKL gluon to the virtual photon are not (see \cite{Bartels04,Bartels02,Bartels01,Bartels00,Fadin02,Fadin01} for work in progress). These are in principle needed to define the NLL results for the splitting functions $P_{qg}$, $P_{qq}$ as well as longitudinal and heavy flavour coefficient functions\footnote{We are using the convention of \cite{WT2} as to what constitutes NLL order, as opposed to e.g. \cite{Catani2}. In our approach, LL order for a particular quantity consists of those terms without which there would be no resummation e.g. $\alpha_S^{n+1}\log^{n-1}(1/x)$ in $xP_{qg}(\alpha_S,x)$. See \cite{WT2} for a discussion of this point.}. However, the LL impact factors with the imposition of the correct kinematical behaviour of the gluon were calculated in \cite{Peschanski}, and in \cite{WT1} were shown to provide a very good approximation to the true NLL impact factors using a comparison with known results from the fixed order expansion. In \cite{WPT} the exact kinematics calculation was extended as far as possible to the case where the virtual photon couples to a heavy quark pair, thus providing all the ingredients necessary for an approximate NLL analysis of scattering data. \\

The paper is laid out as follows. In section 2 we recall the method of \cite{Thorne} for obtaining small $x$ resummed splitting and coefficient functions with running coupling corrections at NLL order. The method is modified somewhat from that paper, and so the discussion in this paper is intended to be self-contained. We compare our results where possible with alternative approaches \cite{ABF,CCSS}, and in section 3 we consider the generalisation of the variable flavour number scheme of \cite{WT2} for dealing with heavy flavours to NLL order. In section 4 we discuss the details of the global fit, and compare the results of a NLL resummed fit with a NLO DIS-scheme fixed order fit. In section 5 we discuss the resummed prediction for the longitudinal structure function, and finally our conclusions are presented in section 6. 
\section{The BFKL equation at NLL order}
\subsection{Fixed coupling solution}
The NLL BFKL kernel is presented in \cite{Fadin} together with the solution of the BFKL equation with fixed $\alpha_S$. We briefly review this here in order to introduce our notation and also facilitate the comparison with the running coupling case to be discussed shortly. First one can introduce the unintegrated gluon density $f(x,k^2)$, related to the DGLAP gluon $g(x,Q^2)$ by:
\begin{equation}
f(x,Q^2)=x\frac{\partial g(x,Q^2)}{dQ^2}.
\label{unintgluon}
\end{equation}
It is convenient to work in Mellin space with respect to the Bjorken $x$ variable, so that convolutions in $x$ are unravelled to form products. We use the definition:
\begin{equation}
{\mathbb M}_N[f(x,Q^2)]\equiv f(N,Q^2)=\int_0^1 dx\,x^N f(x,Q^2),
\label{Mellin}
\end{equation}
where rather than add a tilde to denote the Mellin space quantity, we denote the arguments of each function explicitly. Then the BFKL equation can be written schematically as:
\begin{equation}
Nf(k^2,Q_0^2)=Nf_I(Q_0^2)+\abar(k^2)\int dk'^2\left[{\cal K}_0(k^2,{k'}^2,Q_0^2)+\abar(\mu^2){\cal K}_1(k^2,{k'}^2,Q_0^2)\right]f(k'^2),
\label{NLLBFKL}
\end{equation}
where $\abar=3\alpha_S/\pi$ and the $x$ dependence of the gluon is implicit. The function $f_I(Q_0^2)$ is the non-perturbative initial condition, and we have taken the gluon at the bottom of the BFKL ladder to be off-shell by an amount $Q_0^2$. The quantities $\{{\cal K}_n\}$ are the coefficients of $\abar^n$ in the expansion of the BFKL kernel. There would be an additional term 
$\propto \alpha_S^2 \ln(k^2/\mu^2) {\cal K}_0$ on the right hand side, but
this can be omitted by choosing $\mu^2=k^2$ in the overall power of $\abar$.  
One may deal with the convolution in transverse momentum by introducing the further Mellin transform:
\begin{equation}
f(N,\gamma)=\int dk^2 (k^2)^{-1-\gamma}f(N,k^2),
\label{Mellin2}
\end{equation}
adopting the conventional definition for the Mellin variable. By making the 
(unwarranted) assumption that the coupling $\abar$ in equation (\ref{NLLBFKL}) can be held fixed, the result for the double Mellin transformed BFKL equation is:
\begin{equation}
Nf(N,\gamma)=Nf_I(N,Q_0^2)+\abar[\chi_0(\gamma)+\abar\chi_1(\gamma)]f(N,\gamma),
\label{NLLBFKL2}
\end{equation}
where $\chi_i(\gamma)$ is the Mellin transform of the kernel ${\cal K}_i(k_1^2,k_2^2)$. Thus the fixed coupling BFKL equation reduces in double Mellin space to an algebraic equation, which is easily solved to give:
\begin{equation}
f(N,\gamma)=\frac{Nf_I(N,Q_0^2)}{N-\abar[\chi_0(\gamma)+\abar\chi_1(\gamma)]}.
\label{NLLfixed}
\end{equation}
The behaviour in $x$-space is given after the inverse Mellin transformation:
\begin{equation}
f(x,\gamma)=\frac{1}{2\pi\imath}\int_C dN\frac{Nx^{-N}f_I(N,Q_0^2)}{N-\abar\chi(\gamma)},
\label{Melinvx}
\end{equation}
where $C$ is a vertical contour to the right of all singularities in the complex $N$ plane, and $\chi(\gamma)$ is the BFKL kernel truncated at the required order. Assuming the perturbative pole given by the denominator of equation (\ref{Melinvx}) dominates over possible non-perturbative singularities in the initial condition $f_I(N,Q_0^2)$, one has the leading behaviour as $x\rightarrow 0$:
\begin{equation}
f(x,\gamma)=N_0f_I(N_0,Q_0^2)x^{-\alpha_S\chi(\gamma)}
\label{invmel2}
\end{equation}
with $N_0=\alpha_S\chi(\gamma)$. The behaviour in momentum space is obtained after a second inverse Mellin transformation, which may be written as:
\begin{equation}
f(x,k^2)=\frac{1}{2\pi\imath}\int_{C'}d\gamma\exp\left[\abar\chi(\gamma)\ln\frac{1}{x}+\gamma\ln\frac{k^2}{Q_0^2}\right]N_0,
\label{invmel3}
\end{equation}
where $C'$ is a contour to the right of all singularities in the $\gamma$ plane. For very small $x$, the first term in the exponent will dominate, and in the LL case ($\chi(\gamma)=\chi_0(\gamma)$) one may evaluate the integral in equation (\ref{invmel3}) using a saddle point approximation to obtain the well-known result:
\begin{equation}
f(x,k^2)\sim x^{-4\abar\ln{2}}\left(\frac{k^2}{Q_0^2}\right)\exp\left[-\frac{\log^2(k^2/Q_0^2)}{56\abar\zeta(3)\ln(1/x)}\right].
\label{fxk^2}
\end{equation}
Thus, modulo logarithms in the normalisation, the BFKL gluon density has a power-like growth in $x$ at LL order where the exponent arises from the minimum of the kernel $\chi_0$ at $\gamma=1/2$. Carrying out the same calculation at NLL order, the power in equation (\ref{fxk^2}) receives a large correction of opposing sign. For $n_f=0$ (no active quark flavours):
\begin{equation}
\chi(1/2)=4\abar\ln(2)[1-6.47\abar+\ldots],
\label{chiNLLpow}
\end{equation}
where the ellipsis denotes higher order contributions from the kernel, which will certainly not be known exactly in the near future. For phenomenologically reasonable values of the coupling, the sign of the power-like growth in $x$ of the gluon is changed and hence it was initially concluded that the BFKL expansion is unstable. Instead, the saddle point calculation is not reliable. The shape of the LL+NLL kernel $\chi(\gamma)$ is completely different to the LL result, with two saddle points at complex values of $\gamma$, rather than a single saddle point at $\gamma=1/2$ \cite{Ross}. \\

There are two main solutions to this problem. Firstly, the large correction induced by the NLL kernel can be attributed to poles in $\gamma$, $1-\gamma$ which correspond to collinear logarithms of type $\alpha_S^n\log^m{Q^2/Q_0^2}$ ($n\geq m$) in the splitting functions obtained from the DGLAP gluon density in the limits $Q^2\gg Q_0^2$, $Q_0^2\gg Q^2$. It is possible to resum these contributions \cite{Salam} and thus obtain a kernel with a stable minimum at NLL order at $\gamma=1/2$. Secondly, the calculation above ignores the fact that the QCD coupling $\alpha_S$ runs with energy, which must be taken into account beyond LL order in the resummation. This changes the nature of the BFKL equation so that it is a differential equation as opposed to a purely algebraic one. We will see in this paper (as has already been observed in \cite{Thorne}) that in solving this equation one is no longer concerned with the behaviour of the kernel at $\gamma=1/2$. Alternative approaches to small $x$ resummation of splitting functions \cite{ABF},\cite{CCSS} involve combining the resummation of collinear poles with inclusion of the running coupling. We do not find it necessary to resum the kernel, and consider the solution of the running coupling BFKL equation at NLL order in the following subsection.
This is precisely because the deep inelastic scenario where $Q^2\gg Q_0^2$
focuses on the region $\gamma =0$. Single scale processes, where there is a large scale at  both ends of the gluon ladder, are a different matter.

\subsection{Running Coupling Solution}
Taking $\alpha_S(k^2)$ as the coupling in equation (\ref{NLLBFKL}), one may substitute the LO expression for the coupling:
\begin{equation}
\frac{\alpha_s(k^2)}{4\pi}=\frac{1}{\beta_0\ln(k^2/\Lambda^2)}
\label{LOcoupling}
\end{equation}
(where $\beta_0=(11-2n_f/3)$) and multiply through by $\ln^2(k^2/\Lambda^2)$ before Mellin transforming in transverse momentum to obtain the second order differential equation:
\begin{equation}
\frac{d^2f(\gamma,N)}{d\gamma^2}=\frac{d^2f_I(\gamma,Q_0^2)}{d\gamma^2}-\frac{1}{\bb N}\frac{d(\chi_0(\gamma)f(\gamma,N))}{d\gamma}+\frac{\pi}{3\bb^2 N}\chi_1(\gamma_N)f(\gamma,N),
\label{BFKLNLLrun}
\end{equation}
introducing $\bb=\pi\beta_0/3$. A comment is in order regarding the use of the LO running coupling. Given that our aim is to build a NLL resummation upon a NLO fixed order expansion, one should in principle use the NLO running coupling in the BFKL equation. This leads to a more complicated equation in momentum space, which is not diagonalised by a Mellin transform. Indeed, it is not clear how an analytic solution can proceed in this case. The use of the LO coupling can be justified given that differences between the LO and NLO couplings can be absorbed by a change in $\Lambda$. We discuss in more detail later how to combine resummed quantities with the fixed order expansion. \\

Equation (\ref{BFKLNLLrun}) may be solved by making the ansatz \cite{Colferai}:
\begin{equation}
f(N,\gamma)=\exp\left(-\frac{X_1(\gamma)}{\bb N}\right)\int_\gamma^\infty A(\tilde{\gamma})\exp\left(\frac{X_1(\tilde{\gamma})}{\bb N}\right)d\tilde{\gamma},
\label{ansatz}
\end{equation}
where:
\begin{equation}
X_1=\int_{\frac{1}{2}}^\gamma\chi_{NLO}(\tilde{\gamma},N)d\tilde{\gamma},
\label{X1}
\end{equation}
and $\chi_{NLO}$ and $A(\gamma)$ are to be determined. Substituting this into equation (\ref{BFKLNLLrun}) and choosing $A(\gamma)$ via:
\begin{equation}
\frac{dA(\gamma)}{d\gamma}=\frac{A(\gamma)[\chi_{NLO}(\gamma,N)-\chi_0(\gamma)]}{\bb N}-\frac{d^2\tilde{f}_I(\gamma)}{d\gamma^2},
\label{dadgamma}
\end{equation}
one finds:
\begin{equation}
\bb N\frac{\partial\chi_{NLO}}{\partial\gamma}-(\chi_{NLO}-\chi_0)\chi_{NLO}=N[\bb{\chi'}_0(\gamma)-\chi_1(\gamma)],
\label{diffchi}
\end{equation}
where a factor of $\pi/3$ has been absorbed into $\chi_1(\gamma)$. This is a first order nonlinear differential equation for $\chi_{NLO}$. It may be solved iteratively by expanding $\chi_1(\gamma)$ in powers of $N$ (the ``$\omega$ expansion'' of \cite{Colferai} in their notation):
\begin{equation}
\chi_{NLO}(\gamma,N)=\sum_{n=0}^\infty \xi_n(\gamma)N^n.
\label{chinlopow}
\end{equation}
One then gets:
\begin{align}
\xi_0(\gamma)&=\chi_0(\gamma);\notag\\
\xi_1(\gamma)&=\frac{\chi_1(\gamma)}{\chi_0(\gamma)};\notag\\
\xi_2(\gamma)&=\frac{1}{\chi_0(\gamma)}\left[-\left(\frac{\chi_1(\gamma)}{\chi_0(\gamma)}\right)^2+\bb\left(\frac{\chi_1(\gamma)}{\chi_0(\gamma)}\right)'\right].
\label{xi01}
\end{align}
Note the term involving $\beta_0$ in $\xi_2$. This is the first of a series of terms $\sim\bb^{n-1} N^n$ in the $N$-expansion, arising in the iterative solution of equation (\ref{diffchi}) from:
\begin{equation}
\xi_n\sim\left(\frac{\bb}{\xi_0}\frac{d}{d\gamma}\right)^{n-1}\xi_1+\ldots,
\label{itdiff}
\end{equation}
where the ellipsis denotes sub-leading running coupling corrections. A decision must now be made as to what constitutes NLL order. It could be argued that one must keep the whole of equation (\ref{chinlopow}) in defining the NLL gluon density, which arises from a complete solution of the NLL BFKL equation. However, this is impossible to achieve in practice due to an infinite number of terms. Instead, one might consider truncating at ${\cal O}(N)$ and keeping the set of leading running coupling corrections given by equation (\ref{itdiff}). Again one has the problem of infinitely many terms, and in fact they should not be included, which can be seen as follows \cite{Colferai}. Expanding in $\gamma$ each of the coefficients of the $N$-expansion one has:
\begin{equation}
\chi_{NLO}(\gamma,N)=\sum_{n=1}^{\infty}\frac{k_n N^n}{\gamma}+{\cal O}(\gamma^0)
\label{chin}
\end{equation}
and we will see that this leads to a gluon density of form:
\begin{equation}
{\cal G}\propto t^{1-\sum k_nN^n}
\label{gluprop}
\end{equation}
with $t=\log(k^2/\Lambda^2)$. This will lead to a contribution to the LO anomalous dimension $P_{gg}$ of form:
\begin{equation}
P_{gg}^{(0)}\sim \alpha_S\left(1-\sum k_nN^n\right)
\label{pgg0prop}
\end{equation}
where if the above-mentioned terms are kept, $k_n\propto\bb^{n-1}$. We know, however, that the LO anomalous dimension does not contain higher order terms in $\beta_0$, and so the terms of form $\bb^{n-1}N^n$ in the $N$-expansion must somehow be cancelled. To see how this works one must consider the BFKL equation at higher orders in the high energy expansion. If one expands the kernel to NNLL order, the BFKL equation will become a third order differential equation in Mellin space due to the extra power of $\alpha_S$ in the kernel (assuming the LO coupling is still used). The Mellin transformed NNLL kernel $\chi_2(\gamma)$ contributes to $\xi_2(\gamma)$ in such a way as to cancel the term in $\beta_0$. Similarly, the Mellin space BFKL kernel at N$^n$LL order in double Mellin space is an $(n+1)^{th}$ order differential equation. The higher order kernels contribute to the coefficients of the $N$-expansion in such a way as to cancel the running coupling terms discussed above, as must be the case.\\

Thus, we now have the gluon density given by equation (\ref{ansatz}), where the series for $\chi_{NLO}$ is truncated at ${\cal O}(N)$. For $\vert \gamma 
\vert < 1$ one may expand $X_1$ as:
\begin{equation}
X_1(\gamma,N)=\left(\log{\gamma}+\gamma_E+\sum a_n\gamma^{2n+1}\right)-N\left(c_l\log{\gamma}+c_0+\sum_{n=1}^\infty c_n\gamma^n\right),
\label{X1exp}
\end{equation}
where $\gamma_E$ is Euler's constant, $a_n=2\zeta(2n+1)/(2n+1)$ and the $\{c_i\}$ follow from the NLL BFKL kernel \cite{Thorne}. Thus the integral on the right-hand-side of equation (\ref{ansatz}) has behaviour $\sim \gamma^{(1-c_l\bb N)/\bb N}$ as $\gamma\rightarrow0$. Hence, an integral of this from $0$ to 
$\gamma$ behaves like $\sim \gamma^{1+(1-c_l\bb N)/\bb N}$, cancelling the 
singularity of the prefactor. The leading singularity of the gluon from
this contribution is then at $\gamma=-1$, and so up to power-suppressed 
corrections $\sim\Lambda^2/Q^2$ in momentum space one may shift the lower 
limit of the integral from $\gamma\rightarrow0$ leading to factorisation 
of the gluon density into a perturbatively calculable piece (the prefactor in equation (\ref{ansatz})) and a non-perturbative piece which ultimately is combined with the bare gluon density in the proton. Transforming back to momentum space, the perturbative piece of the integrated gluon density is:
\begin{equation}
{\cal G}_E^1(N,t)=\frac{1}{2\pi\imath}\int_{1/2-\imath\infty}^{1/2+\imath\infty}\frac{f^{\beta_0}}{\gamma}\exp\left[\gamma t-X_1(\gamma,N)/(\bb N)\right]d\gamma
\label{GE1}
\end{equation}
where the inverse power of $\gamma$ comes from considering the integrated gluon, and we have used the same notation as \cite{Thorne}. The factor $f^{\beta_0}$ in equation (\ref{GE1}) is given by:
\begin{equation}
f^{\beta_0}=\exp\left[\int_{1/2}^\gamma\frac{\bb}{2}\left(\chi_0^2(\tilde{\gamma})+\chi'_0(\tilde{\gamma})\right)d\tilde{\gamma}\right]
\label{fbeta0}
\end{equation}
and corresponds to the contribution in the exponent of equation (\ref{GE1}) from the running coupling contribution to the NLL kernel, effectively providing the corrections required from making the simple choice of scale $k^2$ in the 
coupling. We factor this out explicitly following \cite{Thorne}. From now on, the notation $\chi_1(\gamma)$ refers to the kernel without this contribution. \\

\subsection{Solution for the BFKL Anomalous Dimension}
The solution for the perturbative piece of the integrated gluon density of equation (\ref{GE1}) can be evaluated by numerical integration. Alternatively, one may obtain an analytic expansion in momentum space by substituting the expansion of equation (\ref{X1exp}) into equation (\ref{GE1}) and deforming the Mellin inversion contour to enclose the negative real axis. One obtains (taking into account the discontinuity across the cuts on the real axis):
\begin{align}
{\cal G}_E(N,t)&=-\sin\left(\frac{\pi(1-c_lN)}{\bb N}\right)\exp\left(-\frac{\gamma_E+c_0}{\bb N}\right)\int_{-\infty}^0 f^{\beta_0}(\gamma)\gamma^{-(1-c_lN)/(\bb N)-1}\notag\\
&\quad \times \exp\left[\gamma t-\frac{1}{\bb N}\sum_{n=1}^\infty\left(a_n\gamma^{2n+1}+Nc_n\gamma^N\right)\right]d\gamma.
\label{GE12}
\end{align}
Substituting $y=\gamma t$ and using the result:
\begin{align}
\int_{-\infty}^0 dy y^{\lambda-1}\exp(y)y^n&=(-1)^{\lambda-1+n}\Gamma(\lambda+n)\\
&=(-1)^{\lambda-1}\Delta_n(\lambda)\Gamma(\lambda)\label{deltandef}
\end{align}
after Taylor expansion of the integrand, one obtains:
\begin{align}
{\cal G}_E^1(N,t)&=\sin\left(\frac{\pi(1-c_lN)}{\bb N}\right)\exp\left(-\frac{\gamma_E+c_0N}{\bb N}\right)\Gamma\left(-\frac{(1-c_lN)}{\bb N}\right)t^{(1-c_lN)/(\bb N)}\notag\\
&\times\left\{1+\sum_{n=1}^{\infty}\left[\tilde{A}_n(1/(\bb N))+\frac{C_n(1/\bb)}{\bb}\right]t^{-n}\Delta_n\left(-\frac{1-c_lN}{\bb N}\right)\right.\notag\\
&\left.+\sum_{n=1}^\infty\sum_{m=1}^\infty\frac{\tilde{A}_n(1/(\bb N))C_m(1/\bb)}{\bb}t^{-n-m}\Delta\left(-\frac{1-c_lN}{\bb N}\right)\right\},
\label{g1sol}
\end{align}
with the coefficients $\tilde{A}_n$ and $C_m$ defined by:
\begin{align}
1+\sum_{n=1}^\infty \tilde{A}_n(1/(\bb N))\gamma^n&=f^{\beta_0}\exp\left(\frac{1}{\bb N}\sum_{n=1}a_n\gamma^n\right);\label{tildean}\\
1+\sum_{m=1}^\infty C_m(1/\bb)\gamma^m&=\exp\left(\frac{1}{\bb}\sum_{n=1}^\infty c_n\gamma^n\right),
\label{Cm}
\end{align}
The $t$-independent pre-factors in equation (\ref{g1sol}) can be absorbed into the non-perturbative gluon density. From equation (\ref{deltandef}) one may write for the functions $\{\Delta_n\}$ defined in equation (\ref{deltandef}):
\begin{equation}
\Delta_n[(1-c_lN)/(\bb N)]=\sum_{m=0}^{n-1}(-1)^md_{m,n}(\bb N)^{-n+m},
\label{deltanexp}
\end{equation}
with $d_{0,n}=1$ and $d_{m,n+1}=d_{m,n}+nd_{m-1,n}$. Setting $m=n$ in this recurrence relation yields $d_{n,n+1}=nd_{n-1,n}$ and hence:
\begin{equation}
d_{n-1,n}=(n-1)!,
\label{dnn-1}
\end{equation}
and so the series in equation (\ref{g1sol}) has factorially divergent coefficients at high $n$. Thus the series is asymptotic, and must be truncated\footnote{The series is, however, summable in principle due to the oscillatory factor $(-1)^m$ in equation (\ref{deltanexp}).} at some finite order $n=n_0$. In the LL case of \cite{Thorne}, $n_0=5$ was found to be sufficient. 

\begin{figure}
\begin{center}
\scalebox{0.8}{\includegraphics{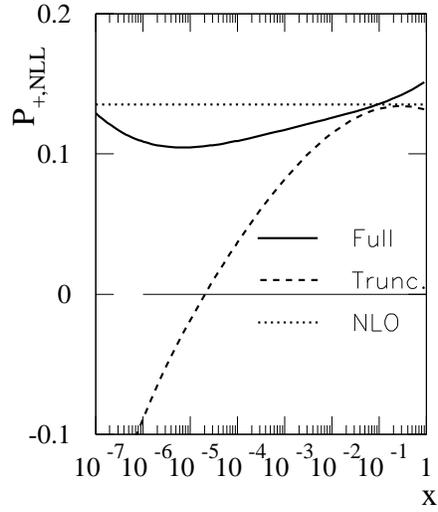}}
\caption{Power-series result for the BFKL anomalous dimension $P_{+,NLL}$ for $t=10$, $n_f=4$ obtained from using the full gluon solution of equation (\ref{g1sol}), together the result obtained after truncating to formal NLL order and the asymptotic NLO (fixed order in $\alpha_S$) result.}
\label{NLLsplits}
\end{center}
\end{figure}

There are other choices that must be made regarding what is meant by NLL order, and in \cite{Thorne} it is noted that equation (\ref{g1sol}) contains NNLL terms coming from the following origins:
\begin{itemize}
\item The $C_m(\bb)$ can be expanded as a power series in $\bb^{-1}$ where terms of ${\cal O}(\bb^{-2})$ are really NNLL contributions.
\item The $\Delta_n((1-c_lN)/(\bb N))$ functions in equation (\ref{g1sol}) can be expanded as a power series in $N$, where terms ${\cal O}(N^2)$ are NNLL contributions. Using Taylor's theorem one may write:
\begin{equation}
\Delta_n\left(-\frac{(1-c_lN)}{\bb N}\right)=\Delta_n\left(-\frac{1}{\bb N}\right)+\frac{c_l}{\bb}\frac{d\Delta_n(-1/\bb N)}{d(-1/\bb N)}+\ldots,
\label{Deltatrunc}
\end{equation}
\end{itemize}
where the ellipsis denotes NNLL terms. In \cite{Thorne} these were discarded, although this is not strictly necessary as one is free to modify a NLL resummation up to NNLL terms. Indeed, making the truncations advocated in \cite{Thorne} leads to resummed splitting functions that are significantly different from the fixed order NLO results at moderate $x$, particularly $P_{qg}$ which shows a pronounced dip \cite{Whiteconf}. Indeed, consideration of the anomalous dimension $\gamma_{+}$ shows that this truncation is incorrect, as described below. Another advantage of keeping the full arguments of the $\Delta_n$ functions is that this argument is very small at $N=1$. When $n_f=4$ for example:
\begin{equation}
-\frac{1-c_lN}{\bb N}=-\frac{1}{\bb N}(1-0.941N)=0.08,\quad N=1.
\label{argN1}
\end{equation}
The splitting functions obtained from the gluon density will also have a small first moment, and hence satisfy the momentum sum rule to a very good approximation. Thus we choose to define the gluon density by equation (\ref{g1sol}) with no further truncation.  The BFKL anomalous dimension (inclusive of running coupling effects) is now given in principle by:
\begin{equation}
\gamma_{+,NLL}=\frac{1}{{\cal G}_E^1(N,t)}\frac{\partial{\cal G}_E^1(N,t)}{\partial t},
\label{gamggdef}
\end{equation}
which corresponds to the leading eigenvalue of the singlet anomalous dimension matrix:
 \begin{equation}
\gamma_+^{NLL}=\gamma_{gg}+\frac{C_F}{C_A}\gamma_{qg}
\label{gam+NLL}
\end{equation}
at high energy\footnote{The alternative approaches \cite{ABF,CCSS} present results for $\gamma_{gg}$. However, these are in the limit $n_f\rightarrow 0$ where one has $\gamma_+\rightarrow\gamma_{gg}$.}. There is now a further choice that has to be made in defining the anomalous dimension. Decomposing the gluon into LL and NLL parts\footnote{The NLL piece contains some higher order terms than strictly NLL order as discussed after equation (\ref{g1sol}).} as ${\cal G}_E^1={\cal G}_E^{LL}+{\cal G}_E^{NLL}$, one has:
\begin{align}
\gamma_{+,NLL}&=\frac{1}{{\cal G}_E^{LL}+{\cal G}_E^{NLL}}\frac{\partial}{\partial t}\left[{\cal G}_E^{LL}+{\cal G}_E^{NLL}\right]\notag\\
&=\left(1+\frac{{\cal G}_E^{NLL}}{{\cal G}_E^{LL}}\right)^{-1}\frac{1}{{\cal G}_E^{LL}}\frac{\partial}{\partial t}\left[{\cal G}_E^{LL}+{\cal G}_E^{NLL}\right].
\label{gam+decomp}
\end{align}
The question now arises: should one keep the whole of the first factor on the right-hand-side, or expand it to NLL order? The latter choice yields:
\begin{equation}
\gamma_{+,NLL}=\left(1-\frac{{\cal G}_E^{NLL}}{{\cal G}_E^{LL}}\right)\frac{1}{{\cal G}_E^{LL}}\frac{\partial}{\partial t}\left[{\cal G}_E^{LL}+{\cal G}_E^{NLL}\right],
\label{gam+decomp2}
\end{equation}
and is a further truncation in addition to those already described in the 
gluon density. The poles in equation (\ref{gam+decomp2}) now only come from 
the zeros in ${\cal G}_E^{LL}$ and  
this ultimately leads to a corrected splitting function behaving like:
\begin{equation}
P_{+,NLL}\sim x^{\eta}[1-\Delta\eta\log(1/x)]
\label{powshift2}
\end{equation}
rather a shifted power-like behaviour:
\begin{equation}
P_{+,NLL}\sim x^{\eta+\Delta\eta},
\label{powshift}
\end{equation}
as argued in \cite{Thorne}. The latter behaviour is physically more sensible, and the argument can be made more compelling by examining the splitting functions obtained using equations (\ref{gam+decomp},\ref{gam+decomp2}). These are shown in figure \ref{NLLsplits}, where the truncated splitting function in this plot corresponds to a full removal of NNLL and higher terms (apart from running coupling corrections) after using equation (\ref{gam+decomp2}) to expand the gluon denominator. There is a marked difference in asymptotic behaviour. 
The NLL result with no truncations dips gently under the NLO fixed order result, ultimately rising at low $x$. The truncated expression decreases far below the fixed order result and never rises at small $x$, consistent with a shift in behaviour given by equation (\ref{powshift2}). Thus to obtain the correct physical behaviour for the structure function, one must use the complete denominator of equation (\ref{gamggdef}).\\

The intermediate possibility exists of truncating the solution (\ref{g1sol}) for the gluon, but still using the complete gluon density in the denominator for the anomalous dimension. However, this introduces spurious terms involving inverse powers of $\bb$, even at orders below that corresponding to the truncation of the gluon density in $\gamma$-space. Such terms are non-physical and can be avoided as far as possible by not truncating equation (\ref{g1sol}). The $\bb^{-1}$ terms introduced by expanding the denominator ultimately cancel with those obtained by keeping the full $\bb$ dependence in the $C_m(\bb)$ coefficients discussed previously. \\ 

There is a power-suppressed correction to the anomalous dimension obtained from equations (\ref{g1sol},\ref{gamggdef}), arising from the fact that the expansion of $X_1(\gamma,N)$ is only valid for $\gamma>-1$ and also from the truncation of the asymptotic series for $X_1(\gamma)$\footnote{This is also the case at LL order with running coupling. See \cite{Thorne}.}. To evaluate this correction, one must evaluate the anomalous dimension from numerical integration of equation (\ref{GE1}). For this purpose the gluon density can be rewritten as:
\begin{equation}
{\cal G}_E^1(N,t)=\frac{1}{2\pi\imath}\int_{1/2-\imath\infty}^{1/2+\imath\infty}d\gamma f^{\beta_0}(\gamma)\gamma^{-1+c_l/\bb}\exp\left[\gamma t-\frac{1}{\bb N}X_0(\gamma)+\frac{1}{\bb}\tilde{X}_1(\gamma)\right],
\label{g1sol2}
\end{equation}
where $X_0(\gamma)=\int_{1/2}^\gamma\chi_0(\tilde{\gamma})d\tilde{\gamma}$ and $\tilde{X}_1(\gamma)$ is that part of $X_1(\gamma)$ that gives $\sum_{n=1}c_n\gamma^n$ when Taylor expanded. A problem arises in implementing this correction, in that a closed result for $\tilde{X}_1(\gamma)$ does not exist as one cannot integrate $\xi_1(\gamma)$ of equation (\ref{xi01}) analytically. Given that the correction is found numerically, however, it is sufficient to parameterise $\xi_1$ along the integration contour. We use the contour:
\begin{equation}
\gamma=\left\{\begin{array}{c}0.4-(\imath-1)w,\quad -\infty<w<0\\0.4+(\imath-1)w,\quad 0\leq w<\infty\end{array},\right.
\label{contourNLL}
\end{equation}
which is a deformation of the Mellin inversion contour that encounters no singularities. In practice it is sufficient to parameterise $\xi_1(\gamma)$ for $0\leq w<3$, and then one has along the upper branch of the contour:
\begin{equation}
X_1(w)=(I-1)\int_0^w \xi_1[0.4+(\imath-1)\tilde{w}]d\tilde{w}.
\label{X1w}
\end{equation}
Then $\tilde{X}_1(w)$ is given along this contour by:
\begin{equation}
\tilde{X}_1(w)=X_1(w)-c_l[\log(0.4+(\imath-1)w)-\log(0.4)]-c_0,
\label{tildeX1}
\end{equation}
where:
\begin{equation}
c_0=\int_{0.4}^0\xi_1(\gamma)d\gamma
\label{c0}
\end{equation}
which can be integrated numerically. After evaluating ${\cal G}_E^1$ and $\partial{\cal G}_E^1/\partial t$, one generates the anomalous dimension according to equation (\ref{gamggdef}). We find that the correction to the power series result is very large. It can be stabilised somewhat by taking fewer orders of $\gamma$ in the expansion of the anomalous dimension in double Mellin space (i.e. $n_0=4$ instead of $n_0=5$), but is still comparable with the power-series result for $t\lesssim11$. Thus, instead of parameterising the correction as in \cite{Thorne}, we choose to parameterise the anomalous dimension directly at lower $t$ values. Using the integral representation of equation (\ref{g1sol2}), one can obtain values of the splitting functions for particular values of  $N$ and $t$ and fit these to a function of form:
\begin{equation}
P(N,t)=\sum_{m=-4}^1\sum_{n=0}^{n_0} p_{nm}\frac{t^{\pm n}}{N^m},
\label{pparam}
\end{equation}
where negative powers of $t$ happen to achieve a better fit for $P_+$ and positive powers improve the fit for other quantities to be discussed in the next section. Several ranges of $t$ are needed to achieve a good fit, and one must also calculate separate parameterisations for each value of $n_f$, due to the dependence on the number of flavours in the NLL BFKL kernel. At high enough $t$ one may use the power series result for the splitting functions. The power series results must be expanded to many orders of $\alpha_S$ (around 20) and due to their length are not presented here.
\subsection{Inclusion of Impact Factors}
\label{impacts}
In order to complete the phenomenology at NLL order for massless quarks, one must calculate the resummed splitting function $P_{qg}$. One also needs the coefficient function $C_{Lg}$ to obtain a prediction of the longitudinal structure function. Thus one needs the impact factors coupling the virtual photon to the gluon evaluated at NLL order. These are not yet available but one may proceed by using the LL factors calculated with the imposition of exact gluon kinematics \cite{Peschanski}, shown in \cite{WT1} to be a good approximation to the complete higher-order results. Their interpretation in terms of physical quantities stems from the $k_T$-factorisation theorem \cite{CollinskT,CatanikT}, which in double Mellin space dictates for the longitudinal structure function:
\begin{equation}
F_L(\gamma,N)=h_L(\gamma,N)g(\gamma,N),
\label{FLkT}
\end{equation}
where the gluon density $g(\gamma,N)$ factorises further into perturbative and non-perturbative contributions as discussed above. From equation (\ref{FLkT}) one may straight-forwardly associate the impact factor $h_L$ with the longitudinal coefficient function $C_{L,g}$ in double Mellin space, where this coefficient is in the ``$Q_0$'' factorisation scheme where the gluon is obtained by solution of the BFKL equation with an off-shell non-perturbative initial condition\footnote{The ``$Q_0$ scheme'' is widely used in the literature to denote a gluon arising from solution of the BFKL equation, whether or not an off-shell regularisation is used in practice. It is thus a type of scheme rather than a specific scheme, due to the number of choices to be made when solving the BFKL equation.}. The case of $F_2$ is not so straightforward. The quantity analogous to $h_L$ in equation (\ref{FLkT}) is the sum of impact factors $h_T(\gamma,N)+h_L(\gamma,N)$ arising from transversely and longitudinally polarised photons respectively, and this has the leading behaviour $\gamma^{-1}$ as $\gamma\rightarrow 0$, corresponding to a collinear divergence of the momentum space impact factor. Put another way, the structure function $F_2$ is proportional to the (non-perturbative) quark singlet distribution at LO in the QCD expansion, and one does not expect perturbation theory to be able to describe this behaviour. Instead one may consider the quantity:
\begin{equation}
{\mathbb M}_{N,\gamma}\left[\frac{\partial F_2}{\partial\log{Q^2}}\right]=h_2(\gamma,N)g(\gamma,N),
\label{F2kT}
\end{equation}
which starts at ${\cal O}(\alpha_S)$ in perturbation theory. The left-hand side denotes the double Mellin transform, and equation (\ref{F2kT}) serves to define the impact factor $h_2$. The interpretation of this quantity depends on the factorisation scheme, and in general represents a mixture of the coefficient function $C_{2,g}(\gamma,N)$ and the anomalous dimension $\gamma_{qg}$. If one chooses the DIS factorisation scheme \cite{Altarelli}, one has the simple identification of $h_2$ with $\gamma_{qg}^{DIS}$ in double Mellin space. \\

The exact kinematics impact factors of \cite{Peschanski} can be expanded in $N$:
\begin{equation}
h_a(\gamma,N)=h_a^{(0)}(\gamma)+Nh_a^{(1)}(\gamma)+{\cal O}(N^2),
\label{hexp}
\end{equation}
corresponding to truncation at NLL order. Then equation (\ref{g1sol2}) can be generalised to give:
\begin{equation}
{\cal F}_E^1(N,t)=\frac{1}{2\pi\imath}\int_{1/2-\imath\infty}^{1/2+\imath\infty}d\gamma h_a(\gamma,N)f^{\beta_0}(\gamma)\gamma^{-1+c_l/\bb}\exp\left[\gamma t-\frac{1}{\bb N}X_0(\gamma)+\frac{1}{\bb}\tilde{X}_1(\gamma)\right].
\label{F1sol2}
\end{equation}
Proceeding analytically as for the gluon density one obtains for the structure function:
\begin{align}
{\cal F}_E^1(N,t)&\propto t^{(1-c_lN)/(\bb N)}\sum_{r=0}^\infty\left[h_{a,r}^{(0)}+Nh_{a,r}^{(1)}\right]\notag\\
&\times \left\{1+\sum_{n=1}^\infty\left[\tilde{A}_n(1/(\bb N))+\frac{C_n(1/\bb)}{\bb}\right]t^{-n}\Delta_{n+r}\left(-\frac{1-c_lN}{\bb N}\right)\right.\notag\\
&\left.+\sum_{n=1}^\infty\sum_{m=1}^\infty\frac{\tilde{A}_n(1/(\bb N)) C_m(1/\bb)}{\bb}t^{-n-m}\Delta_{n+m+r}\left(-\frac{1-c_lN}{\bb N}\right)\right\},
\label{f1sol3}
\end{align}
where the impact factors have been expanded as follows:
\begin{equation}
h_a(\gamma,N)=\sum_{r=0}^\infty \left[h_{a,r}^{(0)}+Nh_{a,r}^{(1)}\right]\gamma^r.
\label{haexp}
\end{equation}
Note that there are NNLL contributions in equation (\ref{f1sol3}) arising from the product of the ${\cal O}(N)$ piece of the impact factor with NLL terms from the gluon. This is not a problem in that they are beyond the NLL nature of the resummation. Truncation of the structure function results, as in the purely gluonic case, in unphysical terms involving inverse powers of $\bb$. In order to avoid such contributions, I choose not to truncate equation (\ref{f1sol3}). Once again the full argument of the ${\Delta_n}$ functions is kept, which will ensure that momentum is conserved by the resummed splitting functions to a very good approximation.\\

Note that the inclusion of impact factors has led to terms $\sim{\cal O}(N)$ in the structure functions, which consequently lead to similar terms in the quantities $P_{qg}$, $C_{Lg}$ in addition to ${\cal O}(N^0)$ terms which are already present at LL order (see \cite{Thorne}).  In $x$-space, these correspond to the Dirac function $\delta(1-x)$ and its first derivative. However, implementation of $\delta'(1-x)$ is numerically cumbersome and is somewhat unnatural given that there are no real terms proportional to $\delta'(1-x)$ in the fixed order splitting and coefficient functions. The ${\cal O}(N)$ terms are an artifact of expanding about $N=0$ in the resummation. Instead of implementing them as Dirac functions, one may choose a suitable function whose expansion about $N=0$ has similar properties.\\

In more detail, consider a resummed splitting or coefficient function that has form:
\begin{equation}
P=\delta_1(t)N+\delta_0(t)+{\cal O}(N^{-1}).
\label{P}
\end{equation}
One wishes to replace the Dirac terms of $P$ with a function whose Taylor expansion about $N=0$ reproduces these first two terms. Furthermore, it is important to preserve the first moment (value at $N$=1) of a splitting function to ensure compliance with the momentum sum rule. We have already seen that the power series results for the splitting functions ought to have a very low first moment, and this property must be preserved when implementing the Dirac terms. One thus has three pieces of information with which to construct a function. There is relatively little sensitivity to the choice, and a suitable form which has no
${\cal O}(N^{-1})$ component 
is\footnote{An alternative form e.g. $P'(x)=k_1(t)+k_2(t)\log(1/x)+k_3(t)\log(1/x)^2$ gives similar results.}:
\begin{equation}
P'(x)=\frac{k_1(t)}{\sqrt{x}}+k_2(t)+k_3(t)\sqrt{x},
\label{P'}
\end{equation}
which in $N$-space becomes:
\begin{align}
P'(N)&=\frac{k_1(t)}{N+1/2}+\frac{k_2(t)}{N+1}+\frac{k_3(t)}{N+3/2}\notag\\
&=\left(2k_1(t)+k_2(t)+\frac{2}{3}k_3(t)\right)+\left(-4k_1(t)-k_2(t)-\frac{4}{9}k_3(t)\right)N+{\cal O}(N^2).
\label{P'2}
\end{align}
The need to reproduce the first two terms of $P$, together with the first moment condition then gives the simultaneous equations:
\begin{align}
2k_1(t)+k_2(t)+\frac{2}{3}k_3(t)&=\delta_0(t);\notag\\
-4k_1(t)-k_2(t)-\frac{4}{9}k_3(t)&=\delta_1(t);\notag\\
\frac{2}{3}k_1(t)+\frac{1}{2}k_2(t)+\frac{2}{5}k_3(t)&=\delta_0(t)+\delta_1(t),
\label{Psimult}
\end{align}
which one may solve to give:
\begin{align}
k_1(t)&=\frac{3}{2}\left(\frac{1}{2}\delta_0(t)+\frac{7}{4}\delta_1(t)\right);\notag\\
k_2(t)&=-8\delta_0(t)-24\delta_1(t);\notag\\
k_3(t)&=\frac{45}{8}[2\delta_0(t)+5\delta_1(t)].
\label{Psimult2}
\end{align}
If one simply omits the ${\cal O}(N)$ terms in the resummed coefficient and splitting functions, it is not possible to achieve a good fit to scattering data as the resulting splitting functions lead to a large violation of momentum conservation in the evolution.\\

One must again parameterise the splitting and coefficient functions for $t\lesssim11$. A function of the form (\ref{pparam}) is sufficient, with positive powers of $t$ for both $P_{qg}$ and $C_{Lg}$. We have so far ignored those quantities which couple to the quark singlet distribution i.e. $P_{gq}$, $P_{qq}$ and $C_{Lq}$. At LL order, these are related to the analogous gluon quantities via the colour-charge relation:
\begin{equation}
F_q=\frac{C_F}{C_A}\left[F_g-F_g^{(0)}\right],
\label{colourcharge}
\end{equation}
where $F_a$ is a LL resummed coefficient or splitting function coupling to singlet species $a$, and $F_a^{(0)}$ its leading order contribution to the fixed order expansion. We also use this relation to define resummed quark singlet quantities at NLL order. This is not formally true at this order, but is a reasonable approximation with a similar uncertainty to that arising from using the exact kinematics estimates of the NLL impact factors. \\

Before presenting results for these quantities, the resummation from the BFKL equation must be combined with the NLO fixed order expansion. This poses a potential problem in that the gluon in the $Q_0$-scheme is in principle different to that in the fixed order DIS scheme. We discuss this in the next subsection.
\subsection{Transformation from the $Q_0$ Scheme}
The resummed splitting functions found so far are in the $Q_0$ scheme, where the gluon is defined via solution of the BFKL equation. However, the gluon in the DIS scheme at fixed order is given in terms of parton densities in the $\msbar$ scheme, which arise from using dimensional regularisation. Thus the two gluons are not the same, and in principle a scheme change is needed before one can add the resummed quantities to their fixed order counterparts. \\

Let us examine this in more detail by looking at the anomalous dimension given by equation (\ref{gamggdef}), which when expanded out in powers of $\alpha_S$ in $N$-space gives:
\begin{align}
\gamma_{+,NLL}&=\abar\left(\frac{1}{N}-0.91667-0.0061728n_f\right)+\abar^2\Big(-\frac{0.10597n_f}{N}+0.097136n_f\notag\\
&+0.00065412n_f^2\Big)+{\cal O}(N).
\label{gamBFKLNLL}
\end{align}
One may compare this with the relevant fixed order quantity, which is $\gamma_{gg}+(4/9)\gamma_{qg}$ as discussed previously. The NLO DIS scheme result, expanded about $N=0$, gives:
\begin{align}
\gamma_+&=\abar\left(\frac{1}{N}-.91667-0.0061728n_f\right)+\abar^2\Big(-\frac{.10597n_f}{N}-.11579n_f+2.1864\notag\\
&+0.0058299n_f^2\Big)+{\cal O}(N).
\label{gam+FO}
\end{align}
The resummed and fixed order results are not equal, and this is partly due to the difference in gluons in the resummed and fixed order frameworks\footnote{There are also NNLL contributions, distinct from the running coupling corrections, not predicted by the resummation.}. The solution is to implement a scheme change in the $k_T$ factorisation formula, making the ansatz in double Mellin space \cite{Ciafaloni06}:
\begin{equation}
{\cal G}(\gamma,N)=R(\gamma,N){\cal G}^{\DIS}(\gamma,N),
\label{transform}
\end{equation}
where ${\cal G}^{\DIS}$ is the fixed order unintegrated DIS scheme gluon density and $R(\gamma,N)$ is a perturbatively calculable transformation factor. A fully consistent NLL resummation consists of implementing the BFKL kernel as above, but also including an all orders in $\gamma$ result for $R(\gamma,N)$ truncated at ${\cal O}(N)$. The ${\cal O}(N^0)$ part of $R(\gamma,N)$ is known \cite{Catani}, but the ${\cal O}(N)$ piece is not. It would be found by formulating the BFKL equation using dimensional regularisation, and solving with running coupling. This does not seem to be analytically tractable at present \cite{Ciafaloni}, and thus a compromise must be reached. One might instead try exploiting the fact that differences between the fixed order and resummed gluons can be neglected at orders of $\alpha_S$ beyond the extent of the fixed order expansion, and thus expand the $R(\gamma,N)$ factor as follows:
\begin{equation}
R(\gamma,N)=r_0+r_1\gamma+{\cal O}(\gamma^2)+[r_2+r_3\gamma+{\cal O}(\gamma^2)]N+{\cal O}(N^2).
\label{Rexpand}
\end{equation}
This corresponds to truncating the NLL resummed scheme transformation factor at NLO in the fixed order expansion. The generalisation of equation (\ref{GE1}) to include the $R$ factor is straightforward, and in practice it mimics the addition of an impact factor. One may find the coefficients $r_i$ by comparing the anomalous dimension $\gamma_{BFKL}$ obtained by inserting equation (\ref{Rexpand}) into equation (\ref{GE1}) with the fixed order $\gamma_+$. However, it is easy to see that $r_0=1$ and $r_2=0$ from the fact that at LO the scheme transformation must be trivial ($R=1$). Also $r_1=0$ as the LL factor $R(\gamma)=1+{\cal O}(\gamma^3)$ \cite{Catani}. Thus up to NLL and NLO orders, one has $R(\gamma,N)=r_3\gamma N$ and the resulting anomalous dimension turns out to be:
\begin{align}
\gamma_{+,NLL}&=\abar\left(\frac{1}{N}-0.91667-0.0061728n_f\right)+\abar^2\Big(-\frac{0.10597n_f}{N}+0.097136n_f\notag\\
&+0.00065412n_f^2-\bb r_3\Big)+{\cal O}(N).
\label{gamBFKLNLL2}
\end{align}
This can be compared with equation (\ref{gam+FO}), and in principle the coefficient $r_3$ can be read off. However, the term at ${\cal O}(N^0)$ at NLO in equation (\ref{gam+FO}) is a sum of a running coupling correction (predicted here by the resummation) and a NNLL term (not completely predicted by the resummation). It is not possible to disentangle the former contribution from the latter, and hence unambiguously determine $r_3$. However, it seems that one can neglect this contribution to the scheme change to a very good approximation. In \cite{WT1} we have compared the fixed coupling results for the NLL resummed $P_{qg}$ (using the exact kinematics approximation) with the exact NLO, and NNLO splitting functions. The ${\cal O}(N^0)$ term in that splitting function at NLO and
the ${\cal O}(N^{-1})$ term at NNLO were estimated extremely well -- suggesting that the scheme change from the $Q_0$ scheme to a fixed order type DIS scheme (where partons are dimensionally regularised) is a very small effect. It is certainly a less significant effect than using the exact kinematics approximation for the NLL impact factors. The issue of transforming from the $Q_0$ scheme to the $\msbar$ scheme, however, is still unresolved. 
\subsection{Matching to the NLO expansion}
So far we have obtained NLL resummed splitting and coefficient functions, using the BFKL equation and the exact kinematics impact factors. In order to use these in a fit to scattering data, one must combine them with the NLO fixed order QCD expansion. Care must be taken to avoid double-counting of terms. Schematically one has a total splitting or coefficient function given by:
\begin{equation}
P^{tot.}=P^{NLL}+P^{NLO}-\left[P^{NLL(0)}+P^{NLL(1)}\right],
\label{doublecount}
\end{equation}
where the bracketed term on the right-hand side subtracts the double-counted terms, and $P^{NLL(n)}$ is the contribution of the resummed quantity to the $n^{\text{th}}$ order term in the fixed order expansion. This is complicated in practice by the fact that the resummed quantities are not known accurately as a power-series in $\alpha_S$, and must be parameterised according to equation (\ref{pparam}). Nevertheless, one can use the LO and NLO terms of the power series solution (from equations (\ref{f1sol3},\ref{g1sol})) for each quantity to 
determine these terms. As an example, consider $\gamma_{qg}$ at $t=6$ and $n_f=4$. This can be found by numerical integration of equations (\ref{GE12}, \ref{F1sol2}) and parameterised by:
\begin{equation}
\gamma_{qg,\, param.}^{LL+NLL}=\frac{\alpha_Sn_f}{3\pi}\left(-.5501N+.2466+\frac{0.2294}{N}-\frac{0.03443}{N^2}+\frac{0.01381}{N^3}\right).
\label{paramqg}
\end{equation}
However, this includes some of the high energy limit of the LO and NLO anomalous dimension, which must be subtracted to avoid double counting. These double counted terms are obtained from the power series solution for $\gamma_{qg}$ up to NLO, which is:
\begin{equation}
\gamma_{qg,\, pow.}^{LL+NLL}=\frac{\alpha_Sn_f}{3\pi}\left[1-1.083N+\abar\left(\frac{2.167}{N}-6.484+4.184N\right)\right]+{\cal O}(\alpha_S^3).
\label{powqg}
\end{equation}
Subtracting (\ref{powqg}) from (\ref{paramqg}) gives the purely higher order corrections to be added onto the fixed order expansion. \\

\begin{figure}
\begin{center}
\scalebox{0.8}{\includegraphics{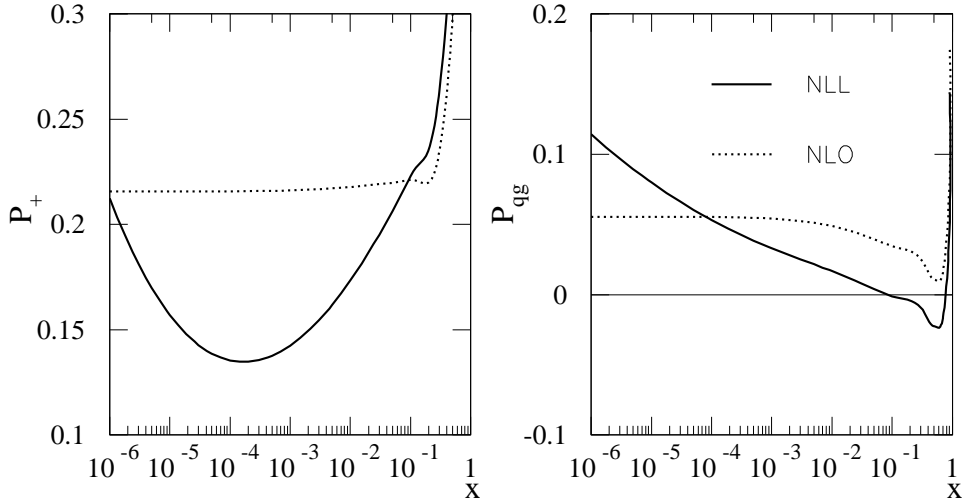}}
\caption{The total NLL resummed splitting functions $P_{qg}$ and $P_+$ for $n_f=4$ and $t=6$, compared with the NLO fixed order results.}
\label{splitsplot2}
\end{center}
\end{figure}

\begin{figure}
\begin{center}
\scalebox{0.8}{\includegraphics{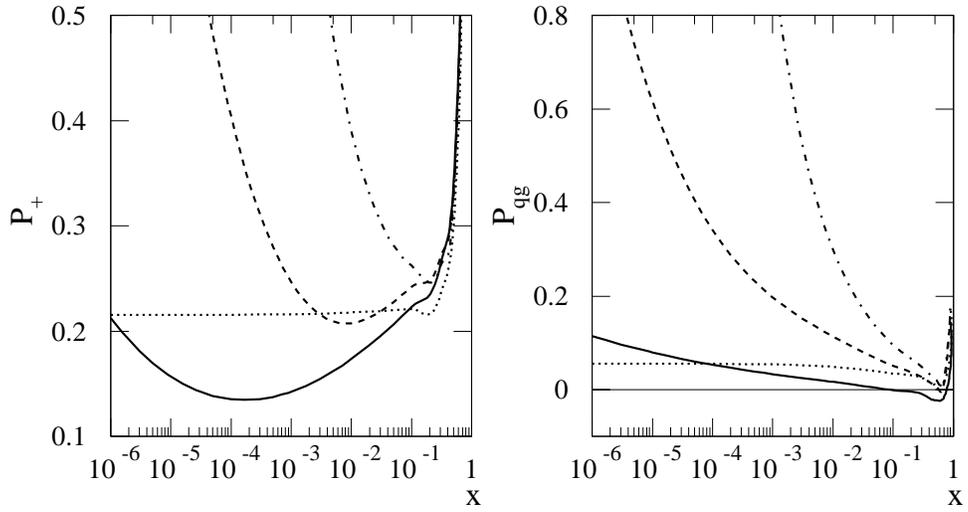}}
\caption{The NLL resummed splitting functions $P_{qg}$ and $P_+$ for $n_f=4$ and $t=6$ (solid), together with the LL result with (dashed) and without (dot-dashed) running coupling corrections. Also shown is the NLO fixed order result (dotted).}
\label{splitsplot}
\end{center}
\end{figure}

\begin{figure}
\begin{center}
\scalebox{0.8}{\includegraphics{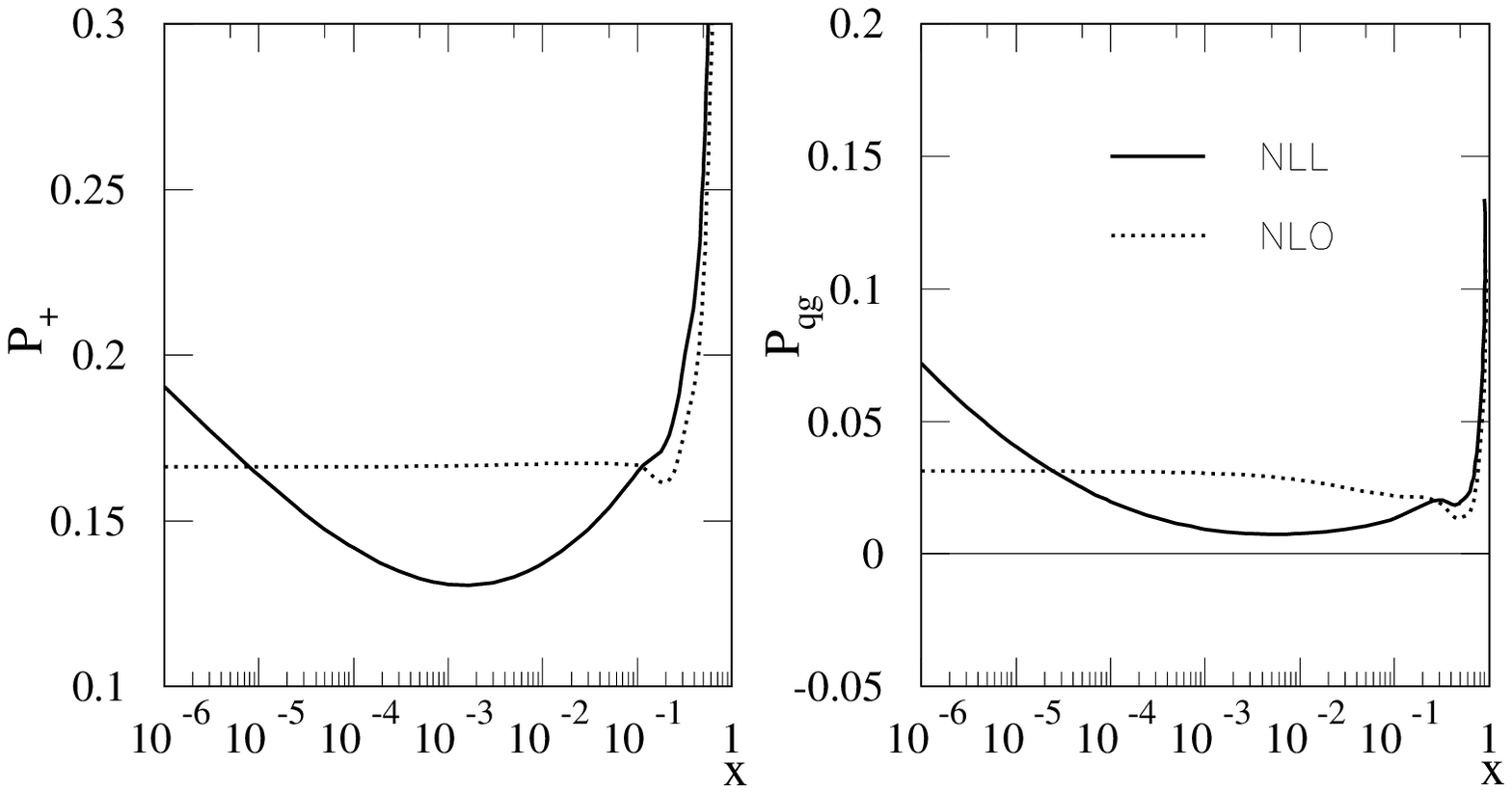}}
\caption{The NLL resummed splitting functions $P_{qg}$ and $P_+$ for $n_f=4$ and $t=8$, compared with the NLO fixed order results.}
\label{splitshight}
\end{center}
\end{figure}

Our results for $P_{qg}$ and $P_{+}$ are shown for $n_f=4$ and $t=6$ in figure \ref{splitsplot2} and alongside their LL counterparts, both with and without the inclusion of running coupling effects, in figure \ref{splitsplot}. 
The inclusion of the running coupling at LL order suppresses the resummed splitting functions, as previously noted \cite{Thorne}. The NLL corrections from the impact factors and kernel then lead to a further suppression of the small $x$ divergence. Both the splitting functions dip below the NLO result, before growing at very small $x$. One must also bear in mind the missing Dirac terms in figure \ref{splitsplot2} and \ref{splitsplot}. These carry a positive first moment in the case of $P_{qg}$, but negative for $P_{gg}$. The NLL corrections delay the onset of the asymptotic behaviour until even lower values of $x$. The same splitting functions at a higher value of $t=8$ are shown in figure \ref{splitshight}.\\

\begin{figure}
\begin{center}
\scalebox{0.8}{\includegraphics{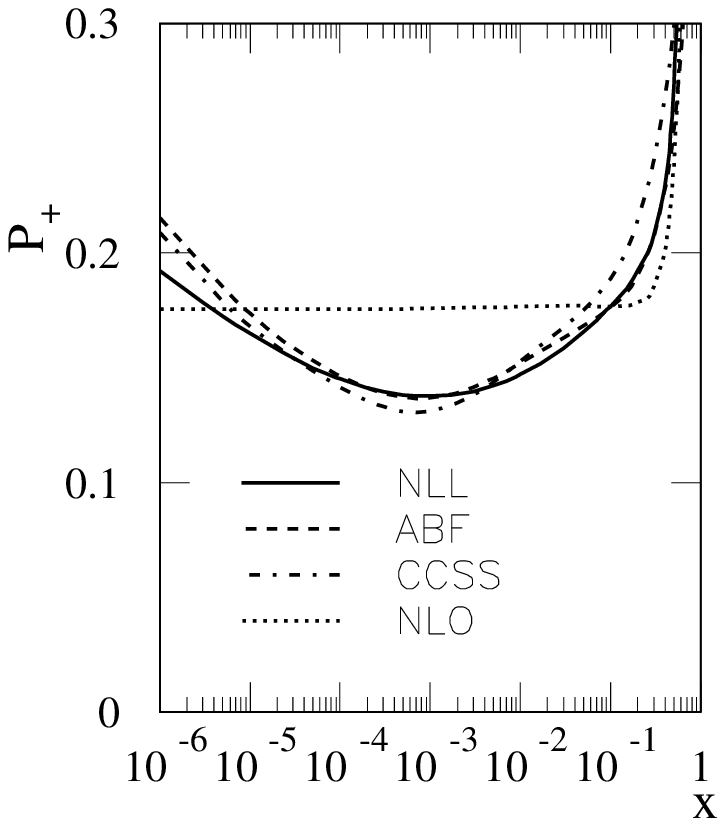}}
\caption{Comparison of the splitting function in figure (\ref{splitsplot}) with the CCSS and ABF results for $\alpha_S=0.2$ ($t\simeq7.5$ for the LO coupling). The CCSS and ABF results have $n_f=0$. Results taken from \cite{ABF}.}
\label{abfplot}
\end{center}
\end{figure}

It is interesting to compare our results for $P_+$ with those of the ABF \cite{ABF} and CCSS \cite{CCSS} approaches, and a comparison is given in figure \ref{abfplot}.
All three splitting functions show the same qualitative behaviour. There is a pronounced dip below the NLO result at moderate $x$, followed by an eventual asymptotic rise as $x\rightarrow 0$. At high $x$ the behaviour is governed by the fixed order DGLAP evolution kernel, as required in each approach. Qualitatively, the CCSS function has a slightly deeper dip, and we note that both the CCSS and ABF results show a slightly higher rise at asymptotically low $x$ than the splitting function presented in this paper. Part of this may be due to the fact that both of the alternative approaches involve the resummation of collinearly enhanced terms in the kernel, which act to suppress the impact of the NLL contribution. We see from the figure that this is not a large effect. However, it must be noted that the CCSS and ABF curves in figure \ref{abfplot} have $n_f=0$. The effect of a higher quark flavour is increase the splitting function at low $x$, implying that resummation of the kernel may in fact be more significant than appears to be the case in figure \ref{abfplot}. It would of course be extremely interesting to see a comparison of the quark splitting functions, as well as an application of all three approaches to experimental data\footnote{A discussion of impact factors in both approaches can be found in \cite{Salam2,ABFimpact}. There is a preliminary investigation of $P_{qg}$ in \cite{Ciafaloni06}.}.\\
\begin{figure}
\begin{center}
\scalebox{0.8}{\includegraphics{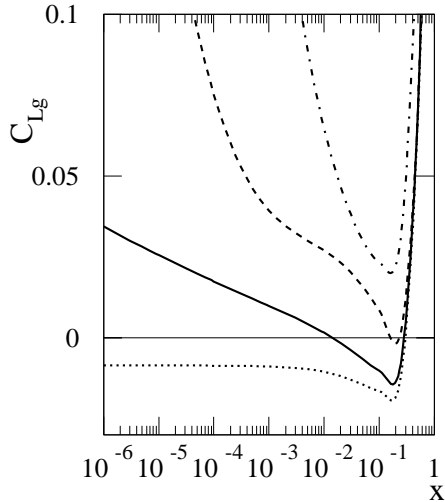}}
\caption{The NLL resummed coefficient $C_{Lg}$ at $t=6$ and $n_f=4$ (solid) together with the LL result with (dashed) and without (dot-dashed) running coupling corrections. Also shown is the NLO result (dotted).}
\label{clgnll}
\end{center}
\end{figure}

\begin{figure}
\begin{center}
\scalebox{0.8}{\includegraphics{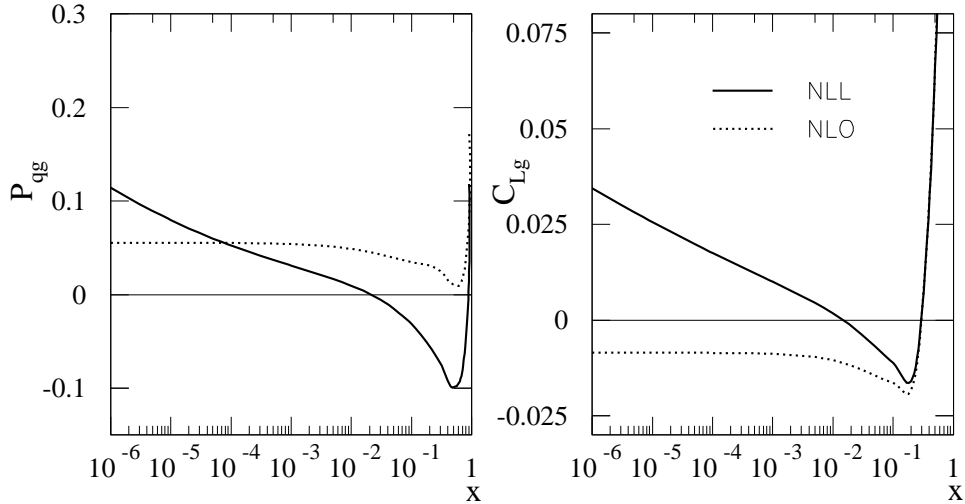}}
\caption{The resummed + NLO results for $P_{qg}$ and $C_{Lg}$ at $t=6$ and $n_f=4$ where the Dirac terms have been implemented according to equations (\ref{diracparam}, \ref{diracparam2}), compared with the NLO results.}
\label{diracplot}
\end{center}
\end{figure}

In figure \ref{clgnll} we show the resummed longitudinal coefficient $C_{Lg}$ at $t=6$ and $n_f=4$. Once again, the NLL corrections suppress the result at small $x$ when compared with the LL result, beyond the suppression already induced by running coupling corrections. The results in figures \ref{splitsplot2}-\ref{clgnll} do not include the effect of the ${\cal O}(N^0)$ and ${\cal O}(N)$ terms from the resummation, which are implemented according to the prescription of equation (\ref{P'}). In the example of $\gamma_{qg}$ at $t=6$ and $n_f=4$ given above, one finds:
\begin{equation}
P'_{qg}=\frac{\alpha_Sn_f}{3\pi}\left[-\frac{.6340}{(N+1/2)}+4.879-\frac{4.213}{(N+3/2)}\right].
\label{diracparam}
\end{equation}
Likewise, for $C_{L,g}$ at this value of $t$ one has:
\begin{equation}
C'_{Lg}=\frac{\alpha_Sn_f}{3\pi}\left[\frac{.4046}{(N+1/2)}-.2850+\frac{.1945}{(N+3/2)}\right].
\label{diracparam2}
\end{equation}
One can examine the form of the resummed + NLO quantities in $x$-space. These are shown in figure \ref{diracplot}.
Comparing this with figures \ref{splitsplot2} and \ref{clgnll}, one sees that the Dirac terms have little visible effect on $C_{Lg}$ (mainly due to the pronounced high $x$ behaviour in the NLO coefficient) whereas in $P_{qg}$ the dip at moderate $x$ is made more significant. However, a direct physical interpretation of figure \ref{diracplot} is not really possible. Other functional forms for implementing the Dirac terms in $N$-space (where they are constrained) can lead to quantitatively different shapes in $x$-space\footnote{The NNLO splitting function $P_{qg}$ does, however, display a narrow but deep dip at high $x$, as can be seen in \cite{WT1}.}, but similar results in convolutions. \\

We now have the complete set of massless resummed splitting and coefficient functions needed for phenomenological investigations. Before a fit can be made, however, one must also consider resummation effects in the heavy flavour sector. This is the subject of the next section. 
\section{Treatment of Heavy Flavours}
In \cite{WT2} we outlined the definition of the $\DIS(\chi)$ scheme for the consistent implementation of small $x$ resummations in the heavy flavour sector alongside a fixed order QCD expansion. We refer the reader to that paper for details, including a full discussion regarding consistency of the variable flavour number scheme between the fixed order and high energy expansions. Here we extend this scheme to NLL order in the resummation as far as is possible. \\

Below the matching scale $Q^2=M^2$ for a heavy quark of mass $M$, the heavy flavour contribution to the structure function $F_2$ is given by the fixed flavour description:
\begin{equation}
F_2^H=\left[C_{2,g}^{FF,LL}(\gamma,N,M^2/Q^2)+C_{2,g}^{FF,NLL}(\gamma,N,M^2/Q^2)\right]g^{(n_f)}(\gamma,N).
\label{f2hffnll}
\end{equation}
Comparing with the $k_T$-factorisation result:
\begin{equation}
F_2^H=\tilde{h}_2(\gamma,N,Q^2/M^2)g(N,\gamma),
\label{F2HkT}
\end{equation}
one identifies the fixed flavour coefficient with the impact factor $\tilde{h}_2(\gamma,N,Q^2/M^2)$ coupling a virtual photon to a gluon via a heavy quark-antiquark pair. A calculation of this quantity to NLL order will certainly not be available in the near future, and by analogy with the massless case one may use the LL factor calculated with exact gluon kinematics which has been presented in \cite{WPT}. \\

The factor $\tilde{h}$ diverges as $Q^2/M^2\rightarrow\infty$ to give the collinear singularity of the massless result, and so when $Q^2>M^2$ one must use the NLL variable flavour description:
\begin{equation}
F_2^H=q_{H+}+\left[C_{2,g}^{VF,LL}(\gamma,N,M^2/Q^2)+C_{2,g}^{VF,NLL}(\gamma,N,M^2/Q^2)\right]g^{(n_f+1)},
\label{f2hvfnll1}
\end{equation}
where $q_{H+}=q_H+\bar{q}_H$ is a parton distribution for the heavy flavour. The $(n_f+1)$-flavour partons are related to their $n_f$-flavour counterparts by resummed heavy matrix elements $A_{ij}$, which at NLL order are contained in the relations\footnote{This formalism was first introduced at fixed order in \cite{Buza}.}:
\begin{align}
q_{H+}(\gamma,N,M^2/Q^2)&=\left[A_{Hg}^{LL}(\gamma,N,M^2/Q^2)+A_{Hg}^{NLL}(\gamma,N,M^2/Q^2)\right]g^{(n_f)};\label{ahgnlldef}\\
g^{(n_f+1)}(\gamma,N,M^2/Q^2)&=A_{gg}^{NLL}(\gamma,N,M^2/Q^2)g^{(n_f)}.\label{aggnlldef}
\end{align}
Substituting these in equation (\ref{f2hvfnll1}) and equating coefficients of the gluon at LL and NLL orders, one finds:
\begin{align}
C_{2,g}^{VF,LL}&=C_{2,g}^{FF,LL}-A_{Hg}^{LL};\label{c2gvfll}\\
C_{2,g}^{VF,NLL}&=C_{2,g}^{FF,NLL}-A_{Hg}^{NLL}-C_{2,g}^{VF,LL}A_{gg}^{NLL}\label{c2gvfnll}.
\end{align}
Equation (\ref{c2gvfll}) has already been derived in \cite{WT2}, and equation (\ref{c2gvfnll}) is its NLL equivalent. The equality must be true for all values of $Q^2/M^2$. Taking $Q^2/M^2\rightarrow\infty$ gives:
\begin{equation}
A_{Hg}^{NLL}=C_{2,g}^{FF,NLL}|_{\frac{Q^2}{M^2}\rightarrow\infty},
\label{ahgnll}
\end{equation}
where we have used the fact that $C_{2,g}^{VF}\rightarrow 0$ as $Q^2/M^2\rightarrow\infty$ in the DIS($\chi$) scheme. A problem now occurs in that $A_{gg}^{NLL}$ and $C_{2,g}^{VF,NLL}$ are both still unknown in equation (\ref{c2gvfnll}). One needs a further equation to fully define the variable flavour scheme, and this must come from a definition of $A_{gg}$. Unfortunately, this is not possible at present, as in principle $A_{gg}$ is determined from a calculation of the gluon density with heavy quark mass effects included. This would involve knowledge of the NLL BFKL kernel with heavy quarks included in the fermionic contributions, and this is not available. One can only neglect the term involving $A_{gg}$ and hope that that it is not significant\footnote{Note that $A_{gg}$ has no small $x$ poles at fixed order until NNLO.}. We thus define the (approximate) NLL variable flavour coefficient to be:
\begin{align}
C_{2,g}^{VF,NLL}&\simeq C_{2,g}^{FF,NLL}-A_{H,g}^{NLL}\notag\\
&=C_{2,g}^{FF,NLL}-C_{2,g}^{FF,NLL}|_{\frac{Q^2}{M^2}\rightarrow\infty},
\label{c2gvfapprox}
\end{align}
which now has the same form as at LL order.\\

\begin{figure}
\begin{center}
\scalebox{0.8}{\includegraphics{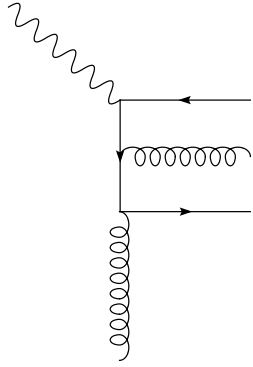}}
\caption{Diagram contributing to the FF coefficient $C_{L,g}^{FF}$ which is not included in the exact kinematics approximation.}
\label{ahgclg}
\end{center}
\end{figure}

One must also consider the heavy flavour contributions to the longitudinal structure function. Equivalence of the VF and FF descriptions demands (in double Mellin space):
\begin{align}
F_L^H&=\left[C_{L,g}^{FF,LL}(\gamma,N,M^2/Q^2)+C_{L,g}^{FF,NLL}(\gamma,N,M^2/Q^2)\right]g^{(n_f)}(\gamma,N)\notag\\
&=C_{L,H}^{LL}(\gamma,N,M^2/Q^2)q_{H+}\left[C_{L,g}^{VF,LL}(\gamma,N,M^2/Q^2)+C_{L,g}^{VF,NLL}(\gamma,N,M^2/Q^2)\right]g^{(n_f+1)}.
\label{flhffvf}
\end{align}
Then use of the relations (\ref{ahgnlldef},\ref{aggnlldef}) yields:
\begin{equation}
C_{L,g}^{VF,NLL}=C_{L,g}^{FF,NLL}-C_{L,H}^{LL}A_{Hg}^{LL}-C_{L,g}^{VF,LL}A_{gg}^{LL}.
\label{clgvfdef}
\end{equation}
The term involving $A_{gg}^{LL}$ must be neglected for the reasons given previously. Furthermore, the term in $A_{Hg}$ presents a problem in the exact kinematics approximation as $Q^2/M^2\rightarrow\infty$. This term diverges as $(Q^2/M^2)^\gamma$ in double Mellin space, whereas the FF coefficient (coming from the exact kinematics massive impact factor) is finite as $Q^2\rightarrow\infty$. One can see this by noting that divergences in the longitudinal impact factor come from diagrams such as that shown in figure \ref{ahgclg}.
Radiation of a gluon from the heavy quark leads to a collinear logarithm of type $\log(M^2/Q^2)$. Such a diagram is a correction to the longitudinal impact factor that is not included by the exact kinematics approximation, and is needed to cancel the similar divergence coming from the term in $A_{Hg}$. The upshot of this is that the right-hand side of equation (\ref{clgvfdef}) is divergent, whereas the left-hand side cannot be as one must have:
\begin{equation}
\lim_{\frac{Q^2}{M^2}\rightarrow\infty}C_{L,g}^{VF,NLL}=C_{L,g}^{NLL},
\label{limclgvf}
\end{equation}
where the right-hand side denotes the massless longitudinal coefficient. Thus given the approximated impact factor does not have the correct asymptotic behaviour to match the term in $A_{Hg}$, this latter term must also be neglected along with the term in $A_{gg}$. The heavy flavour contributions to $F_L$ are in any case very small until higher $Q^2$, where they approach the known massless terms. 
\subsection{Approximate NLL Coefficients}
The results of the previous section can in principle be applied to find the NLL resummed heavy flavour coefficient functions. One uses the impact factors with exact kinematics to estimate the true NLL impact factors. However, as shown in \cite{WPT}, it does not seem possible to obtain a closed analytic form for the exact kinematics impact factors in the case of heavy quarks. Power series results in $N$ and $\gamma$ can be found numerically for given values of $M^2/Q^2$ and used to investigate the NLL resummation. A problem occurs in using the approach described in this paper, where equations (\ref{g1sol}, \ref{f1sol3}) are used without any further truncations. The power series expressions for the coefficient functions are ill behaved up to rather high $t$, and parameterised numerical results cannot be obtained without a full definition of the NLL impact factor (i.e. not just the expansion in $\gamma$). So in this section we investigate the coefficient functions obtained using a gluon and structure function truncated separately to NLL order as described in \cite{Thorne}, but where the full gluon is used in the denominator when defining the coefficient function via:
\begin{equation}
C_{a,g}^{FF,LL}(\gamma,N,M^2/Q^2)=\frac{{\cal F}_{a,E}^H(\gamma,N,M^2/Q^2)}{{\cal G}_E(\gamma,N)}.
\label{coeffheavydef}
\end{equation}
The power suppressed correction for these coefficient functions is much smaller at $t$ values of phenomenological interest, and the conclusions drawn about how to approximate the resummed heavy flavour coefficients should be independent of the choice of NNLL terms. \\

In \cite{WPT} it was noted that a smaller suppression of the heavy coefficient functions is induced by the NLL impact factor corrections than in the massless case. It is useful, in both the massless and massive sectors, to compare this suppression with that induced by the rest of the NLL framework. Overall, there are four factors which act to reduce the resummed splitting and coefficient functions at small $x$:
\begin{itemize}
\item Running Coupling Corrections. 
\item The implementation of the NLL BFKL kernel.
\item The use of the NLL gluon density ${\cal G}_E^1$ instead of the LL result.
\item NLL contributions to the impact factors from exact kinematics.
\end{itemize}
The effect of the running coupling in suppressing small $x$ divergence has already been noted. It is also important to realise that the second and third mechanisms listed above account for most of the further suppression at NLL order, and the impact factors give a smaller correction once the kernel and gluon have been accounted for. \\

\begin{figure}
\begin{center}
\scalebox{0.7}{\includegraphics{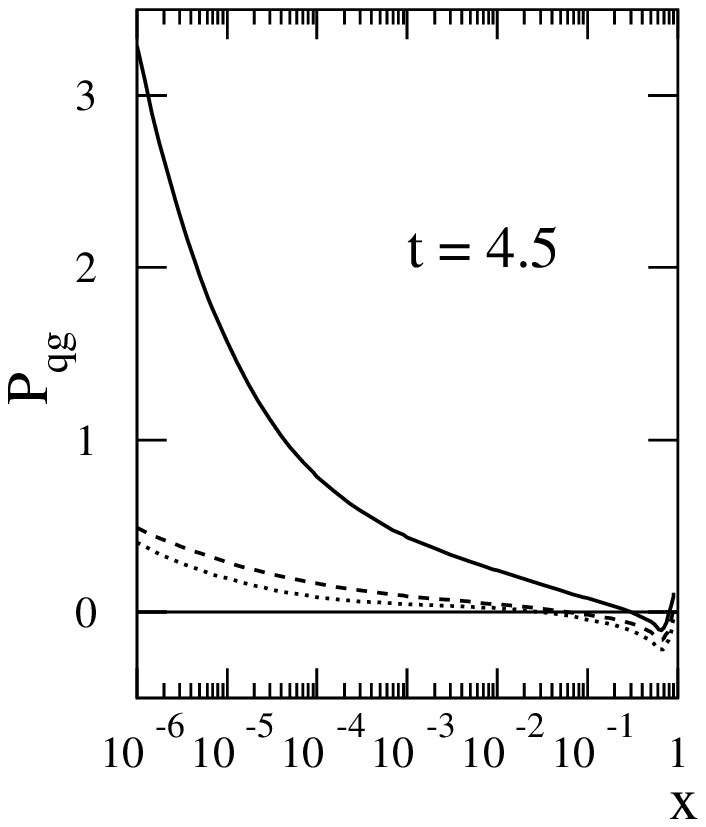}}
\scalebox{0.7}{\includegraphics{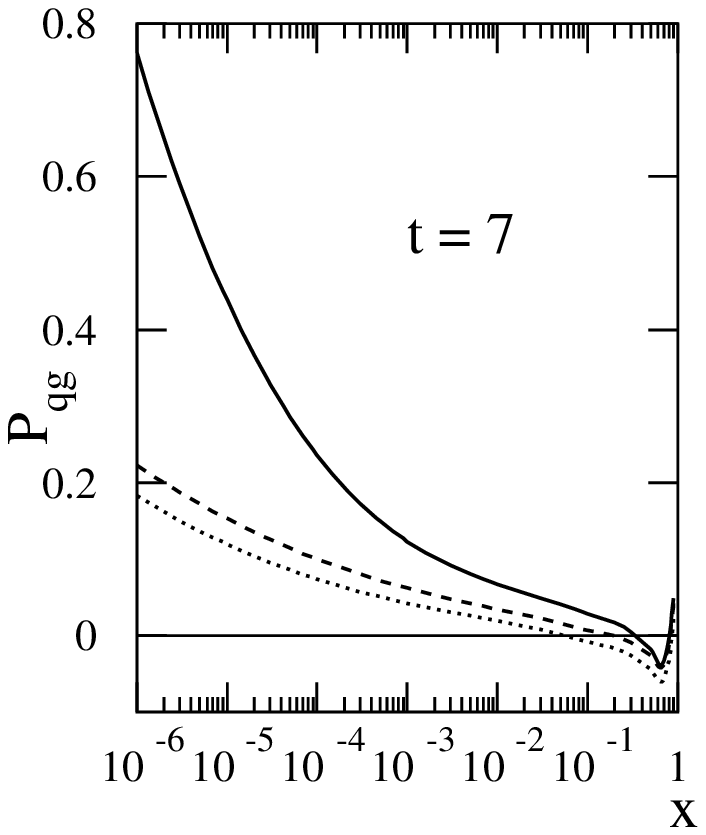}}
\caption{The LL splitting function $P_{qg}$ for $n_f=4$ with running coupling corrections (solid). Also shown is the NLL result with no impact factor correction (dashed) and the NLL resummed result which includes this correction (dotted). Note that the NLO splitting function has been added to give the correct behaviour at high $x$.}
\label{Pqgfig}
\end{center}
\end{figure}

One can see this in the massless case by looking at figure \ref{Pqgfig}, which shows the LL result for $P_{qg}$ with running coupling corrections together with NLL results with and without the impact factor correction\footnote{Note that the splitting function in these plots arises from a truncated gluon and structure function for comparison with the massive results.}. 
At the lower $t$ value, the effect of the NLL term in the impact factor is small compared with the combined suppression resulting from the gluon and kernel. At the higher $t$ value, the impact factor induced suppression is relatively larger, but still reasonably small. Note that such a $t$ value is relevant to the massive case, where one is only interested in the heavy flavour coefficients for $t\lesssim 7$. The VF coefficients decrease rapidly as $t$ increases, as shown in \cite{WT2}.\\

\begin{figure}
\begin{center}
\scalebox{0.7}{\includegraphics{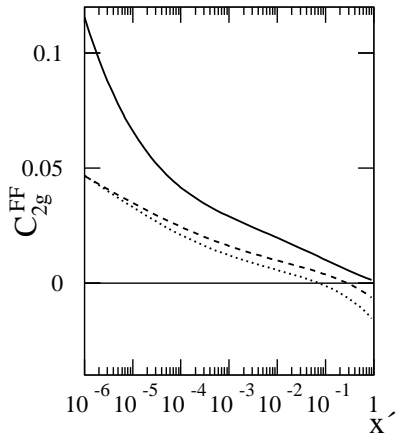}}
\caption{Power series results for the fixed flavour coefficient $C_{2,g}^{FF}$ for $n_f=3$ and $t=7$. The LL result with running coupling corrections is shown (solid), alongside the NLL results with (dotted) and without (dashed) impact factor corrections.}
\label{c2gffnll}
\end{center}
\end{figure}

One can examine the massive case in more detail using the impact factor results of \cite{WPT}. It is not possible, however, to evaluate a power-suppressed correction once the exact kinematics are included as complete closed forms for the approximated impact factors are not known. Thus in order to investigate suppression effects one must choose a reasonably high value of $t$ and hope that the correction is small. We consider the  fixed flavour coefficient $C_{2,g}^{FF}$, as the variable flavour coefficient will be tiny at values of $t$ high enough for the power-suppressed correction to be small. The fixed flavour coefficient is shown in figure \ref{c2gffnll} for $t=7$. The LL power series result with running coupling corrections is again significantly bigger than the NLL result (although not as much as in the massless case), and the correction from the impact factor is much smaller than that due to the change in kernel and gluon density. The exact kinematics give a relatively larger effect at high $x$, but the fixed order NLO coefficient will dominate here.\\

\begin{figure}
\begin{center}
\scalebox{0.7}{\includegraphics{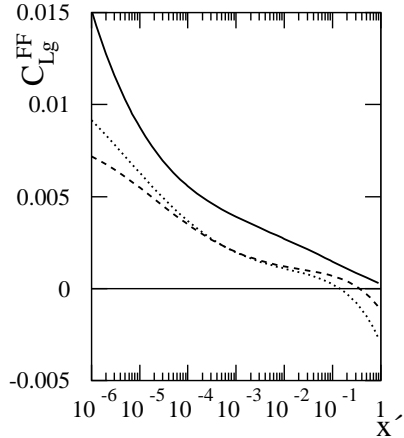}}
\caption{Power series results for the fixed flavour coefficient $C_{L,g}^{FF}$ for $n_f=3$ and $t=7$. The LL result with running coupling corrections is shown (solid), alongside the NLL results with (dotted) and without (dashed) impact factor corrections.}
\label{clgffnll}
\end{center}
\end{figure}

A similar result for $C_{Lg}^{FF}$ is shown in figure \ref{clgffnll}.
Again the change induced by the NLL impact factor is of a size that is small when compared to the suppression induced by the gluon and kernel. Thus, in approximating the heavy flavour coefficient functions, it seems reasonable to ignore the contribution from the exact kinematics impact factors and only include the NLL kernel in the structure function and gluon density. Then one can parameterise the FF and VF coefficients over the whole desired range of $t$. Figure \ref{Pqgfig} suggests that the effect of the exact kinematics is felt more at higher values of $t$. This is encouraging for the approximation proposed here, as the variable flavour coefficients for $F_2$ decay to zero quite rapidly ($\sim M^2/Q^2$) as $t$ increases. \\

\begin{figure}
\begin{center}
\scalebox{0.8}{\includegraphics{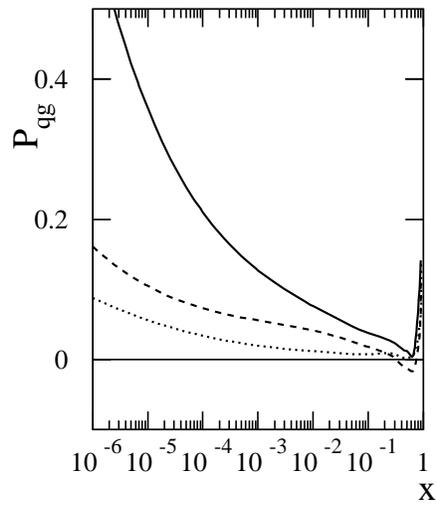}}
\caption{The LL splitting function $P_{qg}$ with running coupling for $t=7$ and $n_f=4$ (solid) compared with the NLL result with (dotted) and without (dashed) impact factor corrections, derived using the procedure outlined in section 2.}
\label{pqgfullsup}
\end{center}
\end{figure}

The results in this section have been derived using the NLL truncated structure function and gluon density. One can, however, check in the massless case that the suppression from the NLL impact factor contribution is also relatively small when the full structure function and gluon density (as described in section 2) are used. Figure \ref{pqgfullsup} shows the resummed $P_{qg}$ at $t=7$ and $n_f=4$, with and without the impact factor contribution. One again sees that the dominant suppression in going to NLL order comes from the kernel and gluon rather than the impact factor. Given also that this procedure for generating the splitting function differs from that used throughout this section only by NNLL terms, the conclusion reached about the approximation for the massive coefficients seems to be robust. Indeed, impact factor corrections have a lesser effect in the heavy flavour sector, as noted previously.\\

With this approximation for the heavy flavour quantities, the FF coefficients can be accurately parameterised in $N$-space by a function of form (\ref{pparam}) with positive powers of $t$. A similar function is sufficient for $C_{2,g}^{VF}$, but with an overall factor of $\exp(-t)$ to reflect the decay of the gluon coefficient as $t\rightarrow\infty$. There is a further subtlety involved in the longitudinal coefficient $C_{L,g}^{VF}$, as it must tend to the massless resummed coefficient as $Q^2\rightarrow\infty$. This will not happen if the NLL impact factor contribution in $C_{L,g}^{VF}$ is ignored. To ensure that the correct asymptotic limit is reached, it is sufficient to model the NLL impact factor contribution by $(1-M^2/Q^2)^{15}h_L^{(1)}$ where $h_L^{(1)}$ is the massless ${\cal O}(N)$ contribution to the impact factor, and the prefactor ensures a smooth matching between the fixed flavour coefficient at $Q^2=M^2$ and the massless coefficient as $Q^2\rightarrow\infty$. Resummed heavy flavour coefficients for the quark singlet distribution for both $F_2^H$ and $F_L^H$ are again approximated by using the colour charge relation (\ref{colourcharge}).

\subsection{NLL Heavy Matrix Element $A_{Hg}$}
The final ingredient for an approximate NLL analysis is the resummed matrix element governing the initial value of heavy quark distributions, evaluated at the matching scale $Q^2=M^2$. This is given in double Mellin space by equation (\ref{ahgnll}) and one could use the asymptotic form of the impact factor, known in power series form from \cite{WPT}, to obtain an analytic $x$ space result according to equation (\ref{f1sol3}). However, at LL order the analytic result was found to suffer from a large power suppressed correction \cite{WT2}. In this paper we have also seen that NLL resummed quantities tend to have larger power suppressed corrections than their LL counterparts, persisting to larger values of $t$. Hence we choose to define the heavy matrix element via the $x$-space equivalent of equation (\ref{c2gvfapprox}) corresponding to the approximate choice for the variable flavour coefficient. This is equivalent to evaluating $A_{Hg}^{NLL}$ by using the LL impact factor only with NLL kernel and gluon density corrections as has been done for the coefficients, and ensures continuity of the structure function across the matching scale in passing between the FF and VF descriptions.\\

\begin{figure}
\begin{center}
\scalebox{0.8}{\includegraphics{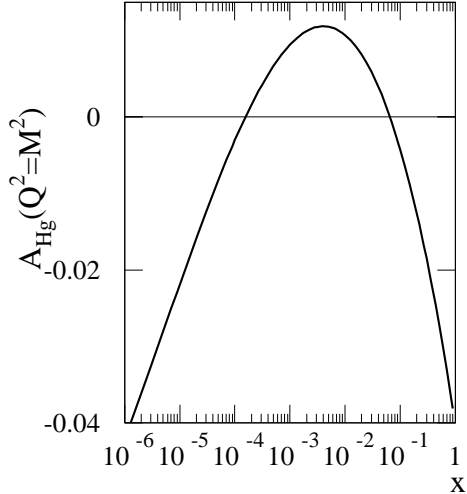}}
\caption{The approximate NLL resummed heavy matrix element $A_{Hg}$ for $n_f=4$ and $Q^2=M^2=2.25\text{GeV}^2$.}
\label{ahgnllplot}
\end{center}
\end{figure}

A plot of the heavy matrix element at $Q^2=M^2$ for $M=1.5\,$GeV is shown in figure \ref{ahgnllplot}, and one sees that it is negative over most of the range of $x$.
This is in contrast to the LL result with running coupling corrections \cite{WT2}, which is positive as $x\rightarrow0$. However, this result may or may not hold true once the complete NLL impact factor corrections (unknown) are included. Even so, the matrix element gives a very small correction to the structure function.\\

This now completes the set of quantities needed to perform an approximate NLL resummed global fit analogous to the LL fit of \cite{WT2}. Further details of this fit and the results obtained are presented in the next section.
\section{A Global Fit at NLL Order}
The results of the previous section allow for a NLL+NLO analysis of DIS and related data. In order to compare with the LL+LO fit results of \cite{WT2}, we use the same data sets as in that paper. We include $F_2$ structure function data from the H1 \cite{H1a,H1b,H1c} and ZEUS \cite{ZEUSa,ZEUSb,ZEUSc} collaborations at HERA; proton data from BCDMS \cite{BCDMSep}, NMC \cite{NMC}, SLAC \cite{SLAC,SLAC2} and E665 \cite{E665}; deuterium data from BCDMS \cite{BCDMSeD}, NMC, SLAC and E665; CCFR data on $F_{2,3}^{\nu(\bar{\nu})N}(x,Q^2)$ \cite{CCFRF2, CCFRF3}; data on the deuterium-proton ratio $F_2^D/F_2^p$ from NMC \cite{NMCrat}; charged current data from H1 \cite{H1a} and ZEUS \cite{ZEUSCC}; data on the charm structure function $F_{2,c}$ from H1 \cite{F2c1} and ZEUS \cite{F2c2}. The non-DIS data sets used are Drell-Yan (DY) data from the E866/NuSea collaboration \cite{DY}; DY asymmetry data from NA51 \cite{DYasym}; data on the DY ratio $\sigma_{DY}^{pD}/\sigma_{DY}^{pp}$ from E866 \cite{DYrat}; W-asymmetry data from CDF \cite{Wasym}. We also check our predictions against the Tevatron jet data \cite{Tevjet1,Tevjet2}, although these are not explicitly included in the fit. In order to compare with the fixed order QCD expansion, we also produce a fit at NLO in the DIS scheme. Before presenting results, we first discuss some implementation issues arising in the fitting procedure.
\subsection{Implementation}
Fixed order cross-sections for each of the data sets mentioned above must be transformed into the DIS scheme. For most quantities, the transformation of splitting and coefficient functions can be carried out analytically. For others, some parameterisation is necessary. For example, the NLO fixed flavour (FF) neutral current coefficient functions for the heavy contribution to the structure functions transform as:
\begin{align}
C_{a,g}^{FF,DIS(2)}=C_{a,g}^{FF,\msbar(2)}+2n_fC_{a,g}^{FF,\msbar(1)}\otimes C_{a,g}^{\msbar(1)};\label{c2gff2trans}\\
C_{a,q}^{FF,DIS(2)}=C_{a,q}^{FF,\msbar(2)}+C_{a,g}^{FF,\msbar(1)}\otimes C_{a,q}^{\msbar(1)},\label{c2qff2trans}
\end{align}
where $a\in\{2,L\}$ and the $n_f$ dependence has been explicitly displayed in the light gluon coefficient. The convolution on the right-hand side cannot be calculated analytically. Furthermore, the NLO $\msbar$ scheme heavy flavour coefficients are also not known analytically \cite{Laenen,Riemersma}, although recently a new parameterisation has been produced \cite{Thorne06}. We parameterise the transformation terms in equations (\ref{c2gff2trans},\ref{c2qff2trans}) by evaluating the convolution in each case numerically for a range of values of $x$ and $M^2/Q^2$ and fitting these to a function of form:
\begin{equation}
\sum_{n,m}\left(\frac{M^2}{Q^2}\right)\left\{[a_n+b_n(x-x_{max})]\log^m\left(\frac{x}{x_{max}}\right)+c_n(x-x_{max})^m\right\},
\label{transparamFF}
\end{equation}
where $x_{max}=(1+4M^2/Q^2)^{-1}$. This functional form then satisfies the kinematic constraint coming from the fact that the partonic centre of mass energy must exceed the threshold for heavy quark pair production. Fortran code for the DIS scheme NLO heavy flavour coefficients is available on request. We define the variable flavour (VF) heavy flavour coefficients according to the DIS($\chi$) scheme \cite{WT2}.\\

The charged current heavy flavour coefficients are more complicated, as there are ${\cal O}(\alpha_S^0)$ contributions to the structure functions resulting from the production of a single charm quark in the final state:
\begin{align}
F_2^c&=2[\cos^2\theta_c\xi s(\xi)+\sin^2\theta_c\xi d(\xi)]\equiv C_{2,s}^{(0)}\otimes s+C_{2,d}^{(0)}\otimes d;\label{F2cCC}\\
xF_3^c&=2[\cos^2\theta_c xs(\xi)+\sin^2\theta_c xd(\xi)]\equiv x[C_{2,s}^{(0)}\otimes s+C_{2,d}^{(0)}\otimes d] \label{F3cCC},
\end{align}
where $\theta_c$ is the Cabibbo mixing angle and $\xi=x/x_{max}^{CC}$. The upper limit:
\begin{equation}
x_{max}^{CC}=\left(1+\frac{m_c^2}{Q^2}\right)^{-1}
\label{xmaxCC}
\end{equation}
is the charged current analogue of the variable $x_{max}$ introduced above for neutral current scattering. We here present results for the transformation of the NLO FF coefficients for charged current scattering, as they have not been given in the literature before. First one must carefully define what is meant by NLO in this case. Denoting the weak eigenstate $[\cos^2\theta_c\xi s+\sin^2\theta_c \xi d]$ by $\tilde{s}$, the production of $\bar{\tilde{s}}$ (the weak eigenstate conjugate to $c$) vanishes at zeroth order for $W^+$ exchange. What is conventionally regarded as NLO in the fixed order expansion includes NLO production of this conjugate eigenstate, and thus includes the ${\cal O}(\alpha_S^2)$ FF coefficient functions. Then results up to this order for the transformation of the FF coefficients are ($i\in\{2,3\}$):
\begin{align}
C_{i(s,d)}^{(1)\DIS}&=C_{i(s,d)}^{(1)\msbar}-C_{i(s,d)}^{(0)}\otimes C_{2q}^{(1)\msbar};\label{cisd1DIS}\\
C_{i,g}^{FF(1)\DIS}&=C_{i,g}^{FF(1)\msbar}-C_{2,g}^{(1)\msbar}\otimes(C_{2s}^{(0)}+C_{2d}^{(0)});\label{cig1DIS}\\
C_{i,g}^{FF(2)\DIS}&=C_{i,g}^{FF(2)\msbar}+2n_fC_{i,g}^{FF(1)\msbar}\otimes C_{2,g}^{(1)\msbar}-C_{i,g}^{(2)\msbar}\otimes(C_{2,s}^{(0)}+C_{2,d}^{(0)})\notag\\
&-2n_f C_{2,g}^{(1)\msbar}\otimes C_{2,g}^{(1)\msbar}\otimes(C_{2,s}^{(0)}+C_{2,d}^{(0)})-C_{2,g}^{(1)\msbar}\otimes(C_{2,s}^{(1)\DIS}+C_{2,d}^{(1)\DIS})\label{cig2DIS}\\
C_{i,q}^{FF(2)\DIS}&=C_{2,q}^{FF(2)\msbar}+C_{i,g}^{FF(1)\msbar}\otimes C_{2q}^{(1)\msbar}-\frac{1}{n_f}C_{2,ps}^{(2)\msbar}-C_{i,g}^{(1)\msbar}\otimes C_{2,q}^{(1)\msbar}\otimes(C_{2,s}^{(0)}+C_{2,d}^{(0)})\label{ciq2DIS}.
\end{align}
Analogous results apply in the case of $W^-$ exchange, where production of ${\tilde{s}}$ vanishes at LO. Some care is needed to obtain the above results, as one must use the $\msbar\rightarrow\DIS$ transformation of a single quark species, which is:
\begin{equation}
q_i^{\DIS}=C_{2,ns}^{\msbar}\otimes q_{i}^{\msbar}+\frac{1}{2n_f}C_{2,ps}^{\msbar}\otimes\Sigma^{\msbar}+C_{2,g}^{\msbar}\otimes g^{\msbar},
\label{ms2disqi}
\end{equation}
with a similar equation for antiquarks. Some simplification of the above results is possible. In \cite{TRCC} it is noted that the NLO correction to single charm production, represented by $C_{i(d,s)}^{(1)\msbar}$, is negligible in practice, and can be approximated by the massless quark coefficient. In the DIS scheme, this is zero and so can be neglected. Note this also means that one can neglect the final term in equation (\ref{cig2DIS}).\\

A slight problem arises in that the NLO quantities $C_{i,a}^{FF(2)\msbar}$ are not known for charged current scattering. In \cite{Thorne06} they are approximated by evaluating the neutral current results with modified threshold behaviour i.e. dependence on $x_{max}$ replaced with dependence on $x_{max}^{CC}$. We follow this approach here, although the transformation terms to the DIS scheme can be parameterised using exact results given that the LO quantities are all known. \\

The variable flavour coefficients are only needed up to ${\cal O}(\alpha_S)$, and adopting the DIS($\chi$) scheme choice as outlined in \cite{WT2} gives:
\begin{align}
C_{i(s,d)}^{VF(1)\DIS}&=C_{i(s,d)}^{(1)\msbar}-C_{2,q}^{(1)\msbar}\otimes C_{2(s,d)}^{(0)};\label{c2s1vfCC}\\
C_{2,g}^{VF(1)\DIS}&=C_{2,g}^{VF(1)\msbar}-C_{2,c}^{(0)}\otimes C_{2,g}^{(1)\msbar}-C_{2,g}^{(1)\msbar}\otimes(C_{2,s}^{(0)}+C_{2,d}^{(0)}).\label{c2g1vfCC}
\end{align}
Approximation of the NLO single charm coefficient by its massless counterpart means that one can neglect the contribution from equation (\ref{c2s1vfCC}) in the DIS scheme. \\

One also needs to specify a scheme for the NLO coupling when heavy flavours are involved. We use the prescription of Marciano \cite{Marciano}. The NLL approximate resummed splitting and coefficient functions have been determined by solving the BFKL equation at NLL order but with the LO running coupling. It is not then correct to use the NLO running coupling in these expressions\footnote{The overall normalisation in the impact factors does contain a power of the NLO coupling, as the scale of this coupling is $Q^2$ and not $k^2$ when solving the BFKL equation (see \cite{Thorne}).}, and in the fitting program one must define the LO coupling by matching with the NLO coupling at some scale. We use $Q^2=16\text{GeV}^2$, as this is guaranteed to correspond to four active quark flavours. One then has:
\begin{equation}
\frac{1}{\beta_0 t^{LO}}=\alpha_S^{NLO}(t^{NLO})
\label{tLO}
\end{equation}
which defines the LO $t$ parameter at this scale, where as usual $t=\log(Q^2/\Lambda^2)$. \\

For the DY data it is conventional to implement the NLO corrections to the cross-section via a multiplicative $K$-factor. Rather than produce a $K$-factor in the DIS scheme, it is sufficient to explicitly transform the parton densities to the DIS scheme in the existing $\msbar$ scheme code. We adopt such an approach here. \\

\begin{figure}
\begin{center}
\scalebox{0.8}{\includegraphics{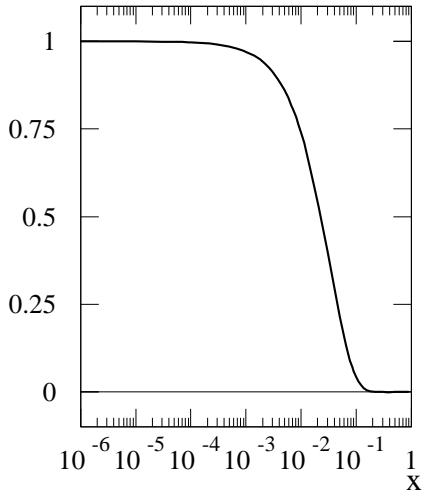}}
\caption{The function $(1-x)^{30}$ used to filter out high $x$ information in the alternative NLL resummed fit.}
\label{damp}
\end{center}
\end{figure}

As noted in section \ref{impacts}, some of the NLL improved splitting and coefficient functions contain terms at ${\cal O}(N^0)$ and ${\cal O}(N)$ in Mellin space, which we model by a suitable function of $x$ in practice. It is perhaps unnatural that the small $x$ resummation should lead to such noticeable effects at large $x$. Thus, for comparison with the standard NLL resummed fit, an alternative fit is also presented in this chapter in which the all resummed quantities have been multiplied by a factor\footnote{One may choose any power of $(1-x)$ for this damping function that is sufficient to filter out the high $x$ behaviour. Note that a function of form $(1-x)^N$ leads to a cutoff of the resummation for $x\gtrsim N^{-1}$. We have checked that $N=20$ gives very similar results.} $(1-x)^{30}$. A plot of this function is shown in figure \ref{damp}, and one can see that its effect is to filter out high $x$ information in the resummed quantities for $x\gtrsim 10^{-2}$.
\subsection{Results}
The $\chi^2$ values for each data set obtained in the NLL and alternative NLL fits are shown in table \ref{NLLresults}, alongside results obtained using a NLO DIS scheme fit with no resummations. We also include for comparison the LL fit values described in \cite{WT2}. The numerical results quoted in that paper are erroneous due to a mistake in the fitting code. The LL values in table \ref{NLLresults} are obtained with the corrected program. 
\begin{table}
\begin{center}
\begin{tabular}{|cc|cccc|}
\hline
Data Set & No. data pts & $\chi^2_{NLL}$ & $\chi^2_{NLL(2)}$ &$\chi^2_{NLO,DIS}$ & $\chi^2_{LL}$\\
\hline
H1 ep & 417 &352 &413 &415 &341\\
ZEUS ep &356 &273 &245 &239 &273\\
$F_2^c$ &27 &25 &43 &40 &25\\
BCDMS $\mu$ p &167 &171 &158 &180 &186\\
BCDMS $\mu$ D &155 &198 &202 &219 &223\\
NMC $\mu$ p &126 &126 &131 &139 &120\\
NMC $\mu$ D &126 &102 &103 &103 &97\\
SLAC ep &53 &65 &78 &76 &92\\
SLAC eD &54 &53 &81 &65 &82\\
E665 $\mu$ p &53 &61 &60 &61 &57\\
E665 $\mu$ D &53 &51 &58 &66 &54\\
CCFR $F_2^{\nu N}$ &74 &74 &75 &62 &97\\
CCFR $F_3^{\mu N}$ &105 &115 &104 &114 &138\\
H1 CC & 28 &29 &31 &32 &32\\
ZEUS CC &30 &44 &47 &49 &44\\
NMC $n/p$ &156 &147 &165 &154 &157\\
E866/ NuSea DY &174 &240 &225 &234 &296\\
NA51 DY asym. &1 &5 &5 &13 &16\\
E866 $\sigma_{DY}^{pD}/\sigma_{DY}^{pp}$ &15 &7 &7 &14 &7\\
CDF $W$ asym. &11 &16 &16 &17 &20\\
\hline
Total &2181 &2154 &2247 &2289 &2357\\
\hline 
\end{tabular}
\caption{The quality of fit from the NLL and modified ($(1-x)^{30}$ weighted, labelled NLL(2)) fits for each dataset, as well as results from a NLO DIS scheme fit and the LL fit of \cite{WT2}.}
\label{NLLresults}
\end{center}
\end{table}
A feature of the previous LL fit, not explicitly noted in \cite{WT2}, is the need for a significant violation of momentum conservation due to the fact that the LL resummed splitting functions do not satisfy the momentum sum rule. The input partons in the LL fit of table \ref{NLLresults} carry a total of $78\%$ momentum, consistent with the fact that the evolution is enhanced over a large range of $x$. All of the next-to-leading fits in table \ref{NLLresults}, on the other hand, are obtained with full momentum conservation imposed. The input partons carry $100\%$ momentum, and we saw earlier that the NLL resummed splitting functions satisfy the momentum sum rule to a very good approximation. \\

All three next-to-leading fits are good all-round fits. Of the two resummed fits, the standard NLL fit clearly outperforms the NLO DIS-scheme fit, whereas the fit using the $(1-x)^{30}$ weighting gives less of an improvement. The fit to the H1 data in particular suffers without the high $x$ part of the resummation, indicating that this is indeed needed for a truly good fit. Comparing the NLL resummed fit with the NLO results, it seems that the introduction of resummations seem to decrease the tension between different data sets -- even though one might expect some tension from not including resummations 
in the coefficient functions for all observables. The NLL fit is much better than the LL fit. One sees a more satisfactory description of the DY data (due to the NLO corrections), and also the fit to the deuterium data improves. This is a hint that the sea and valence distributions have the correct relative behaviour, and thus that the suppression of small $x$ resummations anticipated in \cite{WT2} has indeed occurred. \\

\begin{figure}
\begin{center}
\scalebox{0.7}{\includegraphics{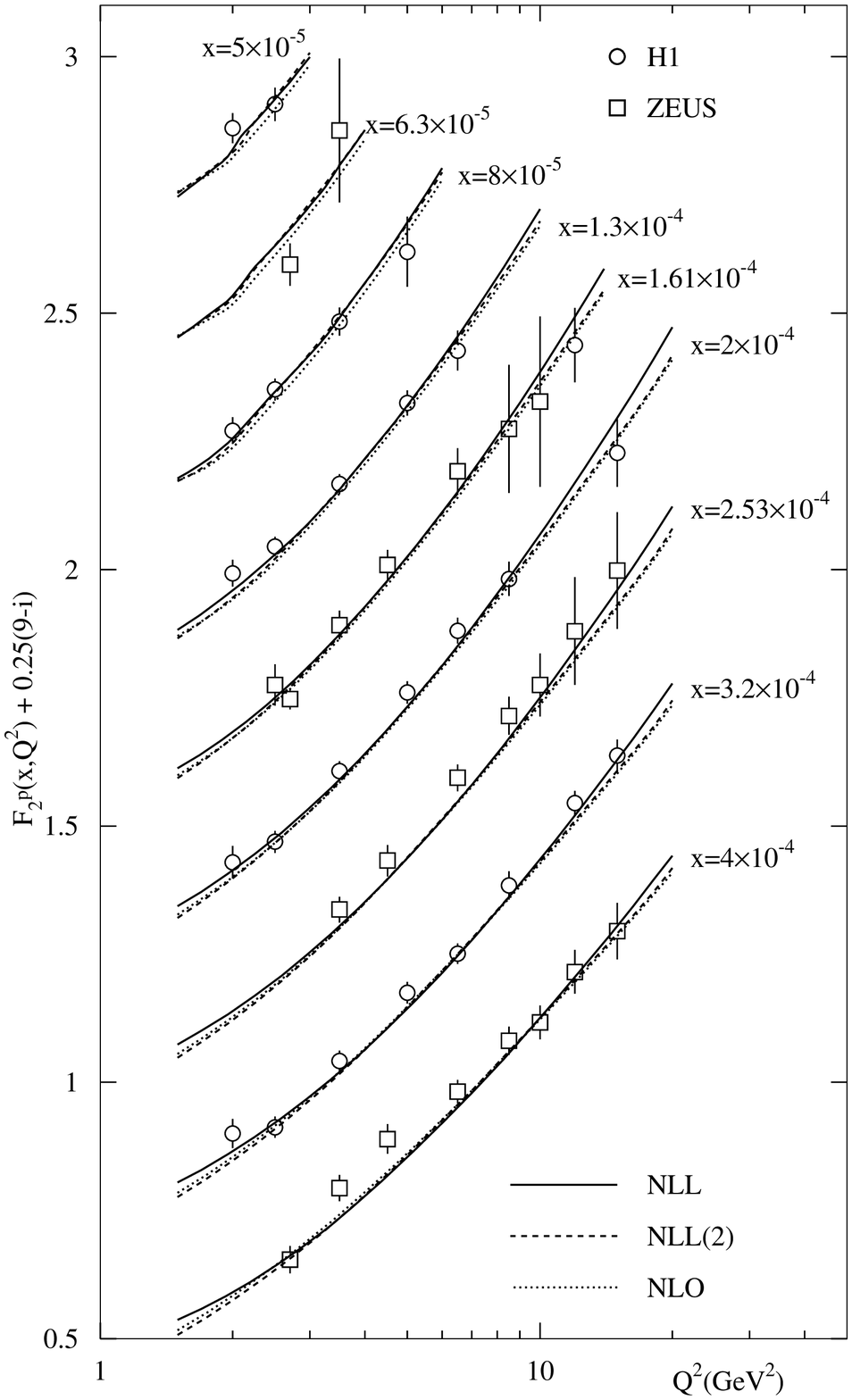}}
\caption{Theoretical predictions for the structure function $F_2$ alongside the data, for $5\times10^{-5}\leq x\leq4\times10^{-4}$.}
\label{dataNLL}
\end{center}
\end{figure}

\begin{figure}
\begin{center}
\scalebox{0.7}{\includegraphics{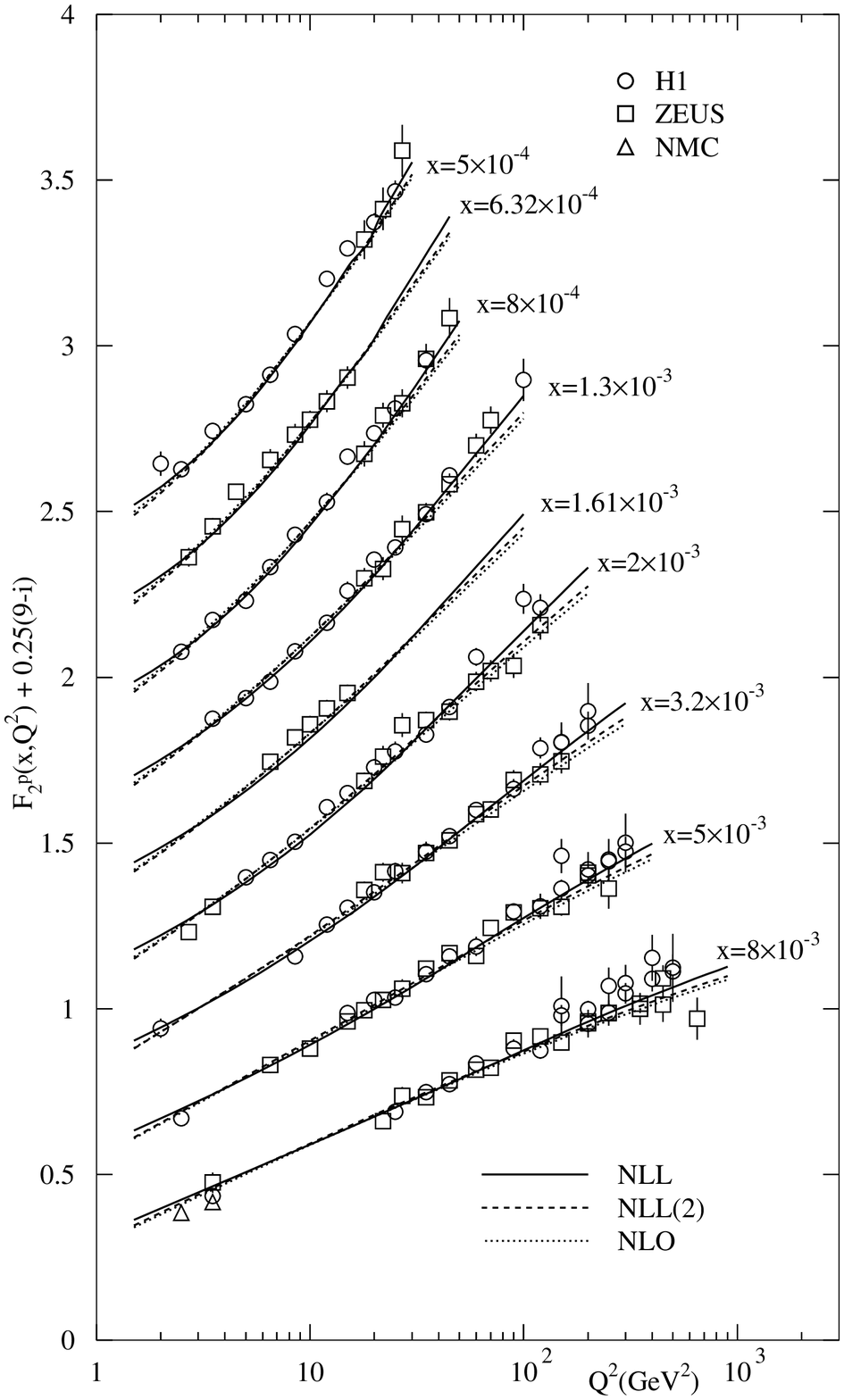}}
\caption{Theoretical predictions for the structure function $F_2$ alongside the data, for $5\times10^{-4}\leq x\leq8\times10^{-3}$.}
\label{data2NLL}
\end{center}
\end{figure}

The $F_{2}$ values from the fit at small $x$ are shown with the data in figures \ref{dataNLL} and \ref{data2NLL}. 
As already noted at LL order \cite{WT2}, one sees that resummations are needed to give the desired slope in $F_2$ as $Q^2$ increases, as is particularly evident in figure \ref{data2NLL}. Curiously, the modified NLL fit with the high $x$ part of the resummation filtered out (labelled NLL(2) in the plots and tables) gives a very similar prediction for $F_2$ to the NLO fit. One expects the theory prediction at $10^{-4}\lesssim x \lesssim 10^{-3}$ to be sensitive to the coefficient and splitting functions at $10^{-2}\lesssim x\lesssim 10^{-1}$ through the convolutions, and so it is perhaps no surprise that filtering out the resummations in this latter regime greatly diminishes their effect at lower $x$. The modified fit only converges with the full NLL fit at very low $x\simeq 10^{-5}$. This suggests why the modified fit is less successful than the full NLL fit - it differs slightly from the NLO approach in the region where the fixed order fit works well, and does not contain enough of the small $x$ resummation to improve matters very well at low $x$. \\

\begin{figure}
\begin{center}
\scalebox{0.7}{\includegraphics{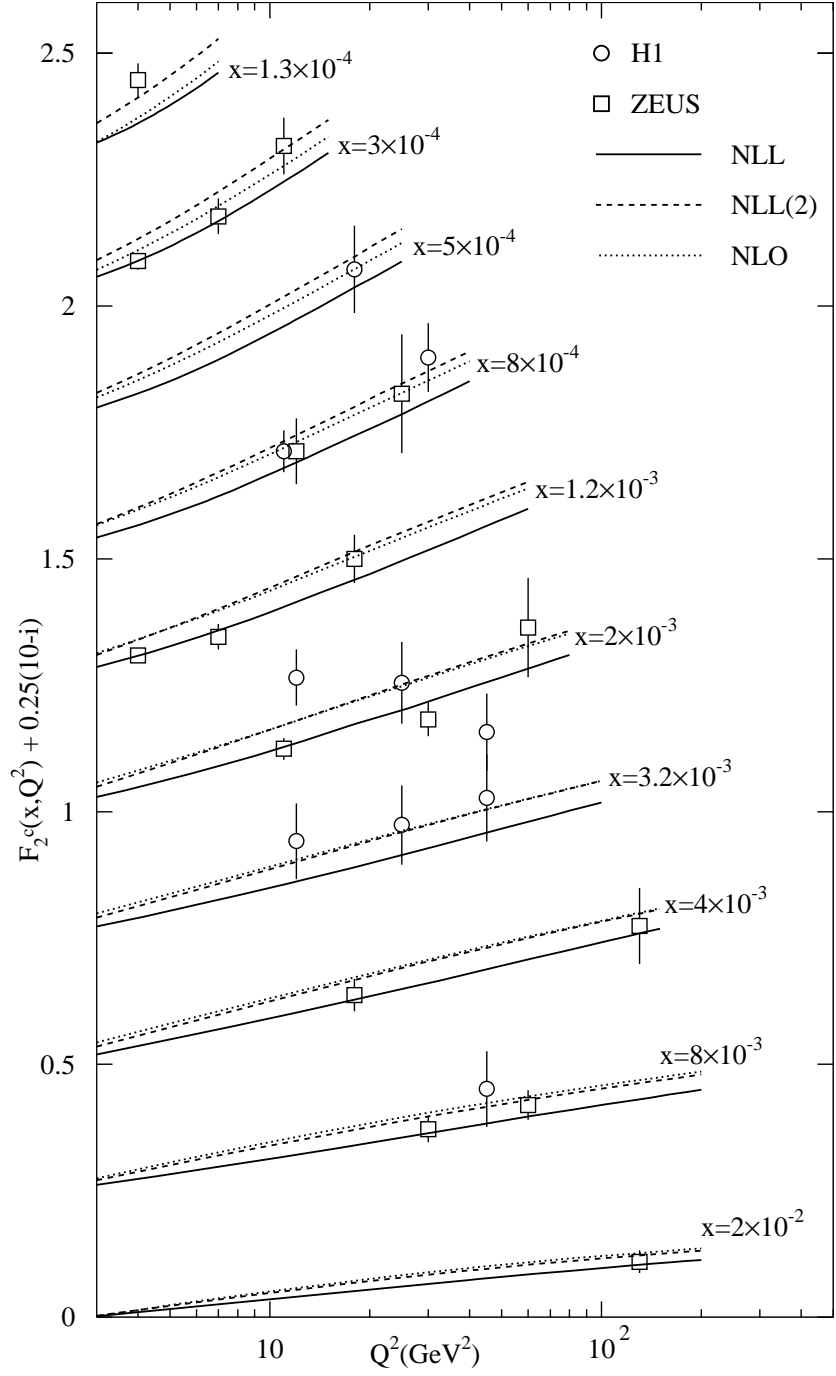}}
\caption{Resummed predictions for the charm structure function $F_2^c$ alongside HERA data, for $1.3\times10^{-4}\leq x\leq 2\times10^{-2}$.}
\label{cdatanll}
\end{center}
\end{figure}

The charm data is shown in figure \ref{cdatanll}.
Once again, the modified resummed and NLO fits give very similar predictions until one reaches very low $x$. The full NLL $F_{2}^c$ lies above the fixed order prediction only at the lowest $x$ values, and otherwise wants to tend towards the lower limit allowed by the data. Nevertheless, the NLL fit to the charm data is excellent as can be seen from table \ref{NLLresults}.\\

The strong couplings $\alpha_S(M_Z^2)$ required by each fit are given in table \ref{couplings}.
\begin{table}
\begin{center}
\begin{tabular}{|c|c|}
\hline
Fit & $\alpha_S(M_Z^2)$\\
\hline
NLL & 0.1184\\
NLL(2) & 0.1130\\
NLO & 0.1181\\
\hline
\end{tabular}
\caption{The values of the strong coupling resulting from each of the next to leading fits. The world average (at NLO) is 0.1187(20) \cite{PDG}.}
\label{couplings}
\end{center}
\end{table}
The NLO and full NLL results are compatible with the world average, whereas the NLL(2) coupling is rather low\footnote{Strictly speaking, the couplings in the resummed and fixed order fits are not the same quantity.}. The latter result is not very precisely determined, however. For example, a value $\alpha_S(M_Z^2)=0.116$ gives a very similar fit (with minor reshuffling of the gluon parameters) in the NLL(2) program. That the modified fit gives the lowest value is expected given that the modified resummed splitting functions have an enhancement at low $x$, but have less of a dip at moderate $x$ as this has been filtered out (see figure \ref{splitsplot}). Thus a lower coupling is needed to fit the data at small $x$.\\

\begin{figure}
\begin{center}
\scalebox{0.8}{\includegraphics{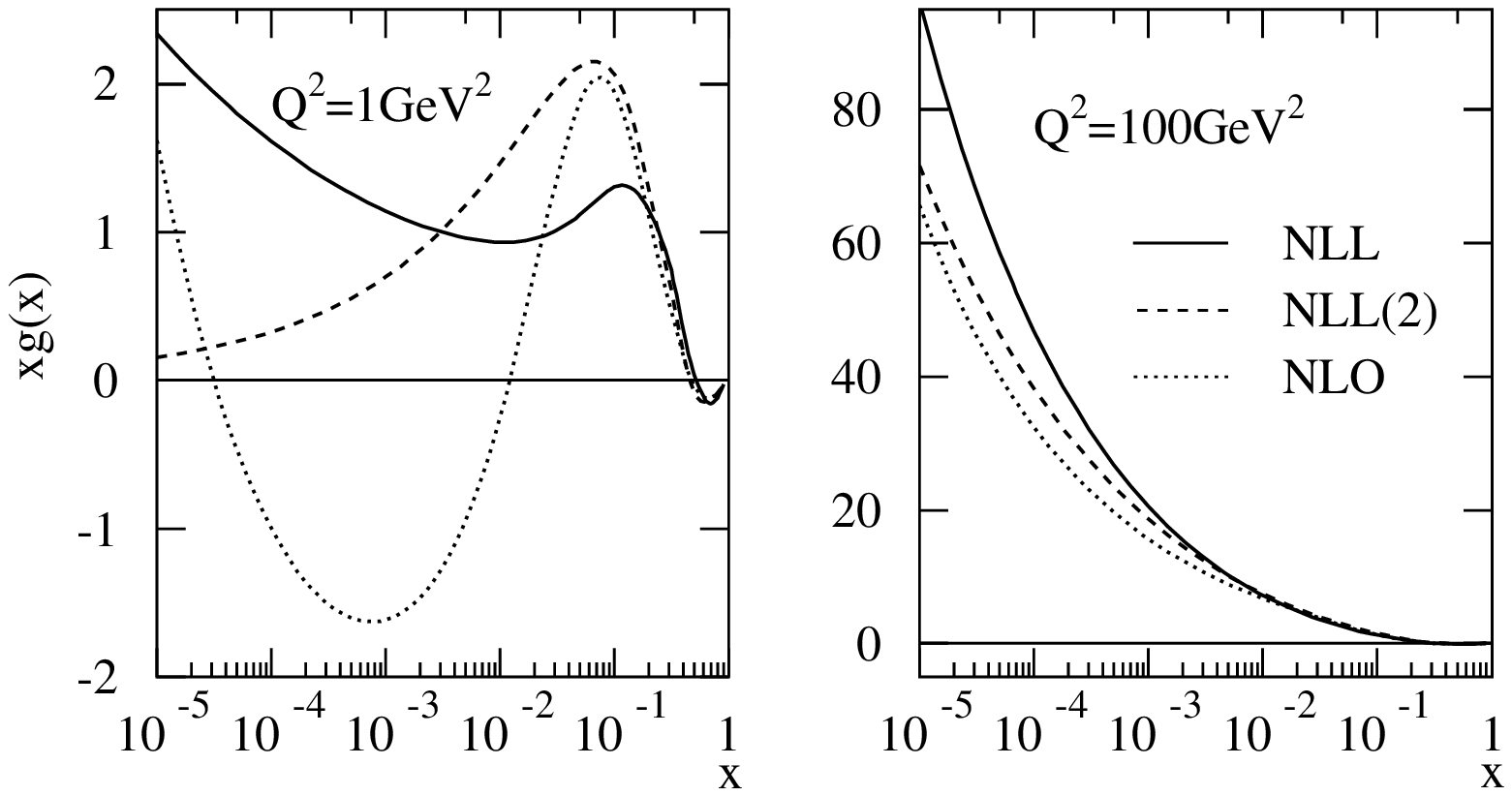}}
\caption{The gluon distributions obtained from each of the next to leading fits at $Q^2=1\text{GeV}^2$ and $Q^2=100\text{GeV}^2$.}
\label{gluonsNLL}
\end{center}
\end{figure}

The improved fit to scattering data suggests that in the NLL fit the resummations are confined to lower values of $x$. To investigate this in more detail, one may examine the gluons obtained in each of the fits. These are shown in figure \ref{gluonsNLL} for $Q^2=1\text{GeV}^2$ and $Q^2=100\,\text{GeV}^2$. 
At the input scale $Q^2=1\text{GeV}^2$ the fixed order DIS-scheme gluon, although positive at very small $x$, is negative over a large range of $x$. On the other hand the resummed gluons are positive in this regime, with the full NLL gluon rising at small $x$. Although positivity of the gluon is not necessary, such a feature is somewhat more physically appealing (for example, structure functions will then be positive at low $Q^2$
as long as the coefficient functions are). At moderately high $x$, the modified NLL gluon is very similar to the NLO result, as expected from filtering out the small $x$ resummations here. The gluon from the modified fit also lies below the full NLL gluon at small $x$, which is to be expected given the increased dip at moderate $x$ in the full NLL splitting functions. \\

Looking at the right-hand panel of figure \ref{gluonsNLL}, one sees that 
at higher $Q^2$ the three gluons are extremely similar at moderate and high $x$, with a noticeable difference occurring only for $x\lesssim 10^{-2}$. The full resummed gluon lies above the fixed order result as $x\rightarrow 0$ due to the increased small $x$ evolution, and the gluon from the modified fit is closer to the fixed order gluon than the full resummed gluon. This is consistent with the above noted fact that the predictions for $F_2$ in the fixed order and modified resummed fits are very similar until very small $x$.\\

\begin{figure}
\begin{center}
\scalebox{0.55}{\includegraphics{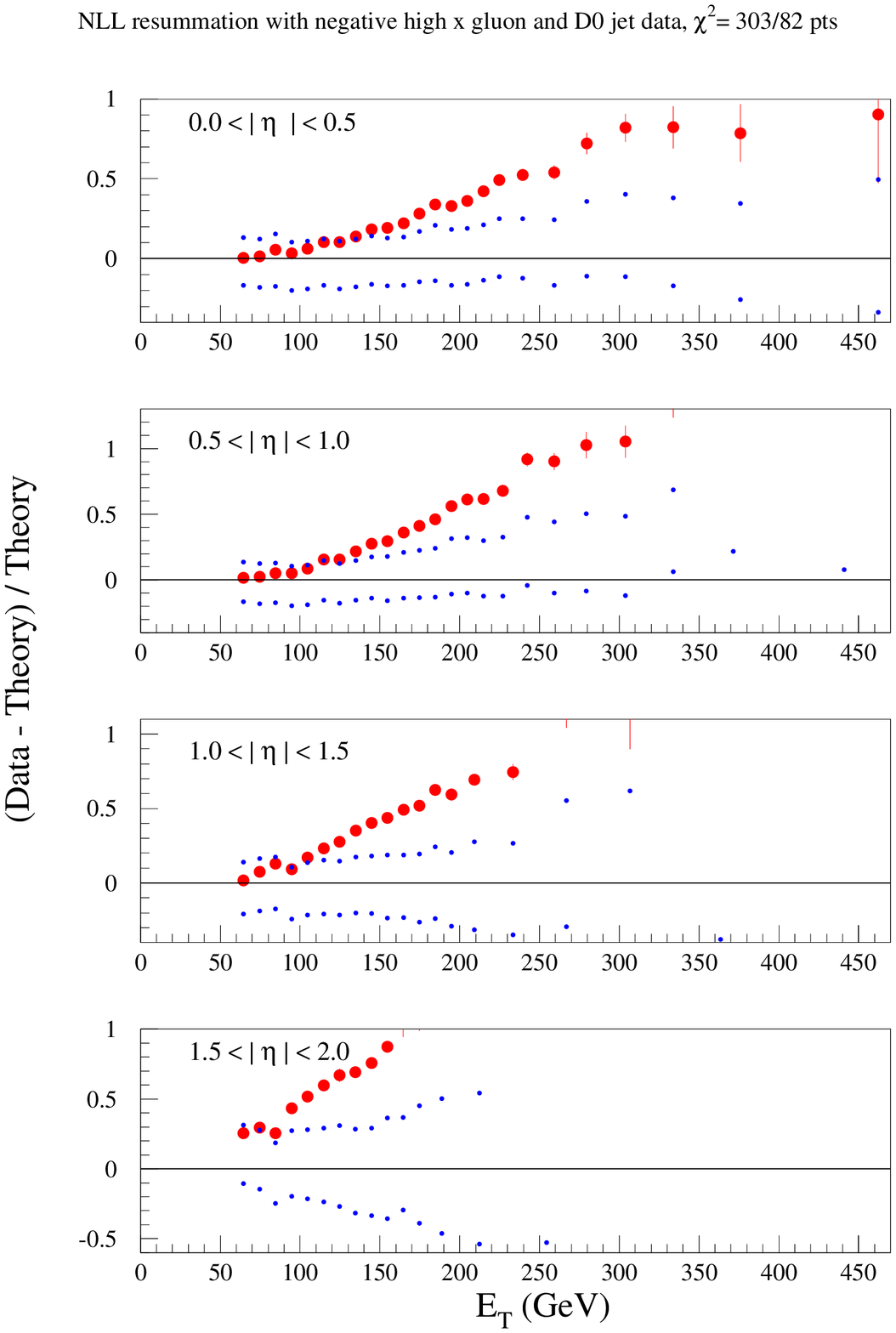}\includegraphics{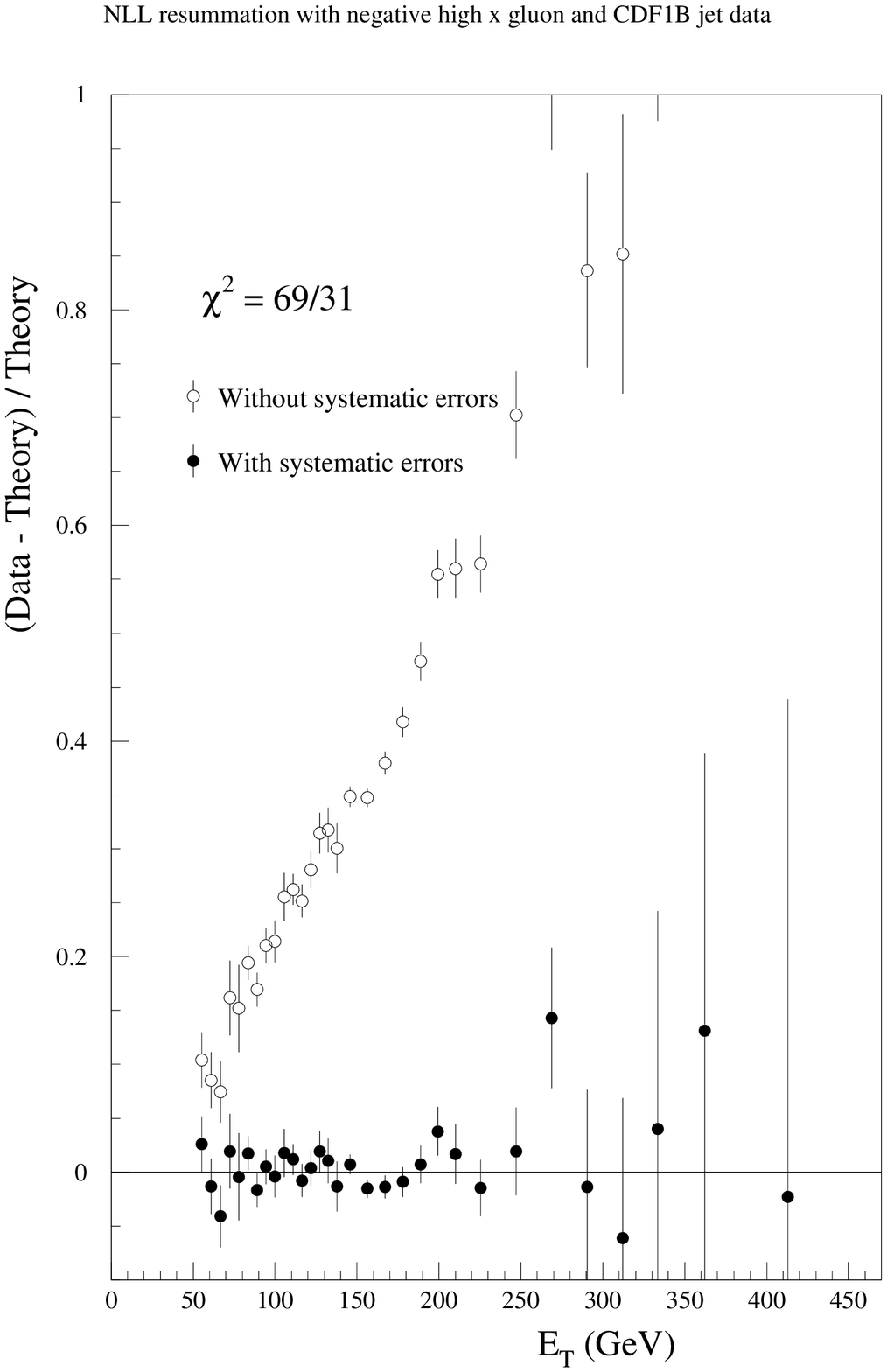}}
\caption{The comparison of theory to the jet data for the NLL parton
distributions using coefficient functions at NLO in $\alpha_S$. For the D0
data \cite{Tevjet1} (left) the blue points represent the size of the 
systematic errors while the uncorrelated errors are shown explicitly. For the 
CDF data \cite{Tevjet2} (right) the movement of theory relative to data using
the systematic errors is shown.} 
\label{jets1}
\end{center}
\end{figure}

Despite the good fits to the DIS and related data from all of the approaches considered here (with the full NLL fit giving a global $\chi^2$ of less than one per point), there is a worrying feature in the input gluon at high $x$. Both the fixed order and resummed gluons show a negative dip at very high $x$. This behaviour of the gluon is not consistent with the Tevatron jet data, which prefers at least a positive semi-definite gluon at high $x$. It was shown in \cite{MRSTgluon} that a positive DIS-scheme gluon at high $x$ automatically gives a good fit to the Tevatron jet data. We now see that a negative gluon certainly does not. The comparison to the jet data achieved by transformation of the gluons from the above fits to the $\msbar$ scheme gives $\chi^2_{jets}\simeq300$ or worse for 113 data points. The quality of the comparison is shown for the NLL
partons in figure \ref{jets1}. The negative high-$x$ gluon in the DIS scheme 
is in fact still slightly negative in the $\msbar$ scheme, and as such 
at high values of $E_T$ and/or high rapidity the theoretical prediction is 
very much below the data. This is similar, but slightly better, for the 
NLL(2) and DIS-scheme NLO partons in this section. In the case of the D0 data
the resulting $\chi^2$ for all the parton sets is very poor indeed, i.e.
$\chi^2 \sim 300$ for 82 points. For the
CDF data it is not so much worse than the standard good fits $\chi^2 \sim 70$
rather than $50$ for 31 points, but this is achieved only by assuming that 
the systematic errors are conspiring to make the true data very much smaller 
than the central value presented. Clearly this type of high-$x$ gluon 
distribution in the DIS scheme is not acceptable. The solution to    
this problem is considered in more detail in the next section. 
\subsection{Global Fit with Positive Large $x$ Gluon}
In the NLO fixed order fit one can constrain the gluon input parameters to ensure a positive high $x$ gluon. Subsequent refits will converge to the local minimum satisfying this condition. In the full NLL resummed fit, fixing the gluon parameters is not sufficient to give the desired effect in the gluon, due to the term proportional to $\delta(1-x)$ in $P_{gg}$ which is negative
in practice. By the colour charge relation of equation (\ref{colourcharge}), this gives a negative delta function contribution to $P_{gq}$ leading to a negative evolution in the high $x$ gluon whose derivative at a given $Q^2$ becomes proportional to the quark singlet. This is a relatively large effect given that the gluon dies away at large $x$ much more quickly than the quarks. However, there are strictly speaking no $\delta(1-x)$ terms in the full fixed order $P_{gq}$ -- the resummed delta function is an artifact of expanding about $N=0$. One can replace the delta function term with a function that has the same first moment and little effect at small $x$. A suitable choice is $N^{-1}$, as the coefficient of this term is sufficiently small to have negligible effect at low $x$. \\

\begin{figure}
\begin{center}
\scalebox{0.7}{\includegraphics{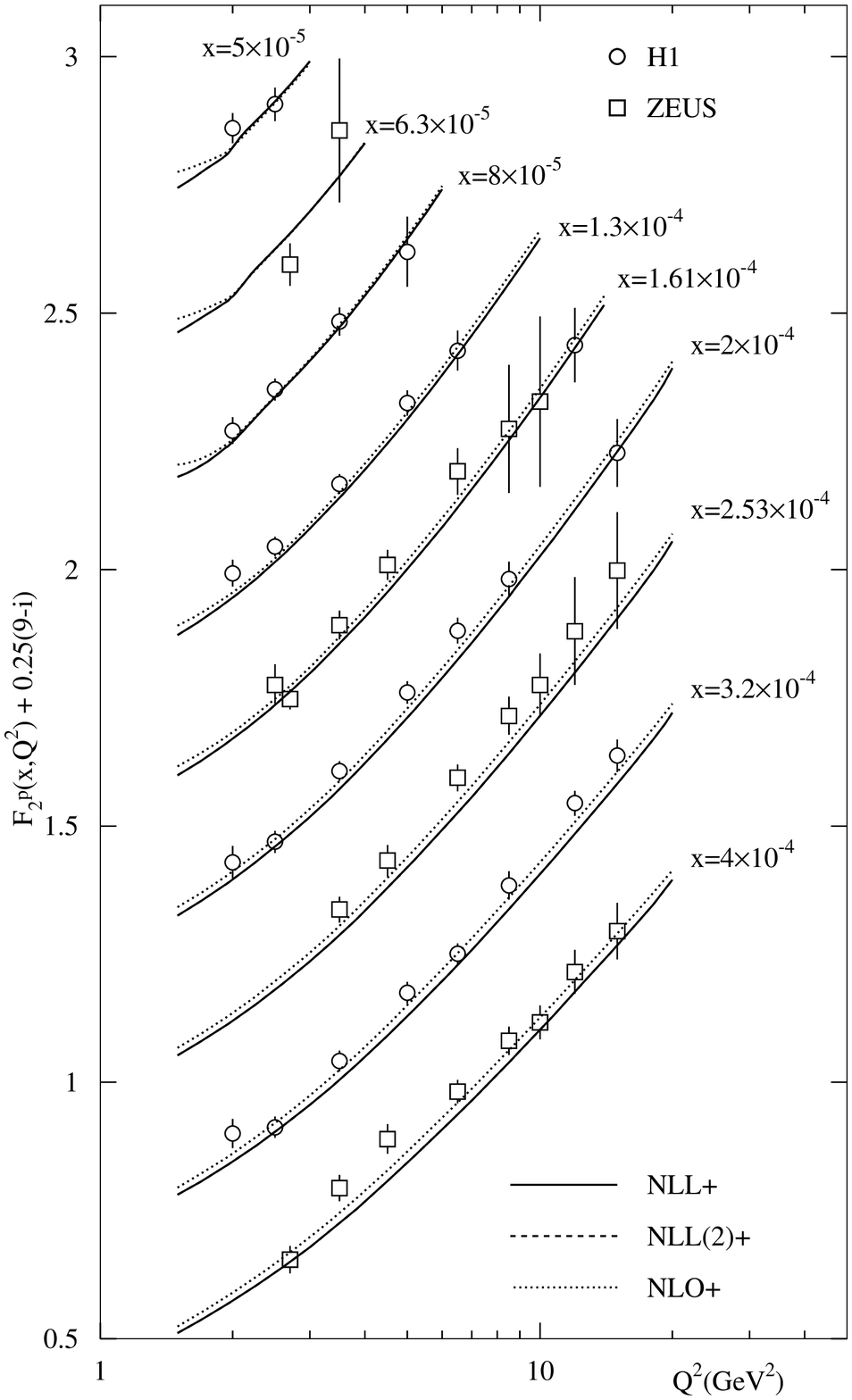}}
\caption{Theoretical predictions for the structure function $F_2$ alongside the data, for $5\times10^{-5}\leq x\leq4\times10^{-4}$, with the gluon constrained to be positive at high $x$.}
\label{dataNLL2}
\end{center}
\end{figure}

\begin{figure}
\begin{center}
\scalebox{0.7}{\includegraphics{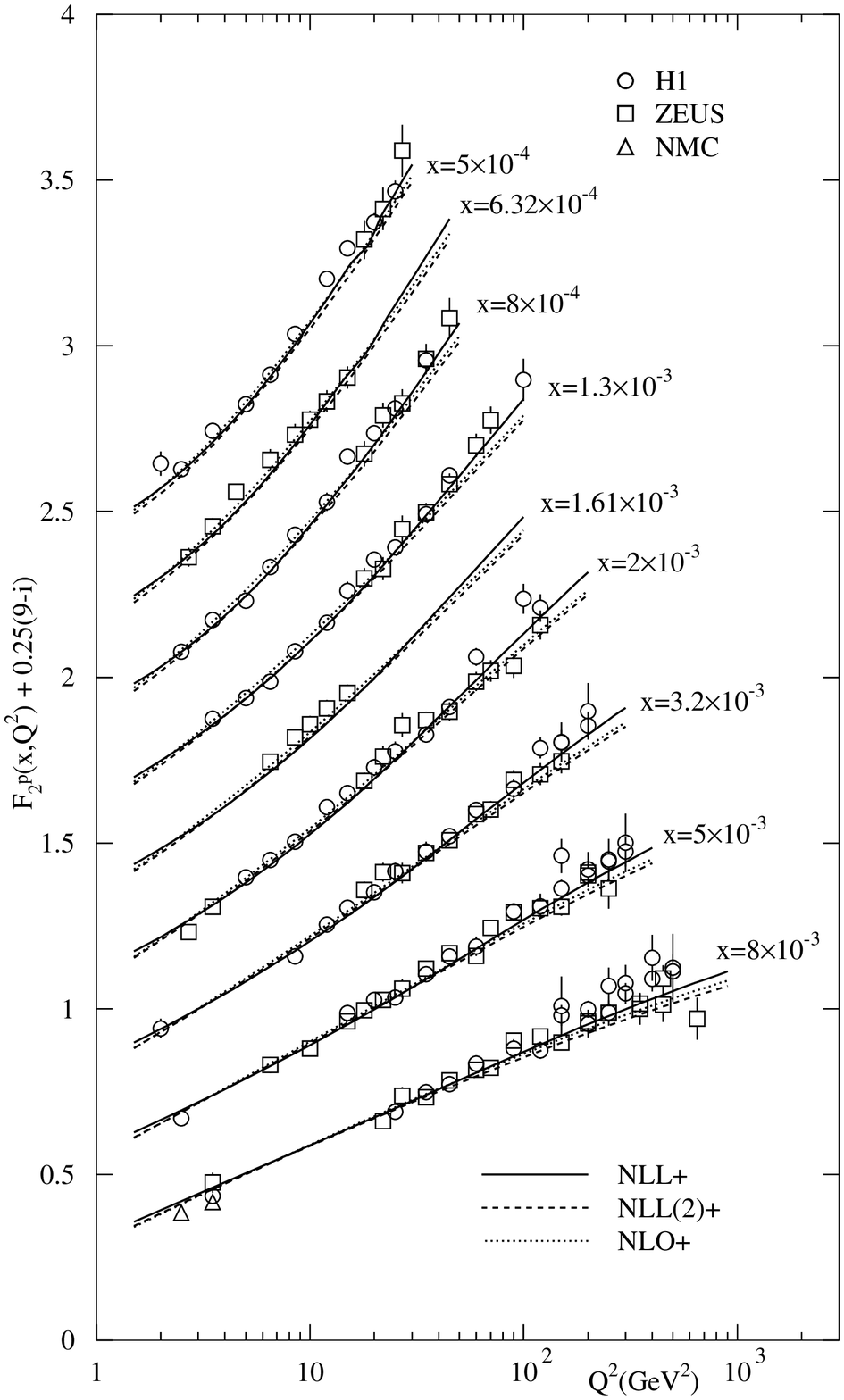}}
\caption{Theoretical predictions for the structure function $F_2$ alongside the data, for $5\times10^{-4}\leq x\leq8\times10^{-3}$, with the gluon constrained to be positive at high $x$.}
\label{data2NLL2}
\end{center}
\end{figure}

\begin{figure}
\begin{center}
\scalebox{0.7}{\includegraphics{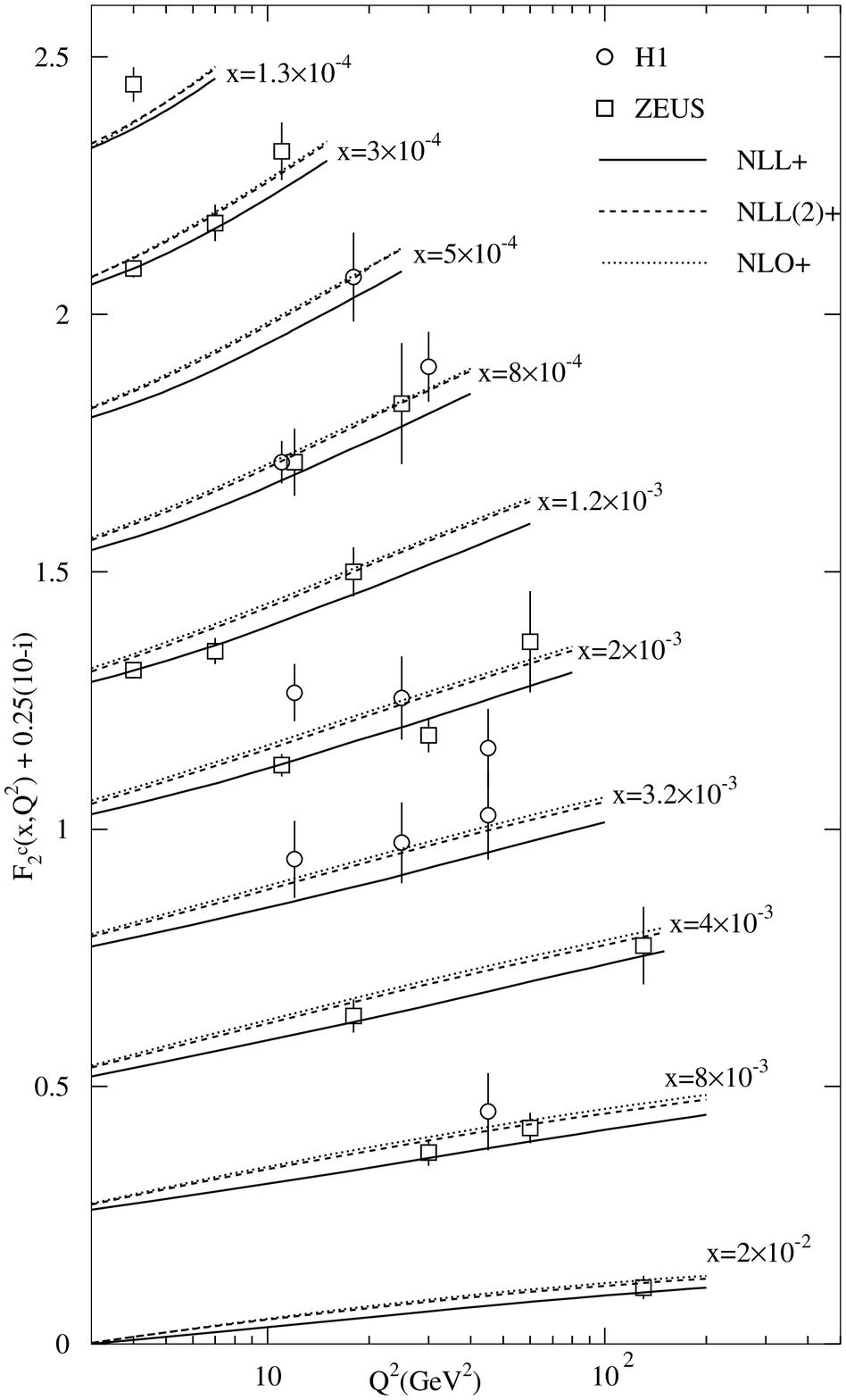}}
\caption{Resummed predictions for the charm structure function $F_2^c$ alongside HERA data, for $1.3\times10^{-4}\leq x\leq 2\times10^{-2}$, where the gluon is constrained to be positive at high $x$}
\label{cdatanll2}
\end{center}
\end{figure}

\begin{figure}
\begin{center}
\scalebox{0.8}{\includegraphics{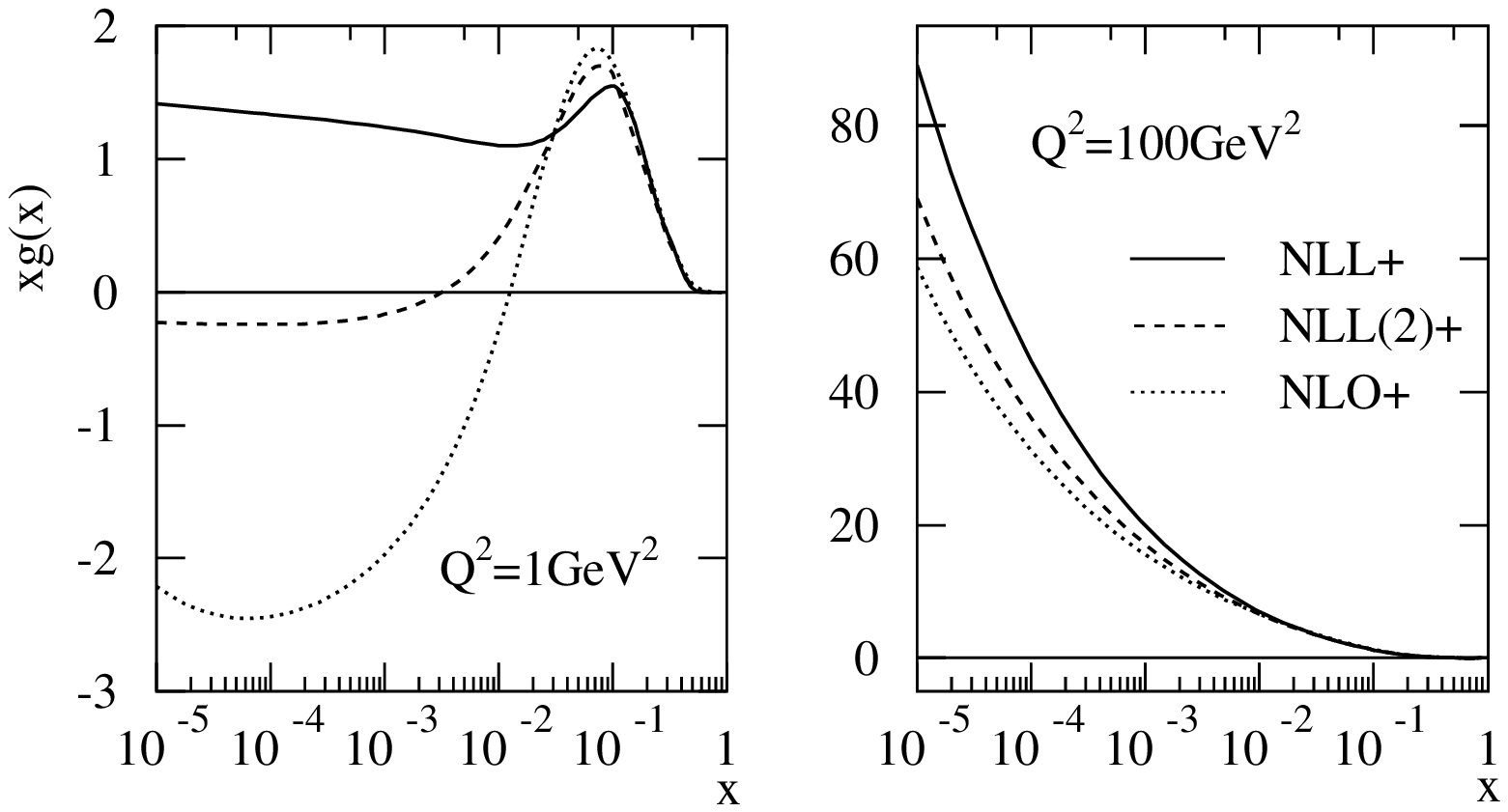}}
\caption{The gluon distributions obtained from fits where the gluon is constrained to be positive at high $x$, for $Q^2=1\text{GeV}^2$ and $Q^2=100\text{GeV}^2$.}
\label{gluonsNLL2}
\end{center}
\end{figure}

We consider the full NLL resummed fit together with a fit with the resummations weighted by a $(1-x)^{30}$ factor as before. The resulting $\chi^2$ values are shown in table \ref{NLLresults+}, where the fits are labelled with a plus sign to distinguish from the previously described next to leading results.
\begin{table}
\begin{center}
\begin{tabular}{|cc|ccc|}
\hline
Data Set & No. data pts & $\chi^2_{NLL+}$ & $\chi^2_{NLL(2)+}$ &$\chi^2_{NLO,DIS+}$\\
\hline
H1 ep &417 &360 &438 &423\\
ZEUS ep &356 &260 &251 &242\\
$F_2^c$ &27 &26 &32 &40\\
BCDMS $\mu$ p &167 &166 &188 &186\\
BCDMS $\mu$ D &155 &228 &217 &231\\
NMC $\mu$ p &126 &126 &140 &130\\
NMC $\mu$ D &126 &95 &106 &100\\
SLAC ep &53 &74 &72 &73\\
SLAC eD &54 &58 &75 &64\\
E665 $\mu$ p &53 &62 &59 &59\\
E665 $\mu$ D &53 &49 &58 &61\\
CCFR $F_2^{\nu N}$ &74 &58 &62 &67\\
CCFR $F_3^{\mu N}$ &105 &160 &124 &153\\
H1 CC & 28 &30 &31 &31\\
ZEUS CC &30 &45 &44 &45\\
NMC $n/p$ &156 &163 &153 &167\\
E866/ NuSea DY &174 &255 &241 &246\\
NA51 DY asym. &1 &10 &5 &7\\
E866 $\sigma_{DY}^{pD}/\sigma_{DY}^{pp}$ &15 &9 &7 &9\\
CDF $W$ asym. &11 &15 &17 &18\\
\hline
Total &2181 &2249 &2318 &2352\\
\hline 
\end{tabular}
\caption{The fit quality from the full NLL resummed, modified resummed and fixed order DIS-scheme fits, where the gluon is constrained to be positive at high $x$ as is required by the Tevatron jet data.}
\label{NLLresults+}
\end{center}
\end{table}
Both the fixed order and full NLL fits are very good overall. The resummed fit performs better, with the main improvement occurring in the H1 (including charged current scattering) and charm data. It is not possible to fit the H1 data as well as in the previous NLL fit once a positive gluon is required - the H1 data would like a lower gluon at high $x$, at the expense of a good fit to the Tevatron jet data\footnote{This has been noticed previously by global fits at NLO e.g. \cite{MRST2001,CTEQ}.}. The improvement in the NLL fit over the fixed order result is considerable, with an overall $\chi^2$ difference of $\simeq 100$ between the two fits. The modified resummed fit again shows much less improvement. The structure function $F_2$ from the fit is compared with the small $x$ data in figures \ref{dataNLL2}--\ref{cdatanll2}.
Again the slope of the resummed prediction is higher at small $x$ as is required by the data, but now the effect is less dramatic. The resummed prediction for the charmed data is again lower for most of the $x$ range, but gives an excellent fit. The modified resummed results are too close to the NLO results until very small $x$, and thus cannot provide as good a description. One may also compare the gluons obtained from the fits described here (shown in figure \ref{gluonsNLL2}) with those of figure \ref{gluonsNLL}.
The gluons at $Q^2=100\text{GeV}^2$ are much the same as before (except for high $x$), and mutually similar for $x\gtrsim10^{-2}$. Pleasingly, the gluons in the left-hand panel of figure \ref{gluonsNLL2} are also very alike at high $x$ suggesting that the small $x$ resummations are indeed not affecting the high $x$ behaviour of the gluon (hence, the structure functions). The NLO gluon is negative for $x\lesssim 10^{-2}$, turning over at very low $x$ but not becoming positive over the range of interest. On the other hand, the full NLL resummed gluon is still positive at small $x$ and growing as $x\rightarrow 0$, albeit very gently. Thus the resummed gluon is now positive definite over the whole range of $x$. The gluon from the modified fit is, as expected, somewhere between the full resummed and fixed order gluons and is now negative (but tending to zero) as $x\rightarrow 0$. This is a further indication of the inadequacy of the modified approach -- it now agrees closely with the NLO framework at high $x$, but has none of the benefits of the small $x$ resummation at moderately low $x$. That the gluons are now qualitatively different in the full and modified resummed fits indicates that the former approach is more appropriate. Note that the gluons in the right-hand panel deviate from each other at $x\simeq0.005$. This is consistent with a previous analysis which added phenomenological $x^{-1}\log^n(1/x)$ terms at ${\cal O}(\alpha_S^4)$ to the splitting functions, and found an input gluon qualitatively similar to that obtained by imposing a cut $x>0.005$ on the data, which thus gives some indication of the value of $x$ at which resummations become important.\\

\begin{figure}
\begin{center}
\scalebox{0.55}{\includegraphics{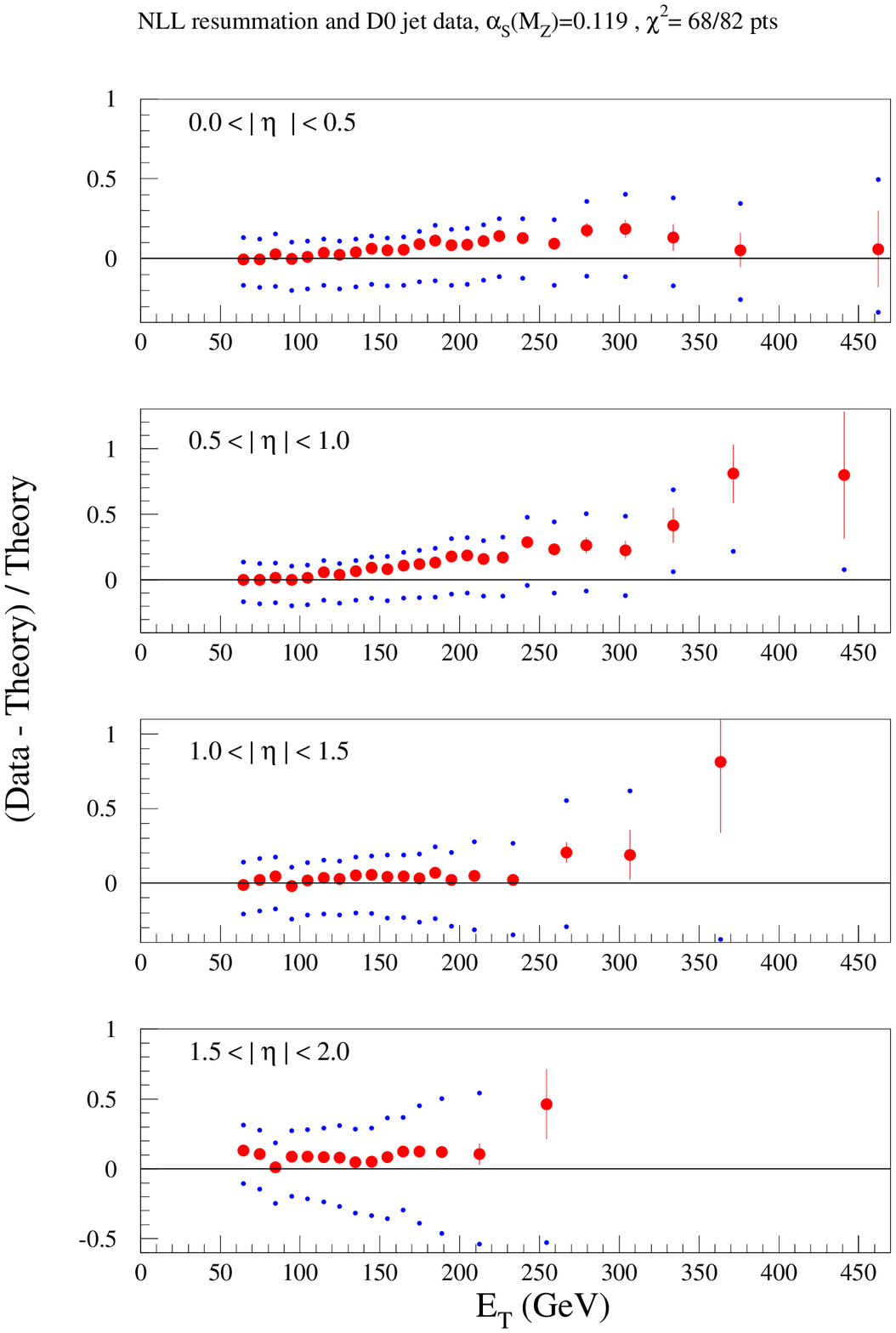}\includegraphics{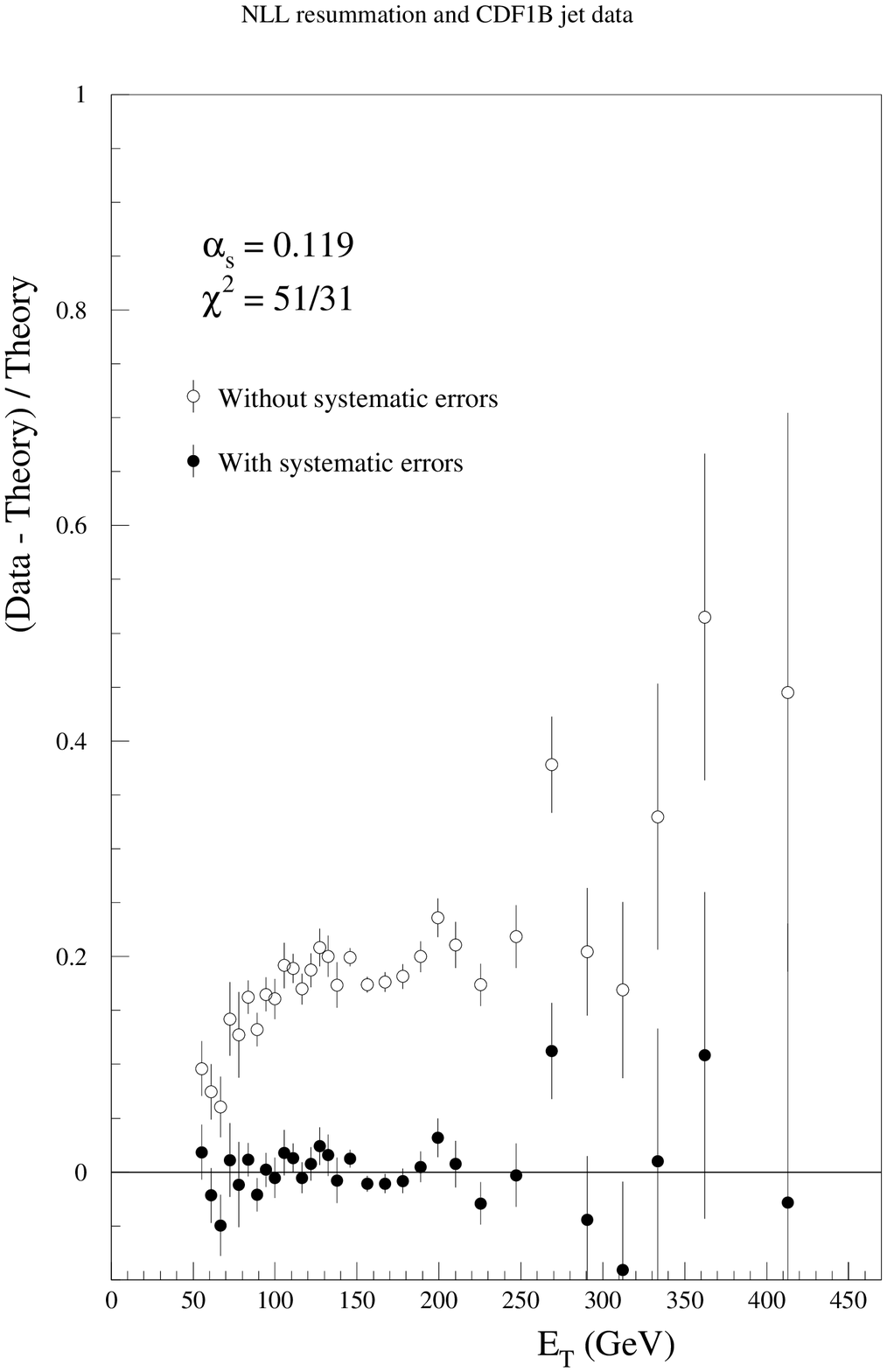}}
\caption{The comparison of theory to the jet data for the NLL+ parton
distributions using coefficient functions at NLO in $\alpha_S$.} 
\label{jets2}
\end{center}
\end{figure}

The fit quality for the jet data is now very good, as expected from the gluon modification. After transforming the gluons to the $\msbar$ scheme, all three fits give $\chi^2\simeq 120$ for 113 points (about as good as is possible in a global fit) with the full NLL fit giving $\chi^2_{jets}=119$. The comparison of the predictions using these partons to the data is shown in \ref{jets2}, and gives $\chi^2=68/82$ for the D0 data and $\chi^2=51/31$ for the CDF data. The situation is very similar for the other two parton sets. We note that the 
quality of the fit to the rest of the data has deteriorated by $50-100$ units 
of $\chi^2$ by the constraint that the high-$x$ DIS scheme gluon be positive,
but the improvement in the fit to the Tevatron jets data is over 200 units 
better, so it is clear that the imposition of the positive high-$x$ DIS scheme
gluon is demanded by a global fit. 
The fact that the jet data has not been explicitly implemented in the fit is further evidence that a positive semi-definite gluon in the DIS scheme automatically leads to a good fit to the Tevatron data in the $\msbar$ scheme \cite{MRSTgluon}. Indeed, the gluon in the DIS scheme is still very small at high $x$,
i.e. $\sim (1-x)^{\eta}$, where $\eta = 7-10$ in the various sets. It 
is then much larger in the $\msbar$ scheme. Actually including the Tevatron 
jet data in the fit can lead to an further improvement of $5-10$ units for 
this data, with a slightly larger high-$x$ gluon. However, this is at the 
expense of a deterioration of $5-10$ units for the rest of the data, so the 
quality of the global fit is much the same in each case.  \\

The values of $\alpha_S$ obtained in the positive gluon fits are shown in table \ref{couplings+}.
\begin{table}
\begin{center}
\begin{tabular}{|c|c|}
\hline
Fit & $\alpha_S(M_Z^2)$\\
\hline
NLL+ & 0.1188 \\
NLL(2)+ & 0.1196 \\
NLO+ & 0.1195\\
\hline
\end{tabular}
\caption{The values of the strong coupling resulting from each of the next to leading fits with the high $x$ gluon constrained to be positive. The world average (at NLO) is 0.1187(20) \cite{PDG}.}
\label{couplings+}
\end{center}
\end{table}
The NLO coupling has increased, which makes sense given the smaller gluon at moderate and small $x$. The full resummed fit result is in excellent agreement with the NLO world average, despite the tendency of resummations to want to decrease the coupling. The value of $\alpha_S(M_Z^2)$ from the modified fit is very similar to that obtained in the NLO fit, as perhaps is expected from filtering out resummations in the moderate and high $x$ regimes. \\
\section{NLL Resummed Prediction for $F_L$}
\begin{figure}
\begin{center}
\scalebox{0.8}{\includegraphics{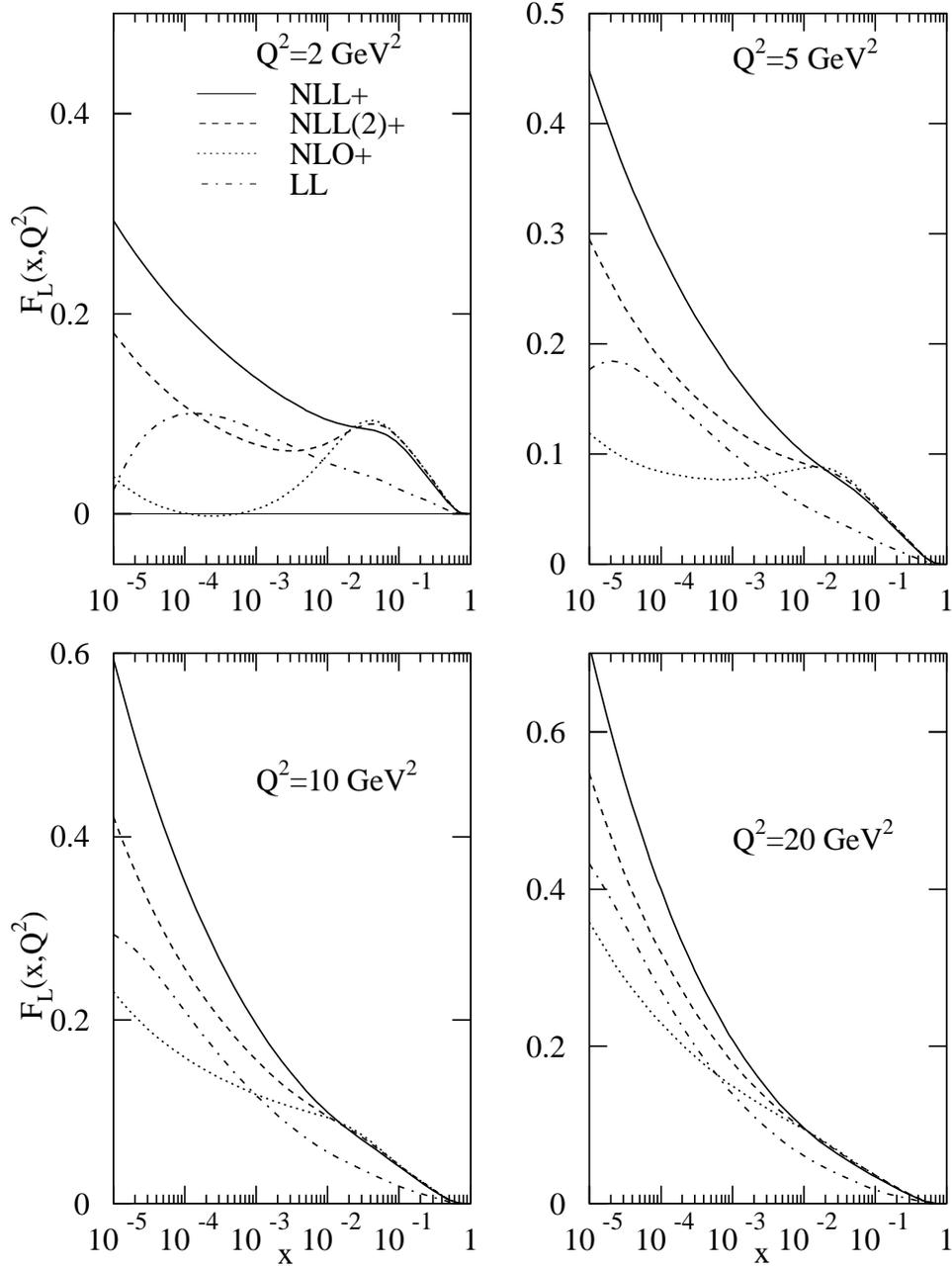}}
\caption{NLL resummed predictions for the longitudinal structure function compared with the NLO DIS-scheme fit.}
\label{flnll}
\end{center}
\end{figure}

With gluons that are more consistent with the Tevatron jet data, we are now in a position to examine the longitudinal structure function. The predictions for $F_L$ using the three global fits of the previous section are shown in figure \ref{flnll}.
The NLO DIS-scheme result, unlike the corresponding $\msbar$-scheme result \cite{ThorneFL}, is positive at low $Q^2$ as $x\rightarrow 0$. However, it still dips slightly negative at moderate $x$ which is unphysical. The NLL resummed result is much more sensible at low $x$, and lies above the NLO result, as can be expected from the increased gluon and coefficient function. The modified resummed approach leads to a prediction for $F_L$ that is very close to the full resummation value at high $x$ as expected, but lower at small $x$ due to the smaller gluon. It catches up with the full resummed result as $Q^2$ increases due to the raised coupling, and shares with the NLO prediction the unsatisfactory feature of being oscillatory (though not nearly as much) at small $Q^2$ due to the negativity of the gluon. This is no longer a problem in the full resummed fit, which predicts that $F_L$ is rising smoothly as $x\rightarrow 0$. \\

We also show in figure \ref{flnll} the LL result (with running coupling corrections) of \cite{WT2}, corrected from that paper due to a mistake in the evolution code. One sees that the LL prediction is turning over as $x\rightarrow0$, becoming negative at low enough $x$ and $Q^2$. This is due to the negativity of the gluon in the LL fit, cured by NLL resummation in the present paper. The LL result dramatically undershoots the fixed order results at moderately high $x$, due to the prevalence of LL small $x$ resummations in that region and to quite large contributions from the NLO coefficient functions. \\

\begin{figure}
\begin{center}
\scalebox{0.8}{\includegraphics{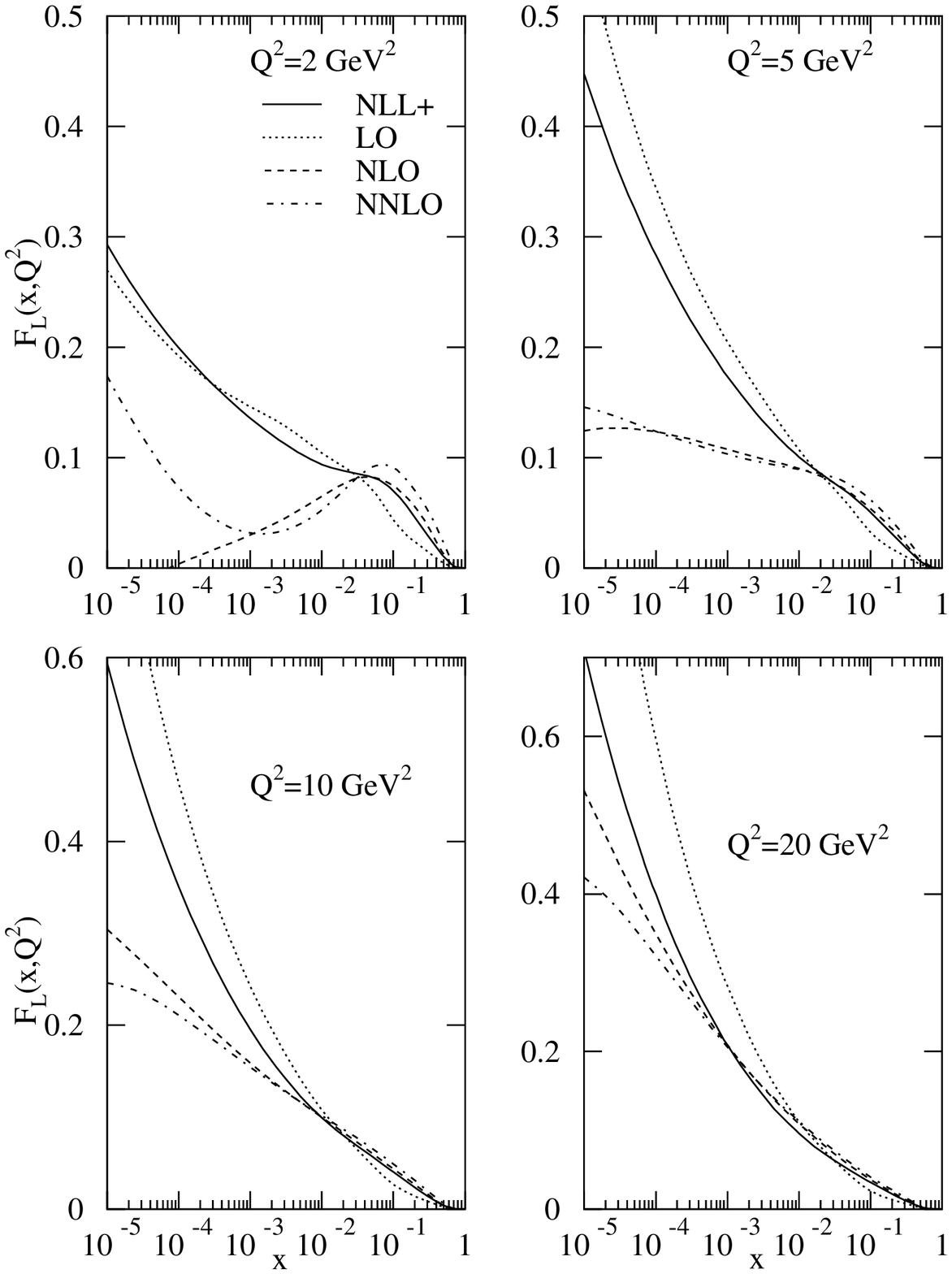}}
\caption{NLL resummed prediction for the longitudinal structure function compared with the fixed order $\msbar$ scheme results.}
\label{flnll2}
\end{center}
\end{figure}

One may also compare the resummed prediction with the $\msbar$ scheme fixed order predictions for $F_L$. Although $F_L$ is formally scheme independent only when considering all orders in the perturbation expansion, comparing the resummed DIS-scheme prediction with the results obtained using the first three orders of the $\msbar$ scheme expansion gives some indication of the stability of the resummed result. This is shown in figure \ref{flnll2}.
The resummed result lies above the $\msbar$ scheme NLO and NNLO results at intermediate $Q^2$, again due to the increased gluon in the moderate $x$ regime (NLO $\msbar$ gluon is negative at moderate and small $x$ -- see \cite{MRST2001}). At higher $Q^2$ it is similar in magnitude to the $\msbar$ scheme NLO result and one can clearly see that the problem of severely undershooting the fixed order results at moderate $x$ has been alleviated by going to NLL order in the resummation\footnote{The fact that the resummed $F_L$ undershoots slightly the fixed order results in figure \ref{flnll2} is due to the different factorisation schemes, as can be seen from figure \ref{flnll} where this effect is lessened considerably.}. \\

\begin{figure}
\begin{center}
\scalebox{0.8}{\includegraphics{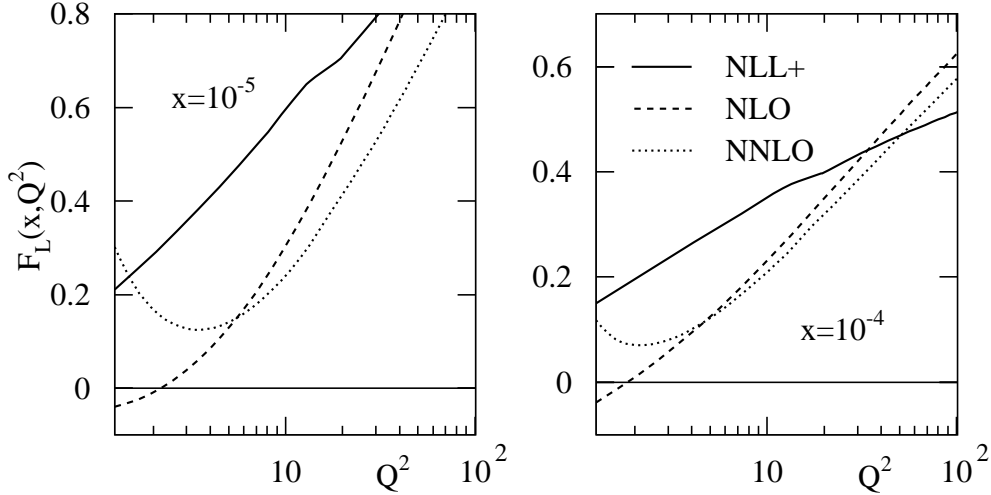}}
\caption{NLL resummed prediction for the longitudinal structure function compared with fixed order $\msbar$ scheme results.}
\label{flnll3}
\end{center}
\end{figure}

The behaviour as $Q^2$ increases at fixed $x$ is shown in figure \ref{flnll3}.
It lies somewhere between the $\msbar$ scheme NLO and NNLO results at very low $Q^2$ and $x$, and is somewhat flatter as $Q^2$ increases due to the 
dip in the gluon splitting function at intermediate $x$, to the reduced 
coupling in the resummed fit, and to the fall off  of the resummed coefficient function.

\section{Discussion and Conclusions}
In this paper the approximate framework for NLL small $x$ resummations introduced in \cite{WT1,WPT} has been used with the approach of \cite{Thorne,WT2} to produce NLL+NLO splitting and coefficient functions in both the massless and heavy sectors. Our results for the splitting function $P_{gg}$ are very similar to those of alternative approaches \cite{ABF,CCSS}, even though we do not resum collinearly enhanced terms in the BFKL kernel \cite{Salam}. We stress, however, that in processes where there are two perturbative scales at either end of the BFKL 4-point function such a kernel resummation would indeed be appropriate. \\

We have successfully applied our results to scattering data, and a new DIS-scheme global fit at NLO in the fixed order QCD expansion has been provided for comparison. An excellent overall fit is obtained with the resummations, and a marked improvement over the fixed order results is observed. Particularly appealing is the positive definite gluon obtained in the NLL fit. Whilst not a necessary condition for the gluon density, a positive gluon avoids the problem of negative structure functions -- exhibited by the NLO prediction for $F_L$, which in both the $\msbar$ and DIS schemes is negative for some range of $x$. Indeed, even if the overall fit quality were the same for the NLL and NLO fits one would probably consider the NLL fit as the better description given the more physical form of the gluon. This is also evident in the prediction for $F_L$, in which the NLO fit shows an oscillatory behaviour at moderate $x$ whereas the resummed prediction grows more sensibly. \\

This more physical gluon also provides a more sensible starting point 
for an investigation of nonlinear corrections to the gluon evolution at small
$x$ \cite{Gribov,Mueller}. 
It seems most appropriate to approach this problem considering 
additional corrections on top of the, in comparison, well understood leading 
twist framework where in most regions of parameters space the partons are 
determined with great precision. This has always proved to be problematic
when using as a starting point the NLO or NNLO gluon distributions from 
global fits since these always become very small, or frequently negative, at 
$Q^2 \sim 2$ GeV$^2$ and $x \sim 10^{-4}$. Consequently saturation due 
to large parton densities is largely counterintuitive. Sometimes LO in 
$\alpha_S$ partons are used, but these are qualitatively inaccurate at 
small $x$ and $Q^2$. The small $x$ gluon produced here represents a 
more complete picture of the leading twist partons at very small $x$ and
should be a more physical quantity to work with when looking at additional 
corrections, particularly since many of the recent approaches are extensions 
of the leading twist physics within the BFKL framework 
\cite{Balitsky,Kovchegov,Weigert,Iancu}. The gluon in this 
paper would also be more appropriate for use in approaches which need 
unintegrated gluon distributions, e.g. \cite{Teubner}, rather than obtaining 
this from fixed order perturbative gluons. The latter have not been obtained
from the global fit in the same manner in which they are being used, whereas 
the gluon in this paper has been obtained by performing $k_T$ integrations
over impact factors in a more compatible manner.     \\

The best fits obtained to the data sets produce a negative gluon at high $x$. Such a feature is not consistent with the Tevatron jet data which wants a positive semi-definite gluon in the DIS scheme at high $x$, leading to a large gluon in the $\msbar$ scheme. One can constrain the gluon at high $x$ to be positive semi-definite, and a very good overall fit is still obtained. The comparison to the jet data is then excellent, even though the Tevatron data is not explicitly included in the global fit. Again, the presence of small $x$ resummations is noticeably preferred. \\

The resummed splitting and coefficient functions contain important high $x$ terms, which are really artifacts of an expansion about $N=0$. Hence it is tempting to filter out these large $x$ terms with a function such as that shown in figure \ref{damp}. However, in each of the fits this approach does not perform well, giving only a slight improvement over the fixed order description. The results are similar to, but slightly different from, the NLO results at moderate and high $x$. Predictions for structure functions do not approach the full NLL results until very low $x$. Thus there is not enough benefit from the small $x$ resummation. Also, in neglecting the Dirac-type terms from the resummed splitting functions, important information is left out. The H1 data in particular benefits from inclusion of the high $x$ terms in the resummed splitting functions. Therefore, it is not really correct to think of this information as being at ``high $x$''. It really is an artifact of a resummed expansion about $N=0$, and all of the resummation thus obtained must be implemented.\\

The NLL resummed fit is also better than the LL resummed fit of \cite{WT2}. The global fit results, gluon distribution and $F_L$ prediction all indicate that the shortcoming of the LL approach -- that the resummations were too prevalent at moderate and high $x$ -- has been eradicated. Indeed, the LL fit is somewhat misleading in that a large momentum violation ($-22\%$) in the input partons was necessary. All of the fits described in this paper have input partons which carry $100\%$ momentum, and the evolution in the NLL evolution conserves momentum to a very good approximation\footnote{We have explicity checked momentum conservation by integrating the partons at various values of $Q^2$. We find that the momentum sum rule is violated by no more than 3.5\%. The effect is 
cumulative and this value is only reached for the highest $Q^2$. 
For much of the 
small $x$ HERA range the $Q^2$ is such that the violation is less than 1\%.}.\\

The fit to the charm data is greatly improved by implementing the (NLL resummed) DIS($\chi$) scheme. The resummed prediction for $F_2^c$ lies towards the lower end of the range allowed by the data except for at very low $x$, but leads to an excellent fit. In fact the features are similar to those seen when 
including the NNLO corrections \cite{Thorne06}.  \\

Further improvements to the NLL resummation are possible. The massless splitting and coefficient functions can in principle be updated once the full NLL impact factors are known. However, we have seen that this is not the most important correction. Furthermore, the massive coefficient functions may benefit from improved modelling of the NLL impact factor effects (although these have a much smaller effect on the overall fit). Nevertheless, the conclusion that small $x$ resummations are beneficial in a global fit to scattering data is a firm one, made even more so by the (slightly) approximate nature of the approach used here. The partons are available from the authors in grid form in the 
direct DIS scheme and also in the $\msbar$ scheme obtained 
via a fixed order transformation. However, these would only be physically meaningful at large $Q^2$ or for a range of (high) 
$x$ where resummation effects in the $\msbar\rightarrow\DIS$ transformation are expected to be less important. \\

To conclude, we believe the form of the gluon together with the $\chi^2$ improvement in the resummed global fit provide extremely compelling evidence that small $x$ resummations are a necessary and beneficial prerequisite for describing the HERA data, and for predicting high energy partons for the LHC. 
\section{Acknowledgments}
CDW is grateful to PPARC for a research studentship and also to the Cavendish Laboratory, Cambridge for hospitality. He would like to thank Jeppe Andersen for useful conversations; also Jeff Forshaw and Albrecht Kyrieleis for discussions, some time ago, about the NLL impact factors.

\bibliography{refs}

\providecommand{\href}[2]{#2}\begingroup\raggedright\begin{thebibliography}{10}

\bibitem{H1a}
{\bf H1} Collaboration, C.~Adloff {\em et al.} {\em Eur. Phys. J.} {\bf C30}
  (2003) 1--32,
\href{http://www.arXiv.org/abs/hep-ex/0304003}{{\tt hep-ex/0304003}}.
%%CITATION = HEP-EX 0304003;%%.

\bibitem{ZEUSa}
{\bf ZEUS} Collaboration, S.~Chekanov {\em et al.} {\em Phys. Rev.} {\bf D70}
  (2004) 052001,
\href{http://www.arXiv.org/abs/hep-ex/0401003}{{\tt hep-ex/0401003}}.
%%CITATION = HEP-EX 0401003;%%.

\bibitem{BFKL}
{L.N. Lipatov, \textit{\it Sov. J. Nucl. Phys.}, \textbf{23} (1976) 338{;}{\\}
  E.A. Kuraev, L.N. Lipatov and V.S. Fadin, \textit{Sov. Phys. JETP},
  \textbf{45} (1977) 199{;}{\\} Ya.Ya. Balitsky and L.N. Lipatov, \textit{Sov.
  J. Nucl. Phys.} \textbf{28} (1978) 822.}

\bibitem{Fadin}
V.~S. Fadin and L.~N. Lipatov {\em Phys. Lett.} {\bf B429} (1998) 127--134,
\href{http://www.arXiv.org/abs/hep-ph/9802290}{{\tt hep-ph/9802290}}.
%%CITATION = HEP-PH 9802290;%%.

\bibitem{Camici}
G.~Camici and M.~Ciafaloni {\em Phys. Lett.} {\bf B412} (1997) 396--406,
\href{http://www.arXiv.org/abs/hep-ph/9707390}{{\tt hep-ph/9707390}}.
%%CITATION = HEP-PH 9707390;%%.

\bibitem{Vogt_c}
J.~A.~M. Vermaseren, A.~Vogt, and S.~Moch {\em Nucl. Phys.} {\bf B724} (2005)
  3--182,
\href{http://www.arXiv.org/abs/hep-ph/0504242}{{\tt hep-ph/0504242}}.
%%CITATION = HEP-PH 0504242;%%.

\bibitem{MRSTerrors}
A.~D. Martin, R.~G. Roberts, W.~J. Stirling, and R.~S. Thorne {\em Eur. Phys.
  J.} {\bf C35} (2004) 325--348,
\href{http://www.arXiv.org/abs/hep-ph/0308087}{{\tt hep-ph/0308087}}.
%%CITATION = HEP-PH 0308087;%%.

\bibitem{ThorneFL}
R.~S. Thorne
\href{http://www.arXiv.org/abs/hep-ph/0511351}{{\tt hep-ph/0511351}}.
%%CITATION = HEP-PH 0511351;%%.

\bibitem{WT2}
C.~D. White and R.~S. Thorne {\em Phys. Rev.} {\bf D74} (2006) 014002,
\href{http://www.arXiv.org/abs/hep-ph/0603030}{{\tt hep-ph/0603030}}.
%%CITATION = HEP-PH 0603030;%%.

\bibitem{Thorne}
R.~S. Thorne {\em Phys. Rev.} {\bf D64} (2001) 074005,
\href{http://www.arXiv.org/abs/hep-ph/0103210}{{\tt hep-ph/0103210}}.
%%CITATION = HEP-PH 0103210;%%.

\bibitem{Bartels04}
J.~Bartels and A.~Kyrieleis {\em Phys. Rev.} {\bf D70} (2004) 114003,
\href{http://www.arXiv.org/abs/hep-ph/0407051}{{\tt hep-ph/0407051}}.
%%CITATION = HEP-PH 0407051;%%.

\bibitem{Bartels02}
J.~Bartels, D.~Colferai, S.~Gieseke, and A.~Kyrieleis {\em Phys. Rev.} {\bf
  D66} (2002) 094017,
\href{http://www.arXiv.org/abs/hep-ph/0208130}{{\tt hep-ph/0208130}}.
%%CITATION = HEP-PH 0208130;%%.

\bibitem{Bartels01}
J.~Bartels, S.~Gieseke, and A.~Kyrieleis {\em Phys. Rev.} {\bf D65} (2002)
  014006,
\href{http://www.arXiv.org/abs/hep-ph/0107152}{{\tt hep-ph/0107152}}.
%%CITATION = HEP-PH 0107152;%%.

\bibitem{Bartels00}
J.~Bartels, S.~Gieseke, and C.~F. Qiao {\em Phys. Rev.} {\bf D63} (2001)
  056014,
\href{http://www.arXiv.org/abs/hep-ph/0009102}{{\tt hep-ph/0009102}}.
%%CITATION = HEP-PH 0009102;%%.

\bibitem{Fadin02}
V.~S. Fadin, D.~Y. Ivanov, and M.~I. Kotsky {\em Nucl. Phys.} {\bf B658} (2003)
  156--174,
\href{http://www.arXiv.org/abs/hep-ph/0210406}{{\tt hep-ph/0210406}}.
%%CITATION = HEP-PH 0210406;%%.

\bibitem{Fadin01}
V.~S. Fadin, D.~Y. Ivanov, and M.~I. Kotsky {\em Phys. Atom. Nucl.} {\bf 65}
  (2002) 1513--1527,
\href{http://www.arXiv.org/abs/hep-ph/0106099}{{\tt hep-ph/0106099}}.
%%CITATION = HEP-PH 0106099;%%.

\bibitem{Catani2}
S.~Catani {\em Proc. DIS 1996:165-171} (1996)
\href{http://www.arXiv.org/abs/hep-ph/9608310}{{\tt hep-ph/9608310}}.
%%CITATION = HEP-PH 9608310;%%.

\bibitem{Peschanski}
A.~Bialas, H.~Navelet, and R.~Peschanski {\em Nucl. Phys.} {\bf B593} (2001)
  438--450,
\href{http://www.arXiv.org/abs/hep-ph/0009248}{{\tt hep-ph/0009248}}.
%%CITATION = HEP-PH 0009248;%%.

\bibitem{WT1}
C.~D. White and R.~S. Thorne {\em Eur. Phys. J.} {\bf C45} (2006) 179--192,
\href{http://www.arXiv.org/abs/hep-ph/0507244}{{\tt hep-ph/0507244}}.
%%CITATION = HEP-PH 0507244;%%.

\bibitem{WPT}
C.~D. White, R.~Peschanski, and R.~S. Thorne {\em Phys. Lett.} {\bf B639}
  (2006) 652--660,
\href{http://www.arXiv.org/abs/hep-ph/0606169}{{\tt hep-ph/0606169}}.
%%CITATION = HEP-PH 0606169;%%.

\bibitem{ABF}
G.~Altarelli, R.~D. Ball, and S.~Forte {\em Nucl. Phys.} {\bf B742} (2006)
  1--40,
\href{http://www.arXiv.org/abs/hep-ph/0512237}{{\tt hep-ph/0512237}}.
%%CITATION = HEP-PH 0512237;%%.

\bibitem{CCSS}
M.~Ciafaloni, D.~Colferai, G.~P. Salam, and A.~M. Stasto {\em Phys. Rev.} {\bf
  D68} (2003) 114003,
\href{http://www.arXiv.org/abs/hep-ph/0307188}{{\tt hep-ph/0307188}}.
%%CITATION = HEP-PH 0307188;%%.

\bibitem{Ross}
D.~A. Ross {\em Phys. Lett.} {\bf B431} (1998) 161--165,
\href{http://www.arXiv.org/abs/hep-ph/9804332}{{\tt hep-ph/9804332}}.
%%CITATION = HEP-PH 9804332;%%.

\bibitem{Salam}
G.~P. Salam {\em JHEP} {\bf 07} (1998) 019,
\href{http://www.arXiv.org/abs/hep-ph/9806482}{{\tt hep-ph/9806482}}.
%%CITATION = HEP-PH 9806482;%%.

\bibitem{Colferai}
M.~Ciafaloni and D.~Colferai {\em Phys. Lett.} {\bf B452} (1999) 372--378,
\href{http://www.arXiv.org/abs/hep-ph/9812366}{{\tt hep-ph/9812366}}.
%%CITATION = HEP-PH 9812366;%%.

\bibitem{Whiteconf}
C.~D. White
\href{http://www.arXiv.org/abs/hep-ph/0605321}{{\tt hep-ph/0605321}}.
%%CITATION = HEP-PH 0605321;%%.

\bibitem{CollinskT}
J.~C. Collins and R.~K. Ellis {\em Nucl. Phys. Proc. Suppl.} {\bf 18C} (1991)
80--85.
%%CITATION = NUPHZ,18C,80;%%.

\bibitem{CatanikT}
S.~Catani, M.~Ciafaloni, and F.~Hautmann {\em Phys. Lett.} {\bf B242} (1990)
97.
%%CITATION = PHLTA,B242,97;%%.

\bibitem{Altarelli}
G.~Altarelli, R.~K. Ellis, and G.~Martinelli {\em Nucl. Phys.} {\bf B143}
  (1978)
521.
%%CITATION = NUPHA,B143,521;%%.

\bibitem{Ciafaloni06}
M.~Ciafaloni, D.~Colferai, G.~P. Salam, and A.~M. Stasto {\em Phys. Lett.} {\bf
  B635} (2006) 320--329,
\href{http://www.arXiv.org/abs/hep-ph/0601200}{{\tt hep-ph/0601200}}.
%%CITATION = HEP-PH 0601200;%%.

\bibitem{Catani}
S.~Catani and F.~Hautmann {\em Nucl. Phys.} {\bf B427} (1994) 475--524,
\href{http://www.arXiv.org/abs/hep-ph/9405388}{{\tt hep-ph/9405388}}.
%%CITATION = HEP-PH 9405388;%%.

\bibitem{Ciafaloni}
M.~Ciafaloni and D.~Colferai {\em JHEP} {\bf 09} (2005) 069,
\href{http://www.arXiv.org/abs/hep-ph/0507106}{{\tt hep-ph/0507106}}.
%%CITATION = HEP-PH 0507106;%%.

\bibitem{Salam2}
M.~Ciafaloni, D.~Colferai, G.~P. Salam, and A.~M. Stasto {\em Phys. Rev.} {\bf
  D68} (2003) 114003,
\href{http://www.arXiv.org/abs/hep-ph/0307188}{{\tt hep-ph/0307188}}.
%%CITATION = HEP-PH 0307188;%%.

\bibitem{ABFimpact}
G.~Altarelli, R.~D. Ball, and S.~Forte {\em Nucl. Phys.} {\bf B599} (2001)
  383--423,
\href{http://www.arXiv.org/abs/hep-ph/0011270}{{\tt hep-ph/0011270}}.
%%CITATION = HEP-PH 0011270;%%.

\bibitem{Buza}
M.~Buza, Y.~Matiounine, J.~Smith, and W.~L. van Neerven {\em Eur. Phys. J.}
  {\bf C1} (1998) 301--320,
\href{http://www.arXiv.org/abs/hep-ph/9612398}{{\tt hep-ph/9612398}}.
%%CITATION = HEP-PH 9612398;%%.

\bibitem{H1b}
{\bf H1} Collaboration, C.~Adloff {\em et al.} {\em Eur. Phys. J.} {\bf C21}
  (2001) 33--61,
\href{http://www.arXiv.org/abs/hep-ex/0012053}{{\tt hep-ex/0012053}}.
%%CITATION = HEP-EX 0012053;%%.

\bibitem{H1c}
{\bf H1} Collaboration, C.~Adloff {\em et al.} {\em Eur. Phys. J.} {\bf C19}
  (2001) 269--288,
\href{http://www.arXiv.org/abs/hep-ex/0012052}{{\tt hep-ex/0012052}}.
%%CITATION = HEP-EX 0012052;%%.

\bibitem{ZEUSb}
{\bf ZEUS} Collaboration, J.~Breitweg {\em et al.} {\em Eur. Phys. J.} {\bf C7}
  (1999) 609--630,
\href{http://www.arXiv.org/abs/hep-ex/9809005}{{\tt hep-ex/9809005}}.
%%CITATION = HEP-EX 9809005;%%.

\bibitem{ZEUSc}
{\bf ZEUS} Collaboration, S.~Chekanov {\em et al.} {\em Eur. Phys. J.} {\bf
  C21} (2001) 443--471,
\href{http://www.arXiv.org/abs/hep-ex/0105090}{{\tt hep-ex/0105090}}.
%%CITATION = HEP-EX 0105090;%%.

\bibitem{BCDMSep}
{\bf BCDMS} Collaboration, A.~C. Benvenuti {\em et al.} {\em Phys. Lett.} {\bf
  B223} (1989)
485.
%%CITATION = PHLTA,B223,485;%%.

\bibitem{NMC}
{\bf New Muon} Collaboration, M.~Arneodo {\em et al.} {\em Nucl. Phys.} {\bf
  B483} (1997) 3--43,
\href{http://www.arXiv.org/abs/hep-ph/9610231}{{\tt hep-ph/9610231}}.
%%CITATION = HEP-PH 9610231;%%.

\bibitem{SLAC}
L.~W. Whitlow, E.~M. Riordan, S.~Dasu, S.~Rock, and A.~Bodek {\em Phys. Lett.}
  {\bf B282} (1992)
475--482.
%%CITATION = PHLTA,B282,475;%%.

\bibitem{SLAC2}
L.~W. Whitlow. SLAC-0357.

\bibitem{E665}
{\bf E665} Collaboration, M.~R. Adams {\em et al.} {\em Phys. Rev.} {\bf D54}
  (1996)
3006--3056.
%%CITATION = PHRVA,D54,3006;%%.

\bibitem{BCDMSeD}
{\bf BCDMS} Collaboration, A.~C. Benvenuti {\em et al.} {\em Phys. Lett.} {\bf
  B237} (1990)
592.
%%CITATION = PHLTA,B237,592;%%.

\bibitem{CCFRF2}
{\bf CCFR/NuTeV} Collaboration, U.-K. Yang {\em et al.} {\em Phys. Rev. Lett.}
  {\bf 86} (2001) 2742--2745,
\href{http://www.arXiv.org/abs/hep-ex/0009041}{{\tt hep-ex/0009041}}.
%%CITATION = HEP-EX 0009041;%%.

\bibitem{CCFRF3}
W.~G. Seligman {\em et al.} {\em Phys. Rev. Lett.} {\bf 79} (1997)
1213--1216.
%%CITATION = PRLTA,79,1213;%%.

\bibitem{NMCrat}
{\bf New Muon} Collaboration, M.~Arneodo {\em et al.} {\em Nucl. Phys.} {\bf
  B487} (1997) 3--26,
\href{http://www.arXiv.org/abs/hep-ex/9611022}{{\tt hep-ex/9611022}}.
%%CITATION = HEP-EX 9611022;%%.

\bibitem{ZEUSCC}
{\bf ZEUS} Collaboration, S.~Chekanov {\em et al.} {\em Eur. Phys. J.} {\bf
  C32} (2003) 1--16,
\href{http://www.arXiv.org/abs/hep-ex/0307043}{{\tt hep-ex/0307043}}.
%%CITATION = HEP-EX 0307043;%%.

\bibitem{F2c1}
{\bf H1} Collaboration, C.~Adloff {\em et al.} {\em Z. Phys.} {\bf C72} (1996)
  593--605,
\href{http://www.arXiv.org/abs/hep-ex/9607012}{{\tt hep-ex/9607012}}.
%%CITATION = HEP-EX 9607012;%%.

\bibitem{F2c2}
{\bf ZEUS} Collaboration, J.~Breitweg {\em et al.} {\em Eur. Phys. J.} {\bf
  C12} (2000) 35--52,
\href{http://www.arXiv.org/abs/hep-ex/9908012}{{\tt hep-ex/9908012}}.
%%CITATION = HEP-EX 9908012;%%.

\bibitem{DY}
{\bf NuSea} Collaboration, J.~C. Webb {\em et al.}
\href{http://www.arXiv.org/abs/hep-ex/0302019}{{\tt hep-ex/0302019}}.
%%CITATION = HEP-EX 0302019;%%.

\bibitem{DYasym}
{\bf NA51} Collaboration, A.~Baldit {\em et al.} {\em Phys. Lett.} {\bf B332}
  (1994)
244--250.
%%CITATION = PHLTA,B332,244;%%.

\bibitem{DYrat}
{\bf FNAL E866/NuSea} Collaboration, R.~S. Towell {\em et al.} {\em Phys. Rev.}
  {\bf D64} (2001) 052002,
\href{http://www.arXiv.org/abs/hep-ex/0103030}{{\tt hep-ex/0103030}}.
%%CITATION = HEP-EX 0103030;%%.

\bibitem{Wasym}
{\bf CDF} Collaboration, F.~Abe {\em et al.} {\em Phys. Rev. Lett.} {\bf 81}
  (1998) 5754--5759,
\href{http://www.arXiv.org/abs/hep-ex/9809001}{{\tt hep-ex/9809001}}.
%%CITATION = HEP-EX 9809001;%%.

\bibitem{Tevjet1}
{\bf D0} Collaboration, B.~Abbott {\em et al.} {\em Phys. Rev. Lett.} {\bf 86}
  (2001) 1707--1712,
\href{http://www.arXiv.org/abs/hep-ex/0011036}{{\tt hep-ex/0011036}}.
%%CITATION = HEP-EX 0011036;%%.

\bibitem{Tevjet2}
{\bf CDF} Collaboration, T.~Affolder {\em et al.} {\em Phys. Rev.} {\bf D64}
  (2001) 032001,
\href{http://www.arXiv.org/abs/hep-ph/0102074}{{\tt hep-ph/0102074}}.
%%CITATION = HEP-PH 0102074;%%.

\bibitem{Laenen}
E.~Laenen, S.~Riemersma, J.~Smith, and W.~L. van Neerven {\em Nucl. Phys.} {\bf
  B392} (1993)
229--250.
%%CITATION = NUPHA,B392,229;%%.

\bibitem{Riemersma}
S.~Riemersma, J.~Smith, and W.~L. van Neerven {\em Phys. Lett.} {\bf B347}
  (1995) 143--151,
\href{http://www.arXiv.org/abs/hep-ph/9411431}{{\tt hep-ph/9411431}}.
%%CITATION = HEP-PH 9411431;%%.

\bibitem{Thorne06}
R.~S. Thorne {\em Phys. Rev.} {\bf D73} (2006) 054019,
\href{http://www.arXiv.org/abs/hep-ph/0601245}{{\tt hep-ph/0601245}}.
%%CITATION = HEP-PH 0601245;%%.

\bibitem{TRCC}
R.~S. Thorne and R.~G. Roberts {\em Eur. Phys. J.} {\bf C19} (2001) 339--349,
\href{http://www.arXiv.org/abs/hep-ph/0010344}{{\tt hep-ph/0010344}}.
%%CITATION = HEP-PH 0010344;%%.

\bibitem{Marciano}
W.~J. Marciano {\em Phys. Rev.} {\bf D29} (1984)
580.
%%CITATION = PHRVA,D29,580;%%.

\bibitem{PDG}
{\bf Particle Data Group} Collaboration, S.~Eidelman {\em et al.} {\em Phys.
  Lett.} {\bf B592} (2004)
1.
%%CITATION = PHLTA,B592,1;%%.

\bibitem{MRSTgluon}
A.~D. Martin, R.~G. Roberts, W.~J. Stirling, and R.~S. Thorne {\em Phys. Lett.}
  {\bf B604} (2004) 61--68,
\href{http://www.arXiv.org/abs/hep-ph/0410230}{{\tt hep-ph/0410230}}.
%%CITATION = HEP-PH 0410230;%%.

\bibitem{MRST2001}
A.~D. Martin, R.~G. Roberts, W.~J. Stirling, and R.~S. Thorne {\em Eur. Phys.
  J.} {\bf C23} (2002) 73--87,
\href{http://www.arXiv.org/abs/hep-ph/0110215}{{\tt hep-ph/0110215}}.
%%CITATION = HEP-PH 0110215;%%.

\bibitem{CTEQ}
J.~Pumplin {\em et al.} {\em JHEP} {\bf 07} (2002) 012,
\href{http://www.arXiv.org/abs/hep-ph/0201195}{{\tt hep-ph/0201195}}.
%%CITATION = HEP-PH 0201195;%%.

\bibitem{Gribov}
L.~V. Gribov, E.~M. Levin, and M.~G. Ryskin {\em Phys. Rept.} {\bf 100} (1983)
1--150.
%%CITATION = PRPLC,100,1;%%.

\bibitem{Mueller}
A.~H. Mueller and J.~W. Qiu {\em Nucl. Phys.} {\bf B268} (1986)
427.
%%CITATION = NUPHA,B268,427;%%.

\bibitem{Balitsky}
I.~Balitsky {\em Nucl. Phys.} {\bf B463} (1996) 99--160,
\href{http://www.arXiv.org/abs/hep-ph/9509348}{{\tt hep-ph/9509348}}.
%%CITATION = HEP-PH 9509348;%%.

\bibitem{Kovchegov}
Y.~V. Kovchegov {\em Phys. Rev.} {\bf D60} (1999) 034008,
\href{http://www.arXiv.org/abs/hep-ph/9901281}{{\tt hep-ph/9901281}}.
%%CITATION = HEP-PH 9901281;%%.

\bibitem{Weigert}
H.~Weigert {\em Nucl. Phys.} {\bf A703} (2002) 823--860,
\href{http://www.arXiv.org/abs/hep-ph/0004044}{{\tt hep-ph/0004044}}.
%%CITATION = HEP-PH 0004044;%%.

\bibitem{Iancu}
E.~Iancu, A.~Leonidov, and L.~D. McLerran {\em Nucl. Phys.} {\bf A692} (2001)
  583--645,
\href{http://www.arXiv.org/abs/hep-ph/0011241}{{\tt hep-ph/0011241}}.
%%CITATION = HEP-PH 0011241;%%.

\bibitem{Teubner}
T.~Teubner. Prepared for Ringberg Workshop on New Trends in HERA Physics 2005,
  Ringberg Castle, Tegernsee, Germany, 2-7 Oct 2005.

\end{thebibliography}\endgroup
\end{document}